\documentclass[letterpaper,twocolumn]{aastex6}
\shorttitle{Properties of K2 Planet Host Stars II: The Planets}
\shortauthors{Dressing et al.}
\usepackage{epsfig}
\usepackage{amsmath}
\usepackage{rotating}
\usepackage{natbib}
\usepackage{enumerate}
\usepackage{hyperref}
\usepackage{url}
\usepackage{longtable}
\usepackage{color}

\begin{document}
\def\mearth{{\rm\,M_\oplus}}                                                    
\def\msun{{\rm\,M_\odot}}                                                       
\def\rsun{{\rm\,R_\odot}}                                                       
\def\rearth{{\rm\,R_\oplus}} ннн
\def\fearth{{\rm\,F_\oplus}}
\def\lsun{{\rm\,L_\odot}}                                                          
\def\kepler {{\emph{Kepler}\,}}                                              
\newcommand{\teff}{\ensuremath{T_{\mathrm{eff}}}}                               
\newcommand{\logg}{\ensuremath{\log g}}

\newcommand{\tc}[2][red]{\textcolor{#1}{\emph{\textbf{#2}}}}


\title{Characterizing K2 Candidate Planetary Systems Orbiting Low-Mass Stars II: Planetary Systems Observed During Campaigns 1--7}
\author{Courtney D. Dressing\altaffilmark{1,2,3}}
\author{Andrew Vanderburg\altaffilmark{4,5}}
\author{Joshua E. Schlieder\altaffilmark{6,7}}
\author{Ian J. M. Crossfield\altaffilmark{2,8}}
\author{Heather A. Knutson\altaffilmark{1}}
\author{Elisabeth R. Newton\altaffilmark{9,10}}
\author{David R. Ciardi\altaffilmark{6}}
\author{Benjamin J. Fulton\altaffilmark{11,12}}
\author{Erica J. Gonzales\altaffilmark{5,13,14}}
\author{Andrew W. Howard\altaffilmark{12}}
\author{Howard Isaacson\altaffilmark{15}}
\author{John Livingston\altaffilmark{16}}
\author{Erik A. Petigura\altaffilmark{1,17}}
\author{Evan Sinukoff\altaffilmark{11, 12}}
\author{Mark Everett\altaffilmark{18}}
\author{Elliott Horch\altaffilmark{19}}
\author{Steve B. Howell\altaffilmark{20}}
\altaffiltext{1}{Division of Geological \& Planetary Sciences, California Institute of Technology, Pasadena, CA 91125, USA}
\altaffiltext{2}{NASA Sagan Fellow}
\altaffiltext{3}{{\tt dressing@caltech.edu}}
\altaffiltext{4}{Harvard-Smithsonian Center for Astrophysics, Cambridge, MA 02138, USA}
\altaffiltext{5}{National Science Foundation Graduate Research Fellow}
\altaffiltext{6}{IPAC-NExScI, Mail Code 100-22, Caltech, 1200 E. California Blvd., Pasadena, CA 91125, USA}
\altaffiltext{7}{NASA Goddard Space Flight Center, Greenbelt, MD 20771, USA}
\altaffiltext{8}{Department of Astronomy and Astrophysics, University of California, Santa Cruz, CA 95064, USA}
\altaffiltext{9}{Department of Physics, Massachusetts Institute of Technology, Cambridge, MA 02139, USA}
\altaffiltext{10}{National Science Foundation Astronomy \& Astrophysics Postdoctoral Fellow}
\altaffiltext{11}{Institute for Astronomy, University of Hawai`i at M\={a}noa, Honolulu, HI 96822, USA}
\altaffiltext{12}{Cahill Center for Astrophysics, California Institute of Technology, Pasadena, CA 91125, USA}
\altaffiltext{13}{Department of Physics, University of Notre Dame, Notre Dame, IN 46556, USA}
\altaffiltext{14}{Deans' Graduate Fellow}
\altaffiltext{15}{Astronomy Department, University of California, Berkeley, CA 94720, USA}
\altaffiltext{16}{The University of Tokyo, 7-3-1 Bunkyo-ku, Tokyo 113-0033, Japan}
\altaffiltext{17}{NASA Hubble Fellow}
\altaffiltext{18}{National Optical Astronomy Observatory, 950 North Cherry Avenue, Tucson, AZ 85719, USA}
\altaffiltext{19}{Department of Physics, Southern Connecticut State University, 501 Crescent Street, New Haven, CT 06515, USA}
\altaffiltext{20}{Space Science and Astrobiology Division, NASA Ames Research Center, Moffett Field, CA 94035, USA}

\vspace{0.5\baselineskip}
\date{\today}
\slugcomment{In preparation}

\begin{abstract}
We recently used near-infrared spectroscopy to improve the characterization of 76~low-mass stars around which K2 had detected 79~candidate transiting planets. Thirty of these worlds were new discoveries that have not previously been published. We calculate the false positive probabilities that the transit-like signals are actually caused by non-planetary astrophysical phenomena and reject five new~transit-like events and three previously reported events as false positives. We also statistically validate 18~planets (eight of which were previously unpublished), confirm the earlier validation of 21~planets, and announce 17~newly discovered planet candidates. Revising the properties of the associated planet candidates based on the updated host star characteristics and refitting the transit photometry, we find that our sample contains 20~planets or planet candidates with radii smaller than $1.25\rearth$, 20~super-Earths ($1.25-2\rearth$), 20~small Neptunes ($2-4\rearth$), three~large Neptunes ($4-6\rearth$), and eight~giant planets ($>6\rearth$). Most of these planets are highly irradiated, but EPIC~206209135.04 (\mbox{K2-72e}, $1.29_{-0.13}^{+0.14}\rearth$), EPIC~211988320.01 ($R_p = 2.86_{-0.15}^{+0.16}\rearth$), and EPIC~212690867.01 ($2.20_{-0.18}^{+0.19}\rearth$) orbit within optimistic habitable zone boundaries set by the ``recent Venus'' inner limit and the ``early Mars'' outer limit. In total, our planet sample includes eight moderately-irradiated $1.5 - 3\rearth$ planet candidates ($F_p \lesssim 20 \fearth$) orbiting brighter stars ($Ks < 11$) that are well-suited for atmospheric investigations with \emph{Hubble}, \emph{Spitzer}, and/or the James Webb Space Telescope. Five validated planets orbit relatively bright stars ($Kp < 12.5$) and are expected to yield radial velocity semi-amplitudes of at least $2\,{\rm m}\,{\rm s}^{-1}$. Accordingly, they are possible targets for radial velocity mass measurement with current facilities or the upcoming generation of red optical and near-infrared high-precision RV spectrographs.
\end{abstract}


\maketitle
\section{Introduction}
Since 2014, the NASA K2 mission has been using the \emph{Kepler} spacecraft to search for transiting planets in 100 sq deg fields along the ecliptic plane. K2 observes $10,000-30,000$ stars in each field for roughly 80 days before moving onto the next field. The placements of each field are driven by the three primary requirements of the extended mission design: (1) K2 must look along the ecliptic plane so that the torque from solar radiation pressure is balanced, (2) sunlight must illuminate the solar panels, and (3) K2 cannot look so close to the Sun that sunlight illuminates the detectors \citep{howell_et_al2014, vancleve_et_al2016}.

During the main \emph{Kepler} mission, the planet search targets were primarily selected by the \emph{Kepler} Science Office with a small contribution from Guest Observer proposals. In contrast, all K2 targets are nominated by members of the community. Although some of the selected target stars are well-characterized, many have poorly estimated properties constrained by only photometry, proper motion, and (when lucky) parallax. The problem of inadequate stellar characterization is particularly dire for the smallest, coolest target stars around which K2 has the highest sensitivity to transiting planets. In order to improve the characterization of low-mass K2 target stars, we are conducting an extensive spectroscopic survey of potentially low-mass K2 target stars. In the first paper in this series \citep[][hereafter D17]{dressing_et_al2017a}, we presented NIR spectra and determined stellar parameters for 144~potentially low-mass stars observed by K2 during Campaigns $0-7$ (30 May 2014 -- 26 December 2015). 

In this paper, we use our previously determined stellar parameters combined with planet candidate lists to generate a catalog of K2 planetary systems orbiting low-mass stars. The structure of the paper is as follows:

In Section~\ref{sec:sources}, we explain the origins of our planet candidate sample and describe the K2 planet candidate lists from which we selected our targets.  We then consult the catalog of stellar parameters presented in D17 and update the host star parameters accordingly in Section~\ref{sec:hosts_update}. Next, we fit the K2 photometry in Section~\ref{sec:transit_fits} to determine the transit parameters for each candidate. In Section~\ref{sec:vespa}, we use the open-source {\tt vespa} software \citep{morton2015} to run a false positive analysis to statistically validate planets and identify likely false positives. In Section~\ref{sec:revised_planets}, we combine the information from our various analyses to revise parameters for K2 planet candidates and validated planets orbiting low-mass dwarfs. We highlight particularly noteworthy individual systems in Section~\ref{sec:discussion} before concluding in Section~\ref{sec:conc}.

\section{Sources of Planet Candidates}
\label{sec:sources}
We obtained NIR spectroscopy of all stars  in our sample because they were initially believed to host transiting planets or because they were close enough to the candidate host star that they might have been responsible for the transit-like signal observed in the light curve. In some cases, subsequent detailed analyses of the K2 photometry or ground-based follow-up observations revealed that the putative transit signals were actually due to false positives. 

The majority of the 74~systems in the cool dwarf sample were selected from unpublished candidate lists provided by A.~Vanderburg (53 stars hosting 59~candidates) or the K2 California Consortium (K2C2; 43~stars hosting 54~candidates); several stars appear on both lists and many of the K2C2 targets were later published in \citet{crossfield_et_al2016}. The cool dwarf sample also includes 20~single-planet systems from \citet{barros_et_al2016}, 16~singles from \citet{pope_et_al2016}, 16~stars hosting 18~candidates reported by \citet{vanderburg_et_al2016}, 3~singles from \citet{adams_et_al2016}, and 5~singles from \citet{montet_et_al2015}, who refined the properties of the planet candidates reported by \citet{foreman-mackey_et_al2015}. Two of the stars in the sample (EPIC~212773309 and EPIC~211694226) are in close proximity to other stars and EPIC~212773309 also displays a clear secondary eclipse.

The 74~systems in our cool dwarf sample host 79~unique K2~Objects of Interest (K2OIs). As of 7 September 2016, 24~of those K2OIs had been confirmed as bona fide planets, 23~had been previously published as planet candidates, two~were classified as false positives, and 30~were new detections. Twenty-seven of the new detections were identified by A.~Vanderburg, eight~were found by K2C2, and five were discovered by both collaborations. In total, our cool dwarf planet sample consists of 60~singles, six~doubles (EPIC~201549860 = \mbox{K2-35}, EPIC~206011691 = \mbox{K2-21}, EPIC~210508766 = \mbox{K2-83}, EPIC~210968143 = \mbox{K2-90}, EPIC~211305568, and EPIC~211331236), one triple (EPIC~211428897) and one quadruple (\mbox{EPIC~206209135 = K2-72}). 

\section{Updates to Planet Host Star Characterization}
\label{sec:hosts_update}
In D17, we applied empirical relations to NIR spectra acquired at IRTF/SpeX and Palomar/TSPEC to revise the classifications of putative low-mass stars harboring potential K2 planet candidates. Of the 144~K2 targets we observed, 49\% were actually contaminating giant stars or hotter dwarfs (typically reddened by interstellar extinction) and 74 (51\%) were bona fide low-mass dwarfs. For the cool dwarfs, we measured a series of equivalent widths and spectral indices and applied empirically-based relations from \citet{newton_et_al2015}, and \citet{mann_et_al2013a, mann_et_al2015} to estimate effective temperatures, radii, masses, metallicities, and luminosities. Our results agree well with those of \citet{martinez_et_al2017}, who used lower resolution NIR spectra from NTT/SOFI to improve the characterization of late-type dwarfs hosting K2 planet candidates. 

In general, we found that our new radii were typically $0.13\rsun$ (39\%) larger than the original estimates provided in the Ecliptic Plane Input Catalog \citep[EPIC,][]{huber_et_al2016}. These changes are unsurprising because, as noted by \citet{huber_et_al2016}, the EPIC values were determined by comparing photometry to stellar models that have been shown to systematically underestimate the temperatures and radii of cool stars \citep[e.g.,][]{kraus_et_al2011, boyajian_et_al2012, feiden+chaboyer2012, spada_et_al2013, vonbraun_et_al2014, newton_et_al2015}. We adopt those new stellar classifications in this paper. 

Although the planet candidate catalogs we consulted typically provided their own estimates of stellar properties, we found that the original stellar classifications provided in the planet candidate catalogs also tended to underestimate the radii and temperatures of the cool dwarfs. For most candidates, the amplitude of the suggested radius change is similar to the $15\%$ radius increase found by \citet{newton_et_al2015} for \emph{Kepler} planet candidates orbiting M~dwarfs with previous radius estimates based on fits to stellar models \citep[e.g.,][]{muirhead_et_al2012b, muirhead_et_al2014, dressing+charbonneau2013, mann_et_al2013c, huber_et_al2014}. 

The exceptions to the general trend of underestimated stellar radii are the parameters provided in \citet{vanderburg_et_al2016} and the associated unpublished Vanderburg lists. Indeed, their estimates are based on empirical relations \citep{casagrande_et_al2008, gonzalez-hernandez+bonifacio2009, boyajian_et_al2013, pecaut+mamajek2013}. For those two sources, the discrepancies between our revised values and their initial values are likely due to the fact that spectroscopic observations help break the degeneracy between stars that are intrinsically red and stars that appear red due to interstellar extinction. 

\section{Improving Transit Fits}
\label{sec:transit_fits}
Some of the planets in our target list have well-determined properties because they were previously published in other catalogs, but others are new. In order to provide a uniform catalog of properties for all of the planets in our sample, we perform our own fits to the K2 photometry to determine updated properties and errors for the full planet sample. 

We began by downloading the K2 Self Flat Fielding (K2SFF) photometry provided by A.~Vanderburg\footnote{\url{https://archive.stsci.edu/prepds/k2sff/}}. The K2SFF pipeline processes K2 photometry by recording the roll angle of the spacecraft during each cadence and removing the correlation between flux and roll angle  \citep{vanderburg+johnson2014, vanderburg_et_al2016}. This procedure yields precision within 60\% of that achieved during the baseline \emph{Kepler} mission for faint stars ($Kp > 12.5$) and \kepler-like performance for brighter stars. Prior to fitting the transits, we re-processed the K2SFF data by simultaneously fitting for the transit light curves, long term variability, and K2 pointing systematics using the procedure described by \citet{vanderburg_et_al2016}.

Next, we ran an MCMC analysis to constrain the time of transit $T_0$, orbital period $P$, planet/star radius ratio $R_p/R_*$, semimajor axis/stellar radius ratio $a/R_*$, inclination $i$, quadratic limb darkening, eccentricity $e$, and longitude of periastron $\omega$. We ran our fits in {\tt Python} and used the {\tt emcee} package \citep{foreman-mackey_et_al2013} to determine the errors on transit parameters. We assessed convergence by measuring the integrated autocorrelation time of our chains and requiring that the MCMC ran for at least ten times longer than the estimated autocorrelation time. 

In order to efficiently sample the full allowed parameter space, we fit the limb darkening parameters using the $q_1, q_2$ coordinate-space defined by \citet{kipping2013c}. We also assumed that the eccentricity distribution of transiting planets followed the Beta distribution found by \citet{kipping2013b} for short period planets ($P < 382$~d, $\alpha = 0.697$, $\beta = 3.27$) and fit for the uniform variate $x_e$ rather than $e$ to enable more efficient sampling of low-eccentricity orbits \citep{kipping2014}.

At each point in the analysis, we computed the likelihood of our transit model by using the {\tt BATMAN} package written by \citet{kreidberg2015} to solve the equations of \citet{mandel+agol2002} and generate a model transit lightcurve. Our K2 light curves were obtained in ``long cadence'' mode using 30-minute integration times, which is relatively long compared to total durations of the transits we model. Accordingly, we employed the ``supersample'' feature of {\tt BATMAN} to generate sample long cadence light curves by modeling the brightness of the star at 1-minute cadence and recording the average of thirty consecutive modeled fluxes.

We restricted our fits to consider $70^\circ < i <  90^\circ$, $0 < R_p/R_* < 1$, $a/R_* > 1$, $0 \leq q_1 \leq 1$, $0 \leq q_2 \leq 1$, $0 < x_e < 1$, and $0^\circ < \omega < 360^\circ$. We further required that K2OIs transit their host stars (i.e., $b \leq 1 + R_P/R_\star$) and imposed priors on the limb darkening coefficients by interpolating the tables produced by \citet{claret_et_al2012} at the temperatures and surface gravities of the host stars.\footnote{\citet{claret_et_al2012} consider multiple methods for computing limb-darkening coefficients; we adopt the coefficients found using the least-square method.} Specifically, we assumed that $u_1$ and $u_2$ were drawn from gaussian distributions with dispersions set by propagating the errors in our stellar parameter estimates. Despite our attention to limb darkening, we note that the dominant contribution to the shape of ingress and egress is the smearing due to the lengthy exposure times used for long cadence \emph{K2} data. 

All of our target stars were spectroscopically characterized by D17. We incorporated our knowledge of the host stars into our transit fits by using Kepler's third law and the estimated host star masses to determine the orbital semimajor axes of planets with the observed orbital periods. We then compared the resulting distances to the estimated stellar radii and set Gaussian priors on the expected $a/R_*$ ratio for each planet \citep{seager+mallen-ornelas2003, sozzetti_et_al2007, torres_et_al2008}. The widths of these priors reflect the uncertainties in the stellar radii and masses, but by imposing this prior we implicitly assume that the candidate transit event is indeed caused by the transit of a planet across the target star rather than a blended transit or eclipse of a contaminating star. The {\tt vespa} false positive analysis discussed in Section~\ref{sec:vespa} should reveal such scenarios.

For a handful of targets, the transit depths were shallow enough that the MCMC sometimes wandered away from the transit signal under consideration. Accordingly, we required that the transit center must be within 6\% of the initial guess (up to a maximum difference of 6 hours) and that the orbital period must be within 0.025~days (36~minutes) of the initial guess. As discussed in Section~\ref{ssec:photcomp}, we later repeated the transit fitting process using the k2phot and K2SC photometry to check for systematic offsets in planet parameters.

\section{Determining K2OI Dispositions}
\label{sec:vespa}
Early in the \emph{Kepler} mission, transiting planet candidates were ``confirmed'' by conducting radial velocity or transit timing variation studies to measure planet masses. However, mass measurements are expensive and not feasible for all systems. As a result, tools like {\tt BLENDER} \citep[e.g.,][]{fressin_et_al2011, torres_et_al2011}, {\tt vespa} \citep{morton2012, morton2015}, and {\tt PASTIS} \citep{diaz_et_al2014} have been adopted to ``validate'' planet candidates in a statistical sense. These tools simulate the range of astrophysical configurations that could generate the observed lightcurve subject to the constraints placed by in-depth analyses of the transit photometry (e.g., shifts in the photocenter during transit; presence or absence of secondary eclipse; variations in the depths of odd and even transits) and subsequent follow-up investigations (e.g., high-contrast imaging, radial velocity searches for additional stars, (a)chromaticity of transit events, analyses of archival observations).

\subsection{Identifying Clear False Positives}
\label{sec:id_fps}
Four of the K2OIs in our sample displayed clear secondary eclipses when we phase folded the light curve of the host star to the assumed orbital period: EPIC~ 212679798.01, 212773272.01, 212773309.01 (also identified as an EB by \citealt{barros_et_al2016}), and  213951550.01. Consulting the ExoFOP-K2 follow-up website\footnote{\url{https://exofop.ipac.caltech.edu/k2/}}, we found that three of these stars were flagged by D.~LaCourse as possible eclipsing binaries (212773309 \& 212679798) or false positives (212773272). K2OIs 212773309.01 and 212773272.01 were identified at exactly the same ephemeris, suggesting that the transit-like events detected in the light curve of EPIC~212773272 are actually due to the eclipses of EPIC~212773309. 

In addition to classifying EPIC~212679798.01, 212773272.01, 212773309.01, and 213951550.01 as false positives due to the presence of secondary eclipses, we also rejected EPIC~211831378.01, 211970234.01, and 212572452.01 due to blended photometry or inconsistent transit depths when fitting data processed by different K2 pipelines. For instance, the K2SFF photometry of EPIC~211831378 displays 650~ppm transits while the k2phot and K2SC photometry reveal events with depths of 9000~ppm and 12\%, respectively. We attribute this discrepancy in event depth to the use of different apertures for each pipeline. As mentioned on the \mbox{ExoFOP} website, archival photometry from DSS, SDSS, 2MASS, and WISE reveals a brighter star $12\farcs3$ away from EPIC~211831378; the K2SFF pipeline incorrectly placed the aperture around this star (EPIC~211831539) rather than around the target star. 

EPIC~211970234 is also in a crowded field and the assigned K2SFF aperture includes contributions from multiple stars, one of which is much brighter than the target star. Similarly, the photometry of EPIC~212572452 is contaminated by light from EPIC~212572439 (2MASS~\mbox{J13374562-1111331}), which is 1~magnitude brighter than the target star and $5\farcs96$ away. 

Finally, we classified EPIC~212628098.01 as a false positive due to the large implied planet radius of $14-24\rearth$ and the presence of a neighboring star only $1\farcs25$ from the target star. The nearby star is only 3.8~magnitudes fainter than the target star and was detected both in Gemini-N/NIRI AO images and in speckle images acquired with DSSI at Gemini-S and WIYN. 

\begin{figure}[tbh]
\centering
\includegraphics[width=0.5\textwidth]{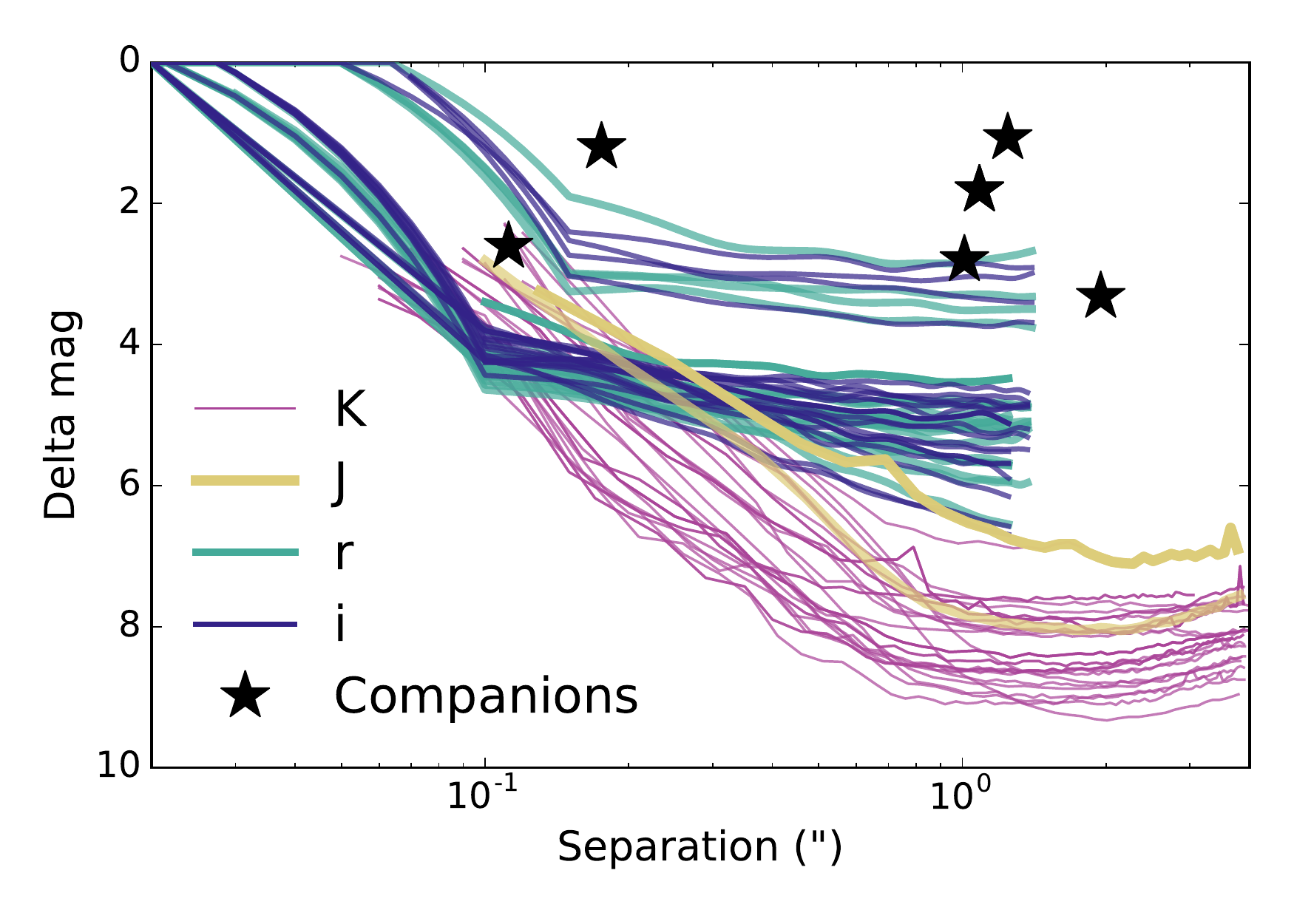}
\caption{Sensitivity of our follow-up imaging observations as a function of separation. The purple and yellow contrast curves mark the limits achieved for our adaptive optics observations at near-infrared wavelengths, while the teal and navy curves display the limits for our speckle imaging at optical wavelengths. The stars indicate the magnitude difference between the K2 target stars and detected nearby ``companion'' stars that may or may not be physically associated. The magnitude differences shown here are those in the band used for the follow-up observations; the flux ratio in the K2 bandpass may be different.}
\label{fig:ao}
\end{figure}

\subsection{Assessing False Positive Probabilities}
\label{sec:id_fps}
For the more promising K2OIs, we used the {\tt vespa} framework to assess the probabilities that each was truly a transiting planet. We first fit the transit photometry as described in Section~\ref{sec:transit_fits} and rescaled the photometric errors so that the adopted transit model has a reduced chi squared of 1 for the segment of the light curve centered on the transit event. Prior to rescaling the errors, we clipped the light curves and kept only the points within six transit durations of transit center. We then searched for secondary eclipses by phase-folding the full data set to the orbital period of the planet in question and measuring the ``eclipse depth'' at multiple points in the light curve. We assumed that the eclipse has the same duration as the primary transit, but we allowed the phase of secondary eclipse to vary from phase=0.3 to phase=0.7 (i.e., we do not assume that the orbit is circular). We recorded the depth of the deepest event as the maximum allowed secondary eclipse depth (``secthresh'' in the {\tt vespa} {\tt fpp.ini} file). 

Next, we referred back to the K2SFF photometry and recorded the radius of the selected photometric aperture as the maximum allowed separation between the target and the source of the transit event (``maxrad''). We then consulted previously published K2 papers and the Exo-FOP K2 follow-up website to check whether there are extant speckle or high-contrast images placing limits on the allowed brightness of nearby stars. If so, we included those contrast curves as additional constraints. For reference, we list the adaptive optics observations used in our false positive analysis in Table~\ref{tab:ao} and display the composite set of contrast curves in Figure~\ref{fig:ao}. These observations were obtained with NIRC2 on the 10-m Keck II telescope, NIRI \citep{hodapp_et_al2003} on the 8-m Gemini-N telescope, PHARO \citep{hayward_et_al2001} on the 200'' Palomar Hale telescope, and DSSI \citep{howell_et_al2011, horch_et_al2012} on the 8-m Gemini-N and Gemini-S telescopes and the 3.5-m WIYN\footnote{The WIYN Observatory is a joint facility of the University of Wisconsin-Madison, Indiana University, the National Optical Astronomy Observatory and the University of Missouri and hosts the NASA-NSF NN-EXPLORE program.} telescope.

\begin{deluxetable*}{cccccccccccc}
\tablecolumns{12}
\tabletypesize{\tiny}
\tablecaption{Speckle \& AO Observations Used in {\tt VESPA} Analysis \label{tab:ao}}
\tablehead{
\colhead{} &
\colhead{} &
\colhead{} &
\colhead{} &
\colhead{Pixel} &
\colhead{PSF} &
\multicolumn{3}{c}{Nearby Star\tablenotemark{a}} &
\colhead{Contrast} &
\colhead{Observation} &
\colhead{Uploaded}\\
\cline{7-9}
\colhead{EPIC} &
\colhead{Telescope} &
\colhead{Instrument} &
\colhead{Filter} &
\colhead{Scale} &
\colhead{(")} &
\colhead{Det?} &
\colhead{$\Delta$mag} &
\colhead{Sep ($"$)} &
\colhead{Achieved\tablenotemark{b}} &
\colhead{Date} &
\colhead{By}
}
\startdata
 201205469 &   Keck2\_10m &      NIRC2 &                      K &  0.009942 &  0.066557 & no & $\cdots$ & $\cdots$ & $\Delta$ 7.91 mag at $0\farcs5$ &    4/7/15 &   Ciardi \\
 201208431 &   Keck2\_10m &      NIRC2 &                      K &  0.009942 &  0.079626 & no & $\cdots$ & $\cdots$ & $\Delta$ 5.86 mag at $0\farcs5$ &   2/19/16 &   Ciardi \\
 201549860 &   Keck2\_10m &      NIRC2 &                      K &  0.009942 &  0.060571 & no & $\cdots$ & $\cdots$ & $\Delta$ 8.10 mag at $0\farcs5$ &    4/1/15 &   Ciardi \\
 201617985 &   Keck2\_10m &      NIRC2 &                      K &  0.009942 &  0.091839 & no & $\cdots$ & $\cdots$ & $\Delta$ 7.84 mag at $0\farcs5$ &    4/1/15 &   Ciardi \\
 201637175 &  GeminiN\_8m &       DSSI &  692nm &  0.011410 &  0.020000 & yes\tablenotemark{c} & $\cdots$ & $\cdots$ & $\Delta$ 4.977 mag at $0\farcs5$ &   1/15/16 &   Ciardi \\
 201637175 &  GeminiN\_8m &       DSSI &  880nm &  0.011410 &  0.020000 & yes\tablenotemark{c} & $\cdots$ & $\cdots$ & $\Delta$ 4.666 mag at $0\farcs5$ &   1/15/16 &   Ciardi \\
 201855371 &   Keck2\_10m &      NIRC2 &                      K &  0.009942 &  0.053694 & no & $\cdots$ & $\cdots$ & $\Delta$ 8.49 mag at $0\farcs5$ &    4/7/15 &   Ciardi \\
 205924614 &   Keck2\_10m &      NIRC2 &                      K &  0.009942 &  0.058868 & no & $\cdots$ & $\cdots$ & $\Delta$ 8.17 mag at $0\farcs5$ &    8/7/15 &   Ciardi \\
 206011691 &   Keck2\_10m &      NIRC2 &                      K &  0.009942 &  0.050896 & no & $\cdots$ & $\cdots$ & $\Delta$ 7.45 mag at $0\farcs5$ &   7/25/15 &   Ciardi \\
 206209135 &   Keck2\_10m &      NIRC2 &                      K &  0.009942 &  0.062750 & no & $\cdots$ & $\cdots$ & $\Delta$ 7.74 mag at $0\farcs5$ &   8/21/15 &   Ciardi \\
 210448987 &  GeminiN\_8m &       NIRI &                      K &  0.021400 &  0.102616 & no & $\cdots$ & $\cdots$ & $\Delta$ 6.86 mag at $0\farcs5$ &  12/14/15 &   Ciardi \\
 210489231 &   Keck2\_10m &      NIRC2 &                      K &  0.009942 &  0.101436 & no & $\cdots$ & $\cdots$ & $\Delta$ 5.78 mag at $0\farcs5$ &  10/28/15 &   Ciardi \\
 210508766 &   Keck2\_10m &      NIRC2 &                      K &  0.009942 &  0.054223 & no & $\cdots$ & $\cdots$ & $\Delta$ 7.40 mag at $0\farcs5$ &  10/28/15 &   Ciardi \\
 210508766 &  GeminiN\_8m &       DSSI &  692nm &  0.011410 &  0.020000 & no & $\cdots$ & $\cdots$ & $\Delta$ 4.994 mag at $0\farcs5$ &   1/13/16 &   Ciardi \\
 210508766 &  GeminiN\_8m &       DSSI &  880nm &  0.011410 &  0.020000 & no & $\cdots$ & $\cdots$ & $\Delta$ 5.248 mag at $0\farcs5$ &   1/13/16 &   Ciardi \\
 210508766 &  Palomar-5m &   PHARO-AO &                K\_short &  0.025000 &  0.159000 &no & $\cdots$ & $\cdots$ & $\Delta$ 4.90 mag at $0\farcs5$ &  10/20/16 &   Ciardi \\
 210558622 &   Keck2\_10m &      NIRC2 &                      K &  0.009942 &  0.049441 &no & $\cdots$ & $\cdots$ & $\Delta$ 7.86 mag at $0\farcs5$ &  10/28/15 &   Ciardi \\
 210558622 &   Keck2\_10m &      NIRC2 &                      K &  0.009942 &  0.050778 &no & $\cdots$ & $\cdots$ & $\Delta$ 8.24 mag at $0\farcs5$ &   2/17/16 &   Ciardi \\
 210558622 &  GeminiN\_8m &       DSSI &  692nm &  0.011410 &  0.020000 & no & $\cdots$ & $\cdots$ & $\Delta$ 5.286 mag at $0\farcs5$ &   1/13/16 &   Ciardi \\
 210558622 &  GeminiN\_8m &       DSSI &  880nm &  0.011410 &  0.020000 & no & $\cdots$ & $\cdots$ & $\Delta$ 5.248 mag at $0\farcs5$ &   1/13/16 &   Ciardi \\
 210558622 &   WIYN\_3.5m &       DSSI &       692nm &  0.022000 &  0.050000 &  no & $\cdots$ & $\cdots$ & $\Delta$ 3.5 mag at $0\farcs2$ &  10/24/15 &  Everett \\
 210558622 &   WIYN\_3.5m &       DSSI &       880nm &  0.022000 &  0.063000 &  no & $\cdots$ & $\cdots$ & $\Delta$ 3.5 mag at $0\farcs2$ &  10/24/15 &  Everett \\
 210707130 &   Keck2\_10m &      NIRC2 &                      K &  0.009942 &  0.051244 &no & $\cdots$ & $\cdots$ & $\Delta$ 7.92 mag at $0\farcs5$ &  10/28/15 &   Ciardi \\
 210707130 &   Keck2\_10m &      NIRC2 &                      K &  0.009942 &  0.053366 &no & $\cdots$ & $\cdots$ & $\Delta$ 6.62 mag at $0\farcs5$ &   2/19/16 &   Ciardi \\
 210707130 &  GeminiN\_8m &       DSSI &  692nm &  0.011410 &  0.020000 &no & $\cdots$ & $\cdots$ & $\Delta$ 5.650 mag at $0\farcs5$ &   1/13/16 &   Ciardi \\
 210707130 &  GeminiN\_8m &       DSSI &  880nm &  0.011410 &  0.020000 &no & $\cdots$ & $\cdots$ & $\Delta$ 5.719 mag at $0\farcs5$ &   1/13/16 &   Ciardi \\
 210750726 &   Keck2\_10m &      NIRC2 &                      K &  0.009942 &  0.111601 & no & $\cdots$ & $\cdots$ & $\Delta$ 5.36 mag at $0\farcs5$ &  10/28/15 &   Ciardi \\
 210750726 &   Keck2\_10m &      NIRC2 &                      K &  0.009942 &  0.080981 &no & $\cdots$ & $\cdots$ & $\Delta$ 6.06 mag at $0\farcs5$ &   2/19/16 &   Ciardi \\
 210750726 &  GeminiN\_8m &       DSSI &  692nm &  0.011410 &  0.020000 &no & $\cdots$ & $\cdots$ & $\Delta$ 4.901 mag at $0\farcs5$ &   1/15/16 &   Ciardi \\
 210750726 &  GeminiN\_8m &       DSSI &  880nm &  0.011410 &  0.020000 &no & $\cdots$ & $\cdots$ & $\Delta$ 5.002 mag at $0\farcs5$ &   1/15/16 &   Ciardi \\
 210838726 &  GeminiN\_8m &       NIRI &                      K &  0.021400 &  0.109663 &no & $\cdots$ & $\cdots$ & $\Delta$ 6.12 mag at $0\farcs5$ &   11/2/15 &   Ciardi \\
 210838726 &  GeminiN\_8m &       NIRI &                      K &  0.021400 &  0.099683 &no & $\cdots$ & $\cdots$ & $\Delta$ 6.83 mag at $0\farcs5$ &  12/14/15 &   Ciardi \\
 210968143 &   Keck2\_10m &      NIRC2 &                      K &  0.009942 &  0.053846 &no & $\cdots$ & $\cdots$ & $\Delta$ 7.59 mag at $0\farcs5$ &  10/28/15 &   Ciardi \\
 210968143 &   Keck2\_10m &      NIRC2 &                      K &  0.009942 &  0.051648 &no & $\cdots$ & $\cdots$ & $\Delta$ 7.91 mag at $0\farcs5$ &   2/17/16 &   Ciardi \\
 210968143 &  GeminiN\_8m &       DSSI &  692nm &  0.011410 &  0.020000 &no & $\cdots$ & $\cdots$ & $\Delta$ 5.279 mag at $0\farcs5$ &   1/13/16 &   Ciardi \\
 210968143 &  GeminiN\_8m &       DSSI &  880nm &  0.011410 &  0.020000 &no & $\cdots$ & $\cdots$ & $\Delta$ 5.089 mag at $0\farcs5$ &   1/13/16 &   Ciardi \\
 210968143 &  Palomar-5m &   PHARO-AO &                K\_short &  0.025000 &  0.165000 &no & $\cdots$ & $\cdots$ & $\Delta$ 4.73 mag at $0\farcs5$ &  10/20/16 &   Ciardi \\
 211077024 &   Keck2\_10m &      NIRC2 &                      K &  0.009942 &  0.061843 &no & $\cdots$ & $\cdots$ & $\Delta$ 7.41 mag at $0\farcs5$ &   2/19/16 &   Ciardi \\
 211077024 &  GeminiN\_8m &       DSSI &  692nm &  0.011410 &  0.020000 &no & $\cdots$ & $\cdots$ & $\Delta$ 4.924 mag at $0\farcs5$ &   1/15/16 &   Ciardi \\
 211077024 &  GeminiN\_8m &       DSSI &  880nm &  0.011410 &  0.020000 &no & $\cdots$ & $\cdots$ & $\Delta$ 5.371 mag at $0\farcs5$ &   1/15/16 &   Ciardi \\
 211331236 &   Keck2\_10m &      NIRC2 &                      K &  0.009942 &  0.079162 &no & $\cdots$ & $\cdots$ & $\Delta$ 6.62 mag at $0\farcs5$ &   1/21/16 &   Ciardi \\
 211428897 &   Keck2\_10m &      NIRC2 &                      J &  0.009942 &  0.114165 &yes & $\cdots$ & $\cdots$ & $\Delta$ 5.41 mag at $0\farcs5$ &   1/21/16 &   Ciardi \\
 211428897 &   Keck2\_10m &      NIRC2 &                      K &  0.009942 &  0.060653 & yes & $\cdots$ & $\cdots$ & $\Delta$ 6.86 mag at $0\farcs5$ &   1/21/16 &   Ciardi \\
 211428897 &  GeminiN\_8m &       DSSI &  692nm &  0.011410 &  0.020000 &yes & 1.8 &  1.1 & $\Delta$ 4.442 mag at $0\farcs5$ &   1/15/16 &   Ciardi \\
 211428897 &  GeminiN\_8m &       DSSI &  880nm &  0.011410 &  0.020000 &yes & 1.2 & 1.1 & $\Delta$ 4.869 mag at $0\farcs5$ &   1/15/16 &   Ciardi \\
 211509553 &  GeminiN-8m &       NIRI &                   open &  0.021400 &  0.098000 &yes & 3.3 & 1.9 & $\Delta$ 5.84 mag at $0\farcs5$ &   2/20/16 &   Ciardi \\
 211770795 &   Keck2\_10m &      NIRC2 &                      K &  0.009942 &  0.065018 &no & $\cdots$ & $\cdots$ & $\Delta$ 7.36 mag at $0\farcs5$ &   2/19/16 &   Ciardi \\
 211770795 &  GeminiN-8m &       NIRI &                   open &  0.021400 &  0.112000 &no & $\cdots$ & $\cdots$ & $\Delta$ 5.57 mag at $0\farcs5$ &   2/20/16 &   Ciardi \\
 211799258 &  GeminiN-8m &       NIRI &                   open &  0.021400 &  0.107000 &no & $\cdots$ & $\cdots$ & $\Delta$ 6.61 mag at $0\farcs5$ &   2/20/16 &   Ciardi \\
 211818569 &  Palomar-5m &   PHARO-AO &                Ks &  0.025000 &  0.122000 &no & $\cdots$ & $\cdots$ & $\Delta$ 5.22 mag at $0\farcs5$ &  10/20/16 &   Ciardi \\
 211831378 &  GeminiN\_8m &       NIRI &                      K &  0.021400 &  0.125410 &no & $\cdots$ & $\cdots$ & $\Delta$ 6.59 mag at $0\farcs5$ &   1/28/16 &   Ciardi \\
 211924657 &  GeminiN-8m &       NIRI &               Br-gamma &  0.021400 &  0.101000 &no & $\cdots$ & $\cdots$ & $\Delta$ 6.26 mag at $0\farcs5$ &   2/20/16 &   Ciardi \\
 211970234 &  GeminiN-8m &       NIRI &                   open &  0.021400 &  0.113000 &no & $\cdots$ & $\cdots$ & $\Delta$ 5.60 mag at $0\farcs5$ &   2/20/16 &   Ciardi \\
 212006344 &   Keck2\_10m &      NIRC2 &                      K &  0.009942 &  0.053540 &no & $\cdots$ & $\cdots$ & $\Delta$ 8.06 mag at $0\farcs5$ &   1/21/16 &   Ciardi \\
 212006344 &  GeminiN\_8m &       DSSI &  692nm &  0.011410 &  0.020000 &no & $\cdots$ & $\cdots$ & $\Delta$ 5.281 mag at $0\farcs5$ &   1/14/16 &   Ciardi \\
 212006344 &  GeminiN\_8m &       DSSI &  880nm &  0.011410 &  0.020000 &no & $\cdots$ & $\cdots$ & $\Delta$ 5.800 mag at $0\farcs5$ &   1/14/16 &   Ciardi \\
 212069861 &   Keck2\_10m &      NIRC2 &                      J &  0.009942 &  0.092989 &no & $\cdots$ & $\cdots$ & $\Delta$ 6.17 mag at $0\farcs5$ &   2/17/16 &   Ciardi \\
 212069861 &   Keck2\_10m &      NIRC2 &                      K &  0.009942 &  0.096097 &no & $\cdots$ & $\cdots$ & $\Delta$ 7.63 mag at $0\farcs5$ &   2/17/16 &   Ciardi \\
 212069861 &  GeminiN-8m &       NIRI &               Br-gamma &  0.021400 &  0.131000 &no & $\cdots$ & $\cdots$ & $\Delta$ 4.93 mag at $0\farcs5$ &   2/20/16 &   Ciardi \\
 212154564 &   Keck2\_10m &      NIRC2 &                      K &  0.009942 &  0.111359 &no & $\cdots$ & $\cdots$ & $\Delta$ 6.06 mag at $0\farcs5$ &   2/19/16 &   Ciardi \\
 212154564 &  GeminiN-8m &       NIRI &                   open &  0.021400 &  0.106000 &no & $\cdots$ & $\cdots$ & $\Delta$ 5.52 mag at $0\farcs5$ &   2/20/16 &   Ciardi \\
 212354731 &  GeminiS\_8m &       DSSI &       692nm &  0.011000 &  0.021000 &    no & $\cdots$ & $\cdots$ & $\Delta$ 5 mag at $0\farcs2$ &   6/29/16 &  Everett \\
 212354731 &  GeminiS\_8m &       DSSI &       880nm &  0.011000 &  0.027000 &    no & $\cdots$ & $\cdots$ & $\Delta$ 5 mag at $0\farcs2$ &   6/29/16 &  Everett \\
 212460519 &   WIYN\_3.5m &       DSSI &       692nm &  0.022000 &  0.050000 &  no & $\cdots$ & $\cdots$ & $\Delta$ 3.5 mag at $0\farcs2$ &   4/21/16 &  Everett \\
 212460519 &   WIYN\_3.5m &       DSSI &       880nm &  0.022000 &  0.063000 &  no & $\cdots$ & $\cdots$ & $\Delta$ 3.5 mag at $0\farcs2$ &   4/21/16 &  Everett \\
 212554013 &  GeminiS\_8m &       DSSI &       692nm &  0.011000 &  0.021000 &    no & $\cdots$ & $\cdots$ & $\Delta$ 5 mag at $0\farcs2$ &   6/22/16 &  Everett \\
 212554013 &  GeminiS\_8m &       DSSI &       880nm &  0.011000 &  0.027000 &    no & $\cdots$ & $\cdots$ & $\Delta$ 5 mag at $0\farcs2$ &   6/22/16 &  Everett \\
 212565386 &  GeminiS\_8m &       DSSI &       692nm &  0.011000 &  0.021000 &    no & $\cdots$ & $\cdots$ & $\Delta$ 5 mag at $0\farcs2$ &   6/22/16 &  Everett \\
 212565386 &  GeminiS\_8m &       DSSI &       880nm &  0.011000 &  0.027000 &    no & $\cdots$ & $\cdots$ & $\Delta$ 5 mag at $0\farcs2$ &   6/22/16 &  Everett \\
 212572452 &  GeminiS\_8m &       DSSI &       692nm &  0.011000 &  0.021000 &    no & $\cdots$ & $\cdots$ & $\Delta$ 5 mag at $0\farcs2$ &   6/22/16 &  Everett \\
 212572452 &  GeminiS\_8m &       DSSI &       880nm &  0.011000 &  0.027000 &    no & $\cdots$ & $\cdots$ & $\Delta$ 5 mag at $0\farcs2$ &   6/22/16 &  Everett \\
 212628098 &  GeminiS\_8m &       DSSI &       692nm &  0.011000 &  0.021000 &    yes & $\cdots$ & $\cdots$ & $\Delta$ 5 mag at $0\farcs2$ &   6/22/16 &  Everett \\
 212628098 &  GeminiS\_8m &       DSSI &       880nm &  0.011000 &  0.027000 &    yes & 3.8 & 1.3 & $\Delta$ 5 mag at $0\farcs2$ &   6/22/16 &  Everett \\
 212628098 &   WIYN\_3.5m &       DSSI &       692nm &  0.022000 &  0.050000 &  yes & $\cdots$ & $\cdots$& $\Delta$ 3.5 mag at $0\farcs2$ &   4/20/16 &  Everett \\
 212628098 &   WIYN\_3.5m &       DSSI &       880nm &  0.022000 &  0.063000 &  yes & $\cdots$ & $\cdots$ & $\Delta$ 3.5 mag at $0\farcs2$ &   4/20/16 &  Everett \\
 212628098 &  GeminiN-8m &       NIRI &               Br-gamma &  0.021400 &  0.107000 & yes & $\cdots$ &  $\cdots$ & $\Delta$ 7.01 mag at $0\farcs5$ &   6/20/16 &   Ciardi \\
 212679181 &  GeminiS\_8m &       DSSI &       692nm &  0.011000 &  0.021000 &    yes & 1.1 & 1.2 & $\Delta$ 5 mag at $0\farcs2$ &   6/21/16 &  Everett \\
 212679181 &  GeminiS\_8m &       DSSI &       880nm &  0.011000 &  0.027000 &    yes & 1.1 & 1.3 & $\Delta$ 5 mag at $0\farcs2$ &   6/21/16 &  Everett \\
 212679181 &   WIYN\_3.5m &       DSSI &       692nm &  0.022000 &  0.050000 & yes & 1.5 & 1.5 & $\Delta$ 3.5 mag at $0\farcs2$ &   4/17/16 &  Everett \\
 212679181 &   WIYN\_3.5m &       DSSI &       880nm &  0.022000 &  0.063000 &  yes & 1.2 & 1.5 & $\Delta$ 3.5 mag at $0\farcs2$ &   4/17/16 &  Everett \\
 212679798 &  GeminiS\_8m &       DSSI &       692nm &  0.011000 &  0.021000 &    yes & $\cdots$ & $\cdots$ & $\Delta$ 5 mag at $0\farcs2$ &   6/22/16 &  Everett \\
 212679798 &  GeminiS\_8m &       DSSI &       880nm &  0.011000 &  0.027000 &    yes & 2.6 & 0.1 & $\Delta$ 5 mag at $0\farcs2$ &   6/22/16 &  Everett \\
 212679798 &  GeminiN-8m &       NIRI &                   open &  0.021400 &  0.110000 &yes & $\cdots$ &$\cdots$ & $\Delta$ 6.62 mag at $0\farcs5$ &   6/20/16 &   Ciardi \\
 212686205 &   WIYN\_3.5m &       DSSI &       692nm &  0.022000 &  0.050000 &  no & $\cdots$ & $\cdots$ & $\Delta$ 3.5 mag at $0\farcs2$ &   4/20/16 &  Everett \\
 212686205 &   WIYN\_3.5m &       DSSI &       880nm &  0.022000 &  0.063000 &  no & $\cdots$ & $\cdots$ & $\Delta$ 3.5 mag at $0\farcs2$ &   4/20/16 &  Everett \\
 212773272 &  GeminiS\_8m &       DSSI &       692nm &  0.011000 &  0.021000 &    no & $\cdots$ & $\cdots$ & $\Delta$ 5 mag at $0\farcs2$ &   6/21/16 &  Everett \\
 212773272 &  GeminiS\_8m &       DSSI &       880nm &  0.011000 &  0.027000 &    no & $\cdots$ & $\cdots$ & $\Delta$ 5 mag at $0\farcs2$ &   6/21/16 &  Everett \\
 212773272 &  GeminiN-8m &       NIRI &               Br-gamma &  0.021400 &  0.109000 &no & $\cdots$ & $\cdots$ & $\Delta$ 7.09 mag at $0\farcs5$ &   6/20/16 &   Ciardi \\
 212773309 &  GeminiS\_8m &       DSSI &       692nm &  0.011000 &  0.021000 &    yes & 2.8 & 1.0 & $\Delta$ 5 mag at $0\farcs2$ &   6/21/16 &  Everett \\
 212773309 &  GeminiS\_8m &       DSSI &       880nm &  0.011000 &  0.027000 &    yes & 2.0 & 1.0 & $\Delta$ 5 mag at $0\farcs2$ &   6/21/16 &  Everett \\
 212773309 &   WIYN\_3.5m &       DSSI &       692nm &  0.022000 &  0.050000 &  yes & 2.8 & 1.2 & $\Delta$ 3.5 mag at $0\farcs2$ &   4/24/16 &  Everett \\
 212773309 &   WIYN\_3.5m &       DSSI &       880nm &  0.022000 &  0.063000 &  no & 2.0 & 1.2 & $\Delta$ 3.5 mag at $0\farcs2$ &   4/24/16 &  Everett \\
 213951550 &  GeminiN-8m &       NIRI &                   open &  0.021400 &  0.116000 &yes & $\cdots$ & 0.2 & $\Delta$ 6.41 mag at $0\farcs5$ &   7/15/16 &   Ciardi \\
 216892056 &  GeminiS\_8m &       DSSI &       692nm &  0.011000 &  0.021000 &    no & $\cdots$ & $\cdots$ & $\Delta$ 5 mag at $0\farcs2$ &   6/29/16 &  Everett \\
 216892056 &  GeminiS\_8m &       DSSI &       880nm &  0.011000 &  0.027000 &    no & $\cdots$ & $\cdots$ & $\Delta$ 5 mag at $0\farcs2$ &   6/29/16 &  Everett \\
 216892056 &  GeminiN-8m &       NIRI &               Br-gamma &  0.021400 &  0.110000 &no & $\cdots$ & $\cdots$ & $\Delta$ 6.86 mag at $0\farcs5$ &   6/15/16 &   Ciardi \\
 217941732 &  GeminiS\_8m &       DSSI &       692nm &  0.011000 &  0.021000 &    no & $\cdots$ & $\cdots$ & $\Delta$ 5 mag at $0\farcs2$ &   6/19/16 &  Everett \\
 217941732 &  GeminiS\_8m &       DSSI &       880nm &  0.011000 &  0.027000 &    no & $\cdots$ & $\cdots$ & $\Delta$ 5 mag at $0\farcs2$ &   6/19/16 &  Everett \\
\enddata
\tablenotetext{a}{We use the ``Nearby Star Det?'' column to indicate whether any follow-up image revealed a companion to the star. The values in the ``$\Delta$ mag'' and ``Sep'' columns refer to the magnitude difference and separation measured in the specific image described on the corresponding line of the table.}
\tablenotetext{b}{These point sensitivity estimates provide a rough view of whether the image provides deep or shallow limits on the presence of nearby companions. As shown in Figure~\ref{fig:ao}, the contrast achieved improves with increasing separation from the target star. We use the full separation-dependent contrast curves for our false positive probability estimates.}
\tablenotetext{c}{The Gemini/DSSI speckle images of EPIC~201637175 did not reveal any companions, but Subaru/HSC imaging displayed a second star roughly 12\% as bright as EPIC~201637175. The separation of the two stars is approximately $2"$ \citep{sanchis-ojeda_et_al2015}.  }
\end{deluxetable*}

Finally, we filled in the host star properties (coordinates, magnitudes, and spectroscopic fits from D17) and ran {\tt vespa} to compute the false positive probability (FPP). Recognizing that {\tt vespa} FPP are statistical and depend on the assumed planet radius, we ran the analysis twice for each planet using $R_p/R_\star$ ratios set to the 16th and 84th percentiles of the posterior distribution.  As in \citet{crossfield_et_al2016}, we adopted threshold of FPP~$< 1$\% for validation, but we adopted a less forgiving false positive probability cut of FPP~$>90$\%. When classifying K2OIs, we required that both FPP estimates were below 1$\%$ or above $90\%$ in order to label K2OIs as validated planets or false positives, respectively. We classified the remaining systems as planet candidates.

We summarize our new K2OI dispositions in Table~\ref{tab:disp}. In total, 44 K2OIs met the formal criteria for validation, but six orbit stars with nearby companions and therefore cannot be validated with {\tt vespa}. Of the remaining 38 K2OIs with low FPPs, twenty were previously validated by \citet{crossfield_et_al2016}, eight are new detections, and ten were previously classified as planet candidates. As discussed in Section~\ref{sec:id_fps}, we rejected eight~K2OIs as false positives based on visual inspection of their light curves. No additional K2OIs were classified as false positives based on {\tt vespa} FPPs alone. The remaining 28~K2OIs had ambiguous FPPs between $1\%$ and $90\%$. The ambiguous sample includes three previously confirmed planets, ten previously announced planet candidates, and fifteen new detections. The three confirmed planets that failed to meet our $1\%$ FPP threshold are EPIC~201345483.01 ($R_p = 10.4\rearth$, $\rm{FPP} = 15\%$),  EPIC~201635569.01 ($R_p = 7.7\rearth$, $\rm{FPP} = 4-7\%$),  and EPIC~210508766.02 ($R_p = 2.2\rearth$, $\rm{FPP} = 1.9\%$). One other previously confirmed planet  \citep[EPIC~201637175.01 = K2-22b,][]{sanchis-ojeda_et_al2015} met our FPP cut for validation, but is listed as a planet candidate in Table~\ref{tab:disp} due to the presence of a nearby star. We do not dispute the previous validation of K2-22b, but our {\tt vespa} analysis is not sufficiently sophisticated to validate planets orbiting stars with nearby companions.

The most likely explanation for why we were unable to validate EPIC~201345483.01 is that our estimates of the stellar radius ($0.69_{-0.04}^{+0.06}\rsun$, D17) and planet radius ($10.4_{-0.7}^{+0.9}\rearth$) are much larger than the values of $0.445\pm0.066\rsun$ and $6.71\rearth$ assumed by \citet{crossfield_et_al2016} when validating the system. In contrast, our estimates agree well with those adopted by \citet{vanderburg_et_al2016} in their discovery paper ($0.66\rsun$, $11\rearth$). Larger planets are rarer than smaller planets, which would have caused the planet prior to be higher in the \citet{crossfield_et_al2016} analysis than in our analysis. In the future, including AO or speckle imaging would be useful for discriminating between the planetary interpretation and remaining false positive scenarios. 

\begin{deluxetable}{c|ccc|c}
\tablecolumns{5}
\tabletypesize{\normalsize}
\tablecaption{Breakdown of K2OI Dispositions\label{tab:disp}}
\tablehead{
\colhead{Previous} &
\multicolumn{4}{c}{Updated Disposition\tablenotemark{1}}\\
\cline{2-5}
\colhead{Disposition} &
\colhead{CP} &
\colhead{PC}  &
\colhead{FP} &
\colhead{\textbf{All}}
}
\startdata
 CP & 21 & 3\tablenotemark{2} & 0 & \textbf{24} \\
 PC & 10 & 12 & 1\tablenotemark{3} & \textbf{23} \\
 FP & 0 & 0 & 2 & \textbf{2} \\
 UK & 8 & 17 & 5 & \textbf{30} \\
 \hline
 \textbf{All} & \textbf{39} & \textbf{32} & \textbf{8} & \textbf{79} \\ 
\enddata
\tablenotetext{1}{CP = Confirmed Planet, PC = Planet Candidate, FP = False Positive, UK = Unknown}
\tablenotetext{2}{The previously confirmed planets that we cannot validate with {\tt vespa} are EPIC~201345483.01 ($\rm{FPP} = 15\%$),  EPIC~201635569.01 ( $\rm{FPP} = 4-7\%$), and EPIC~201637175.01 (=K2-22b, nearby star detected). See Section~\ref{sec:id_fps} for details.}
\tablenotetext{3}{The planet candidate rejected as a false positive is EPIC~212572452.01, which was announced by \citet{pope_et_al2016} as a candidate with $R_p/R_\star = 0.174$ and an orbital period of 2.6~days. As discussed in Section~\ref{sec:id_fps}, the K2 photometry for this target is contaminated by light from the nearby, brighter star EPIC 212572439.}
\end{deluxetable}

EPIC~201635569.01 was previously validated by \cite{montet_et_al2015} as a $4.81\pm0.42\rearth$ planet with \mbox{FPP = $4.9\times10^{-3}$} and by \citet{crossfield_et_al2016} as a $4.48\pm0.52\rearth$ planet with \mbox{FPP = $0.6\%$}. Our inability to confirm the validation of this planet may be due to our larger estimate of the planet radius, which is in turn caused by the larger stellar radius found by D17. We estimated a revised radius of $0.62\pm0.03\rsun$, which is significantly larger than the values of $0.45\pm0.01\rsun$ and $0.39\pm0.04\rsun$ assumed by \citet{montet_et_al2015} and \citet{crossfield_et_al2016}, respectively. Our new FPP estimate of $4-7\%$ is still consistent with the planetary interpretation of the transit-like event, but indicates that additional observations such as AO imaging would be useful to rule out the remaining false positive scenarios.

For EPIC~210508766.02, our initial FPP estimate of 1.9\% is only slightly above our validation threshold of 1\% and does not consider the fact that transit-like events detected in candidate multiple planet systems are more likely to be bona fide planets \citep{lissauer_et_al2012}. Given that EPIC~210508766 also hosts 210508766.01, we can apply a multiplicity boost to reduce the FPP for 210508766.02 by a factor of 30 \citep{sinukoff_et_al2016, vanderburg_et_al2016b}. We therefore support the previous validation of EPIC~210508766.02 (K2-83c) by \citet{crossfield_et_al2016} and classify that K2OI as a validated planet while categorizing all of the other K2OIs with FPP between $1\%$ and $90\%$ as planet candidates. The final disposition breakdown for our K2OI sample is 39~validated planets, 8~false positives, and 32~planet candidates.

We list the estimated FPPs and resulting dispositions for individual K2OIs in Table~\ref{tab:fpp}. For conciseness, Table~\ref{tab:fpp} also includes estimates of the orbital periods and mid-points of transit-like events. We present the remaining transit parameters and the corresponding physical parameters in Table~\ref{tab:fits}. As part of our classification and transit fitting process, we produced an array of vetting plots for each candidate. We will upload all of these plots to the ExoFOP-K2 website and provide examples in Appendix~\ref{sec:appendix}. 

As indicated in Figure~\ref{fig:ao} and Table~\ref{tab:fpp}, six of the planet candidates  are associated with K2 targets for which our follow-up imaging observations revealed nearby companions. The companions might be physically associated with the target star or simply background stars that fall along the same line of sight. Regardless, the close proximity of the additional star dilutes the depth of the transit-like events in the K2 light curves and causes the planets to appear smaller than their true size. In general, the radii of planet candidates orbiting stars with stellar companions are underestimated by 6\% if they orbit the target stars and by a factor of three if they orbit the companion stars \citep{furlan_et_al2017}. Assessing whether the companion stars are bound to the target stars and determining the source of the transit events will require additional scrutiny of the K2 photometry and follow-up imagery \citep[see][and references therein]{furlan_et_al2017}. We will discuss these systems in more detail in Gonzalez et al. (\emph{in prep}), an upcoming catalog paper describing the results of our follow-up images of candidate K2 planet host stars of all spectral types.

\begin{deluxetable*}{cccccccccccccccc}
\tablecolumns{16}
\tabletypesize{\tiny}
\tablecaption{K2OI False Positive Probabilities \& Dispositions \label{tab:fpp}}
\tablehead{
\colhead{} &
\colhead{} &
\multicolumn{2}{c}{Disposition} &
\multicolumn{2}{c}{{\tt vespa} FPP\tablenotemark{a}} &
\colhead{Nearby} &
\multicolumn{3}{c}{$R_p/R_\star$} &
\multicolumn{3}{c}{$P$ (days)\tablenotemark{c}}& 
\multicolumn{3}{c}{$t0$ (BKJD)\tablenotemark{c}}\\
\cline{3-4}
\cline{8-10}
\cline{14-16}
\colhead{EPIC} &
\colhead{K2OI} &
\colhead{Old} & 
\colhead{New} &
\colhead{Small $R_p$} &
\colhead{Big $R_p$} &
\colhead{Star?\tablenotemark{b}} &
\colhead{K2SFF} &
\colhead{k2phot} & 
\colhead{K2SC} &
\colhead{Val} &
\colhead{-Err} &
\colhead{+Err} &
\colhead{Val} &
\colhead{-Err} &
\colhead{+Err}
}
\startdata
201205469 &     1 &       CP &   CP &   3.6e-08 &   3.6e-08 &       no &  0.074 &        0.072 &          $\cdots$ &   3.47134 &  0.00016 &  0.00016 &      1976.881 &             0.002 &             0.002 \\
 201208431 &     1 &       CP &   CP &   3.7e-07 &   2.0e-06 &       no &  0.034 &        0.035 &         $\cdots$  & 10.00339 &  0.00099 &  0.00100 &      1982.524 &             0.004 &             0.004 \\
 201345483 &     1 &       CP &   PC &   1.5e-01 &   1.5e-01 &      $\cdots$ &  0.140 &        0.136 &         $\cdots$   & 1.72926 &  0.00001 &  0.00001 &      1976.526 &             0.000 &             0.000 \\
 201549860 &     1 &       CP &   CP &   2.3e-05 &   3.4e-05 &       no &  0.029 &        0.029 &          $\cdots$ &   5.60836 &  0.00034 &  0.00033 &      1979.119 &             0.002 &             0.002 \\
 201549860 &     2 &       CP &   CP &   1.5e-09 &   6.9e-08 &       no &  0.020 &        0.019 &          $\cdots$ &   2.39996 &  0.00020 &  0.00020 &      1977.584 &             0.004 &             0.004 \\
 201617985 &     1 &       PC &   PC &   5.0e-02 &   8.3e-02 &       no &  0.033 &        0.034 &          $\cdots$ &   7.28116 &  0.00058 &  0.00055 &      1979.641 &             0.004 &             0.004 \\
 201635569 &     1 &       CP &   PC &   3.6e-02 &   6.7e-02 &      $\cdots$ &  0.111 &        0.105 &          $\cdots$ &   8.36879 &  0.00018 &  0.00018 &      1978.447 &             0.001 &             0.001 \\
 201637175 &     1 &       CP &   PC &   2.4e-03 &   9.1e-03 &      yes &  0.076 &        0.074 &          $\cdots$ &   0.38108 &  0.00000 &  0.00000 &      2034.139 &             0.000 &             0.000 \\
 201717274 &     1 &       PC &   PC &   1.1e-01 &   9.3e-02 &      $\cdots$ &  0.038 &        0.039 &         $\cdots$ &   3.52674 &  0.00032 &  0.00030 &      1976.915 &             0.004 &             0.004 \\
 201855371 &     1 &       CP &   CP &   6.2e-06 &   1.4e-05 &       no &  0.030 &        0.030 &          $\cdots$ &  17.96901 &  0.00141 &  0.00137 &      1984.944 &             0.003 &             0.003 \\
 205924614 &     1 &       CP &   CP &   6.0e-11 &   7.9e-11 &       no &  0.056 &        0.055 &          0.056 &   2.84927 &  0.00003 &  0.00003 &      2150.423 &             0.000 &             0.000 \\
 206011691 &     1 &       CP &   CP &   4.1e-07 &   1.0e-08 &       no &  0.026 &        0.026 &          0.026 &   9.32502 &  0.00039 &  0.00039 &      2156.422 &             0.001 &             0.001 \\
 206011691 &     2 &       CP &   CP &   2.5e-08 &   8.0e-13 &       no &  0.036 &        0.033 &          0.034 &  15.50189 &  0.00095 &  0.00093 &      2155.471 &             0.002 &             0.002 \\
 206119924 &     1 &       PC &   CP &   2.6e-03 &   2.2e-03 &      $\cdots$ &  0.009 &        0.009 &          0.009 &   4.65541 &  0.00048 &  0.00049 &      2146.948 &             0.004 &             0.004 \\
 206209135 &     1 &       CP &   CP &   1.2e-04 &   1.2e-05 &       no &  0.030 &        0.030 &          0.030 &   5.57722 &  0.00043 &  0.00042 &      2177.376 &             0.002 &             0.002 \\
 206209135 &     2 &       CP &   CP &   8.4e-05 &   1.1e-04 &       no &  0.032 &        0.031 &          0.029 &  15.18903 &  0.00320 &  0.00309 &      2156.465 &             0.005 &             0.005 \\
 206209135 &     3 &       CP &   CP &   7.2e-04 &   5.1e-04 &       no &  0.028 &        0.026 &          0.030 &   7.76013 &  0.00147 &  0.00148 &      2151.788 &             0.007 &             0.008 \\
 206209135 &     4 &       CP &   CP &   8.4e-05 &   1.1e-05 &       no &  0.036 &        0.038 &          0.034 &  24.15872 &  0.00386 &  0.00372 &      2154.054 &             0.005 &             0.005 \\
 206312951 &     1 &       PC &   PC &   2.3e-02 &   2.1e-02 &      $\cdots$ &  0.022 &        0.021 &          0.410 &   1.53402 &  0.00017 &  0.00018 &      2147.159 &             0.005 &             0.005 \\
 206318379 &     1 &       PC &   PC &   1.0e-01 &   4.6e-02 &      $\cdots$ &  0.075 &        0.076 &          0.076 &   2.26043 &  0.00003 &  0.00003 &      2147.250 &             0.000 &             0.000 \\
 210448987 &     1 &       CP &   CP &   2.8e-07 &   7.5e-08 &       no &  0.029 &        0.028 &          0.027 &   6.10230 &  0.00040 &  0.00039 &      2237.441 &             0.002 &             0.002 \\
 210508766 &     1 &       CP &   CP &   5.1e-04 &   3.9e-04 &       no &  0.029 &        0.028 &          0.029 &   2.74723 &  0.00013 &  0.00013 &      2234.060 &             0.002 &             0.002 \\
 210508766 &     2 &       CP &   CP &   1.8e-02 &   1.9e-02 &       no &  0.035 &        0.034 &          0.034 &   9.99744 &  0.00060 &  0.00062 &      2233.271 &             0.002 &             0.002 \\
 210558622 &     1 &       PC &   CP &   2.2e-06 &   2.1e-10 &       no &  0.035 &        0.033 &          0.033 &  19.56325 &  0.00242 &  0.00264 &      2250.779 &             0.003 &             0.003 \\
 210564155 &     1 &       UK &   PC &   2.0e-02 &   1.5e-02 &      $\cdots$ &  0.035 &        0.034 &          0.034 &   4.86406 &  0.00028 &  0.00028 &      2263.759 &             0.001 &             0.001 \\
 210707130 &     1 &       CP &   CP &   1.4e-05 &   3.4e-05 &       no &  0.019 &        0.020 &          0.019 &   0.68456 &  0.00002 &  0.00002 &      2232.367 &             0.001 &             0.001 \\
 210750726 &     1 &       CP &   CP &   5.6e-04 &   3.8e-04 &       no &  0.047 &        0.045 &          0.047 &   4.61228 &  0.00017 &  0.00017 &      2233.218 &             0.001 &             0.001 \\
 210838726 &     1 &       CP &   CP &   9.8e-04 &   8.1e-04 &       no &  0.020 &        0.018 &          0.020 &   1.09598 &  0.00006 &  0.00006 &      2233.008 &             0.002 &             0.002 \\
 210968143 &     1 &       CP &   CP &   2.8e-13 &   9.8e-12 &       no &  0.037 &        0.036 &          0.037 &  13.73490 &  0.00081 &  0.00079 &      2245.659 &             0.001 &             0.001 \\
 210968143 &     2 &       CP &   CP &   2.6e-04 &   1.5e-04 &       no &  0.019 &        0.018 &          0.018 &   2.90070 &  0.00032 &  0.00034 &      2233.740 &             0.004 &             0.004 \\
 211077024 &     1 &       CP &   CP &   1.9e-05 &   4.1e-06 &       no &  0.035 &        0.032 &          0.032 &   1.41960 &  0.00006 &  0.00006 &      2232.430 &             0.002 &             0.002 \\
 211305568 &     1 &       UK &   PC &   5.6e-01 &   1.0e+00 &      $\cdots$ &  0.041 &        0.043 &          0.038 &  11.55059 &  0.00109 &  0.00121 &      2336.344 &             0.002 &             0.002 \\
 211305568 &     2 &       UK &   PC &  -1.0e+03 &  -1.0e+03 &      $\cdots$ &  0.015 &        0.016 &          0.017 &   0.19785 &  0.00001 &  0.00001 &      2343.970 &             0.002 &             0.001 \\
 211331236 &     1 &       PC &   CP &   4.7e-08 &   2.8e-08 &       no &  0.037 &        0.037 &          0.036 &   1.29151 &  0.00004 &  0.00004 &      2309.776 &             0.001 &             0.001 \\
 211331236 &     2 &       UK &   CP &   2.2e-06 &   3.5e-06 &       no &  0.038 &        0.037 &          0.037 &   5.44481 &  0.00042 &  0.00040 &      2310.560 &             0.003 &             0.004 \\
 211336288 &     1 &       UK &   PC &   1.9e-01 &   1.2e-01 &      $\cdots$ &  0.018 &        0.015 &          0.017 &   0.22181 &  0.00002 &  0.00002 &      2344.090 &             0.002 &             0.002 \\
 211357309 &     1 &       PC &   PC &   7.6e-02 &   6.7e-02 &      $\cdots$ &  0.017 &        0.018 &          0.018 &   0.46395 &  0.00002 &  0.00002 &      2368.458 &             0.001 &             0.001 \\
 211428897 &     1 &       PC &   PC &   7.0e-06 &   5.5e-08 &      yes &  0.024 &        0.025 &          0.023 &   1.61092 &  0.00006 &  0.00006 &      2309.275 &             0.001 &             0.001 \\
 211428897 &     2 &       UK &   PC &   7.4e-04 &   2.5e-05 &      yes &  0.021 &        0.019 &          0.021 &   2.17807 &  0.00012 &  0.00012 &      2310.647 &             0.002 &             0.002 \\
 211428897 &     3 &       UK &   PC &   1.2e-03 &   5.2e-04 &      yes &  0.021 &        0.021 &          0.020 &   4.96883 &  0.00037 &  0.00038 &      2340.523 &             0.002 &             0.002 \\
 211509553 &     1 &       PC &   PC &   1.0e-03 &   1.3e-03 &      yes &  0.182 &        0.176 &          0.177 &  20.35954 &  0.00032 &  0.00034 &      2318.412 &             0.001 &             0.001 \\
 211680698 &     1 &       UK &   CP &   5.6e-03 &   7.6e-03 &      $\cdots$ &  0.032 &        0.029 &          0.032 &  50.92099 &  0.00519 &  0.00556 &      2327.473 &             0.004 &             0.004 \\
 211694226 &     1 &       UK &   PC &   2.9e-01 &   6.2e-01 &      yes &  0.021 &        0.017 &          0.449 &   1.91828 &  0.00017 &  0.00019 &      2342.946 &             0.002 &             0.002 \\
 211762841 &     1 &       UK &   PC &   9.8e-02 &   1.2e-01 &      $\cdots$ &  0.032 &        0.261 &          0.191 &   1.56493 &  0.00009 &  0.00009 &      2343.261 &             0.001 &             0.001 \\
 211770795 &     1 &       PC &   CP &   1.2e-05 &   7.6e-05 &       no &  0.031 &        0.031 &          0.029 &   7.72857 &  0.00070 &  0.00073 &      2315.826 &             0.003 &             0.003 \\
 211791178 &     1 &       UK &   CP &   1.7e-03 &   1.5e-03 &      $\cdots$ &  0.028 &        0.028 &          0.029 &   9.56276 &  0.00069 &  0.00068 &      2342.604 &             0.002 &             0.002 \\
 211799258 &     1 &       UK &   PC &   9.0e-01 &   9.4e-01 &       no &  0.274 &        0.275 &          0.316 &  19.53405 &  0.00078 &  0.00078 &      2320.146 &             0.001 &             0.001 \\
 211817229 &     1 &       UK &   PC &   3.9e-01 &   9.5e-01 &      $\cdots$ &  0.307 &        0.307 &          0.217 &   2.17693 &  0.00006 &  0.00005 &      2342.525 &             0.001 &             0.001 \\
 211818569 &     1 &       PC &   CP &   1.4e-04 &   6.1e-04 &       no &  0.110 &        0.109 &          0.109 &   5.18575 &  0.00020 &  0.00019 &      2310.560 &             0.001 &             0.001 \\
 211822797 &     1 &       UK &   CP &   6.7e-04 &   7.6e-04 &      $\cdots$ &  0.032 &        0.030 &          0.029 &  21.16987 &  0.00171 &  0.00175 &      2332.577 &             0.002 &             0.002 \\
 211826814 &     1 &       UK &   PC &   5.6e-01 &   9.6e-01 &      $\cdots$ &  0.036 &        0.348 &          0.059 &   1.53453 &  0.00012 &  0.00014 &      2343.198 &             0.002 &             0.002 \\
 211831378 &     1 &       UK &   FP &   2.4e-05 &   1.9e-06 &       no &  0.015 &        0.074 &          0.606 &   3.48928 &  0.00049 &  0.00048 &      2310.755 &             0.005 &             0.005 \\
 211924657 &     1\tablenotemark{d} &       PC &   PC &   5.3e-01 &   6.4e-01 &       no &  0.052 &        0.051 &          0.051 &   2.64484 &  0.00011 &  0.00011 &      2311.641 &             0.001 &             0.001 \\
 211965883 &     1 &       PC &   PC &   3.4e-01 &   3.8e-01 &      $\cdots$ &  0.044 &        0.040 &          0.040 &  10.55632 &  0.00065 &  0.00067 &      2334.605 &             0.001 &             0.001 \\
 211969807 &     1 &       PC &   PC &   8.6e-02 &   9.5e-02 &      $\cdots$ &  0.037 &        0.036 &          0.037 &   1.97424 &  0.00011 &  0.00011 &      2342.915 &             0.001 &             0.001 \\
 211970234 &     1 &       UK &   FP &   1.6e-01 &   1.9e-01 &       no &  0.068 &        0.105 &          0.201 &   1.48350 &  0.00003 &  0.00004 &      2310.371 &             0.001 &             0.001 \\
 211988320 &     1 &       UK &   PC &   7.5e-02 &   7.2e-02 &      $\cdots$ &  0.041 &        0.036 &          0.034 &  63.84825 &  0.00598 &  0.00562 &      2309.721 &             0.004 &             0.005 \\
 212006344 &     1 &       PC &   CP &   1.5e-04 &   1.7e-04 &       no &  0.020 &        0.019 &          0.020 &   2.21940 &  0.00007 &  0.00007 &      2311.048 &             0.001 &             0.001 \\
 212069861 &     1 &       PC &   CP &   1.2e-04 &   2.9e-06 &       no &  0.043 &        0.043 &          0.040 &  30.95679 &  0.00266 &  0.00258 &      2314.491 &             0.003 &             0.003 \\
 212154564 &     1 &       PC &   CP &   6.8e-07 &   9.7e-07 &       no &  0.070 &        0.067 &          0.069 &   6.41354 &  0.00025 &  0.00025 &      2309.182 &             0.002 &             0.002 \\
 212398486 &     1 &       UK &   CP &   7.5e-03 &   1.7e-03 &      $\cdots$ &  0.050 &        0.050 &          0.045 &  21.75025 &  0.00197 &  0.00200 &      2410.088 &             0.002 &             0.002 \\
 212443973 &     1 &       PC &   PC &   1.6e-01 &   8.9e-01 &      $\cdots$ &  0.023 &        0.356 &          0.165 &   0.77923 &  0.00005 &  0.00005 &      2423.710 &             0.002 &             0.002 \\
 212460519 &     1 &       PC &   CP &   4.2e-13 &   1.1e-11 &       no &  0.027 &        0.028 &          0.027 &   7.38707 &  0.00026 &  0.00026 &      2390.794 &             0.001 &             0.001 \\
 212554013 &     1 &       PC &   CP &   1.0e-03 &   1.6e-03 &       no &  0.118 &        0.115 &          0.117 &   3.58816 &  0.00001 &  0.00001 &      2390.926 &             0.000 &             0.000 \\
 212572452 &     1 &       PC &   FP &   3.3e-07 &   4.3e-08 &       no &  0.073 &        0.068 &          0.176 &   2.58148 &  0.00001 &  0.00001 &      2390.028 &             0.000 &             0.000 \\
 212628098 &     1 &       FP &   FP &   9.2e-02 &   3.0e-02 &      yes &  0.228 &        0.385 &          0.288 &   4.35244 &  0.00003 &  0.00003 &      2390.348 &             0.000 &             0.000 \\
 212634172 &     1 &       UK &   PC &   3.0e-01 &   1.0e+00 &      $\cdots$ &  0.088 &        0.074 &          0.079 &   2.85177 &  0.00010 &  0.00010 &      2421.669 &             0.001 &             0.001 \\
 212679181 &     1 &       PC &   PC &   4.2e-04 &   6.4e-04 &      yes &  0.027 &        0.025 &          0.027 &   1.05459 &  0.00001 &  0.00001 &      2423.570 &             0.000 &             0.000 \\
 212679798 &     1 &       UK &   FP &   2.1e-01 &   9.5e-01 &      yes &  0.489 &        0.673 &          0.192 &   1.83473 &  0.00002 &  0.00002 &      2389.389 &             0.002 &             0.002 \\
 212686205 &     1 &       UK &   CP &   1.1e-07 &   4.2e-06 &       no &  0.017 &        0.016 &          0.015 &   5.67581 &  0.00042 &  0.00041 &      2422.503 &             0.002 &             0.002 \\
 212690867 &     1 &       UK &   PC &   2.9e-02 &   3.5e-02 &      $\cdots$ &  0.049 &        0.046 &          0.049 &  25.86125 &  0.00311 &  0.00299 &      2422.464 &             0.003 &             0.003 \\
 212773272 &     1 &       UK &   FP &   9.4e-01 &   9.8e-01 &       no &  0.630 &        0.205 &          0.172 &   4.68189 &  0.00005 &  0.00005 &      2389.666 &             0.001 &             0.001 \\
 212773309 &     1 &       FP &   FP &   8.5e-01 &   9.4e-01 &      yes &  0.754 &        0.740 &          0.784 &   4.68199 &  0.00007 &  0.00006 &      2389.665 &             0.001 &             0.001 \\
 212773309 &     1 &       FP &   FP &   8.5e-01 &   9.4e-01 &      yes &  0.754 &        0.740 &          0.784 &   4.68199 &  0.00007 &  0.00006 &      2389.665 &             0.001 &             0.001 \\
 213951550 &     1 &       UK &   FP &   1.0e+00 &   1.0e+00 &       no &  0.661 &        0.640 &          $\cdots$ &   1.11704 &  0.00002 &  0.00002 &      2478.230 &             0.001 &             0.001 \\
 214254518 &     1 &       UK &   PC &   9.8e-03 &   1.2e-02 &      $\cdots$ &  0.015 &        0.015 &         $\cdots$ &   5.05900 &  0.00053 &  0.00053 &      2506.630 &             0.003 &             0.003 \\
 214522613 &     1 &       UK &   PC &   5.1e-02 &   4.9e-02 &      $\cdots$ &  0.024 &        0.022 &          $\cdots$ &  10.99044 &  0.00127 &  0.00126 &      2506.314 &             0.002 &             0.002 \\
 214787262 &     1 &       UK &   CP &   6.1e-04 &   1.4e-03 &      $\cdots$ &  0.027 &        0.026 &          $\cdots$ &   8.23949 &  0.00029 &  0.00030 &      2502.009 &             0.001 &             0.001 \\
 216892056 &     1 &       UK &   PC &   6.1e-01 &   9.8e-01 &       no &  0.342 &        0.061 &          $\cdots$ &   2.78592 &  0.00005 &  0.00005 &      2478.406 &             0.001 &             0.001 \\
 217941732 &     1 &       UK &   CP &   1.2e-06 &   2.9e-07 &       no &  0.015 &        0.015 &          $\cdots$ &   2.49413 &  0.00013 &  0.00013 &      2507.612 &             0.001 &             0.001 \\\enddata
\tablenotetext{a}{The ``Small $R_p$'' and ``Big $R_p$'' values refer to {\tt vespa} false positive probability estimates made using the 16th and 84th percentiles of the $R_p/R_\star$ posterior probability distribution, respectively. Note that {\tt vespa} operates in a statistical fashion and that the reported FPP may occasionally be higher for the small $R_p$ case than for the big $R_p$ case due to changes in the simulated population of stars. }
\tablenotetext{b}{Indicates whether a nearby star was revealed in the follow-up AO and speckle imagery (see Table~\ref{tab:ao}). Rows without definitive answers mark stars without follow-up images posted to the ExoFOP-K2 website. Note that planet radius estimates are not corrected for flux dilution due to the presence of nearby stars.}
\tablenotetext{c}{Values and errors for orbital period and transit center are from fits using K2SFF photometry.}
\tablenotetext{d}{This K2OI displays transit timing variations, but our fits assumed a linear ephemeris.}
\end{deluxetable*}

\begin{deluxetable*}{ccccccccccccccccccc}
\tablecolumns{19}
\tabletypesize{\tiny}
\tablecaption{Planet Properties\tablenotemark{a} \label{tab:fits}}
\tablehead{
\colhead{} &
\colhead{} &
\colhead{$P$\tablenotemark{b}} & 
\multicolumn{3}{c}{$R_p/R_\star$} & 
\colhead{} &
\colhead{} &
\multicolumn{2}{c}{Limb Darkening} & 
\colhead{} &
\colhead{} &
\multicolumn{3}{c}{$R_p$ ($\rearth$)} & 
\colhead{$a$} &
\multicolumn{3}{c}{$F_p$ ($\fearth$)} \\ 
\cline{4-6}
\cline{9-10}
\cline{13-15}
\cline{17-19}
\colhead{EPIC} &
\colhead{K2OI} &
\colhead{(d)} &
\colhead{Val} &
\colhead{-Err} &
\colhead{+Err} &
\colhead{$a/R_\star$} &
\colhead{$i$} & 
\colhead{$q_1$} & 
\colhead{$q_2$} & 
\colhead{$e$} & 
\colhead{$\omega$} & 
\colhead{Val} &
\colhead{-Err} &
\colhead{+Err} &
\colhead{(au)} &
\colhead{Val} &
\colhead{-Err} &
\colhead{+Err} }
\startdata
201205469 &     1 &   3.47134 &  0.074 &      0.002 &      0.002 &             13.51 &          88.88 &          0.47 &          0.31 &            0.1 &              207 &   4.76 &   0.33 &   0.34 &  0.038 &    46.4 &      16.3 &      23.0 \\
 201208431 &     1 &  10.00339 &  0.034 &      0.001 &      0.001 &             27.72 &          89.51 &          0.45 &          0.33 &            0.1 &              229 &   2.10 &   0.19 &   0.20 &  0.078 &    15.6 &       6.2 &       9.2 \\
 201345483 &     1 &   1.72926 &  0.140 &      0.002 &      0.003 &              8.07 &          87.77 &          0.45 &          0.43 &            0.1 &              128 &  10.44 &   0.70 &   0.90 &  0.025 &   380.5 &     150.1 &     219.7 \\
 201549860 &     1 &   5.60836 &  0.029 &      0.001 &      0.001 &             19.13 &          88.51 &          0.55 &          0.44 &            0.1 &              151 &   1.94 &   0.12 &   0.13 &  0.055 &    68.0 &      10.5 &      12.1 \\
 201549860 &     2 &   2.39996 &  0.020 &      0.001 &      0.001 &             10.54 &          88.92 &          0.55 &          0.44 &            0.1 &              258 &   1.32 &   0.08 &   0.08 &  0.031 &   211.0 &      32.5 &      37.4 \\
 201617985 &     1 &   7.28116 &  0.033 &      0.003 &      0.003 &             26.25 &          88.28 &          0.47 &          0.27 &            0.2 &              127 &   1.79 &   0.18 &   0.20 &  0.060 &     9.2 &       2.6 &       3.3 \\
 201635569 &     1 &   8.36879 &  0.111 &      0.004 &      0.004 &             23.05 &          88.20 &          0.51 &          0.33 &            0.2 &              251 &   7.52 &   0.48 &   0.48 &  0.069 &     5.5 &       3.2 &       6.1 \\
 201637175 &     1 &   0.38108 &  0.076 &      0.004 &      0.004 &              3.19 &          77.25 &          0.48 &          0.30 &            0.1 &              146 &   4.81 &   0.37 &   0.34 &  0.009 &   737.9 &     196.7 &     242.8 \\
 201717274 &     1 &   3.52674 &  0.038 &      0.002 &      0.004 &             15.58 &          88.32 &          0.59 &          0.29 &            0.1 &              178 &   1.32 &   0.25 &   0.26 &  0.026 &    15.0 &       3.3 &       4.1 \\
 201855371 &     1 &  17.96901 &  0.030 &      0.002 &      0.002 &             40.27 &          89.08 &          0.51 &          0.37 &            0.2 &              135 &   2.04 &   0.16 &   0.18 &  0.117 &    10.5 &       2.9 &       3.7 \\
 205924614 &     1 &   2.84927 &  0.056 &      0.001 &      0.002 &             10.44 &          88.01 &          0.53 &          0.44 &            0.1 &              172 &   4.40 &   0.26 &   0.29 &  0.035 &   141.3 &      23.5 &      28.8 \\
 206011691 &     1 &   9.32502 &  0.026 &      0.001 &      0.001 &             25.40 &          88.97 &          0.51 &          0.41 &            0.1 &              147 &   1.85 &   0.10 &   0.11 &  0.076 &    10.0 &       1.7 &       2.1 \\
 206011691 &     2 &  15.50189 &  0.036 &      0.002 &      0.001 &             35.62 &          88.84 &          0.50 &          0.41 &            0.2 &              123 &   2.51 &   0.20 &   0.15 &  0.107 &     5.1 &       0.9 &       1.1 \\
 206119924 &     1 &   4.65541 &  0.009 &      0.000 &      0.000 &             15.37 &          89.04 &          0.54 &          0.44 &            0.1 &              211 &   0.69 &   0.04 &   0.04 &  0.048 &    78.6 &      10.5 &      12.2 \\
 206209135 &     1 &   5.57722 &  0.030 &      0.001 &      0.002 &             24.44 &          89.16 &          0.54 &          0.27 &            0.1 &              184 &   1.08 &   0.11 &   0.11 &  0.040 &     8.5 &       1.1 &       1.2 \\
 206209135 &     2 &  15.18903 &  0.032 &      0.002 &      0.002 &             48.09 &          89.53 &          0.54 &          0.27 &            0.1 &              173 &   1.16 &   0.13 &   0.13 &  0.078 &     2.2 &       0.3 &       0.3 \\
 206209135 &     3 &   7.76013 &  0.028 &      0.002 &      0.002 &             30.22 &          89.26 &          0.54 &          0.27 &            0.1 &              181 &   1.01 &   0.12 &   0.12 &  0.050 &     5.4 &       0.7 &       0.8 \\
 206209135 &     4 &  24.15872 &  0.036 &      0.002 &      0.002 &             65.01 &          89.69 &          0.54 &          0.27 &            0.1 &              177 &   1.29 &   0.13 &   0.14 &  0.106 &     1.2 &       0.2 &       0.2 \\
 206312951 &     1 &   1.53402 &  0.022 &      0.002 &      0.002 &              9.44 &          87.18 &          0.47 &          0.27 &            0.1 &              155 &   1.13 &   0.10 &   0.11 &  0.021 &   120.2 &      16.8 &      19.0 \\
 206318379 &     1 &   2.26043 &  0.075 &      0.002 &      0.004 &             16.57 &          88.40 &          0.54 &          0.27 &            0.1 &              150 &   2.30 &   0.26 &   0.28 &  0.020 &    30.1 &       3.8 &       4.5 \\
 210448987 &     1 &   6.10230 &  0.029 &      0.001 &      0.001 &             19.79 &          89.30 &          0.54 &          0.45 &            0.1 &              209 &   1.97 &   0.11 &   0.12 &  0.059 &    62.9 &       8.4 &       9.2 \\
 210508766 &     1 &   2.74723 &  0.029 &      0.001 &      0.001 &             12.62 &          88.68 &          0.47 &          0.29 &            0.1 &              188 &   1.71 &   0.10 &   0.10 &  0.032 &    38.8 &       5.9 &       6.3 \\
 210508766 &     2 &   9.99744 &  0.035 &      0.001 &      0.001 &             29.67 &          89.47 &          0.47 &          0.29 &            0.1 &              208 &   2.12 &   0.12 &   0.12 &  0.076 &     6.9 &       1.0 &       1.1 \\
 210558622 &     1 &  19.56325 &  0.035 &      0.001 &      0.001 &             38.63 &          89.74 &          0.53 &          0.43 &            0.5 &              272 &   2.59 &   0.15 &   0.17 &  0.125 &    13.2 &       2.1 &       2.3 \\
 210564155 &     1 &   4.86406 &  0.035 &      0.002 &      0.002 &             27.35 &          89.03 &          0.53 &          0.26 &            0.1 &              156 &   1.09 &   0.13 &   0.13 &  0.036 &     7.7 &       1.0 &       1.2 \\
 210707130 &     1 &   0.68456 &  0.019 &      0.001 &      0.001 &              4.30 &          80.09 &          0.55 &          0.45 &            0.2 &              124 &   1.39 &   0.12 &   0.10 &  0.013 &  1069.1 &     144.7 &     163.5 \\
 210750726 &     1 &   4.61228 &  0.047 &      0.003 &      0.003 &             20.05 &          87.74 &          0.48 &          0.27 &            0.2 &              126 &   2.35 &   0.22 &   0.22 &  0.042 &    16.4 &       2.0 &       2.2 \\
 210838726 &     1 &   1.09598 &  0.020 &      0.001 &      0.001 &              7.35 &          85.64 &          0.47 &          0.28 &            0.1 &              137 &   1.08 &   0.08 &   0.09 &  0.017 &   144.4 &      18.0 &      20.2 \\
 210968143 &     1 &  13.73490 &  0.037 &      0.001 &      0.001 &             33.90 &          89.38 &          0.54 &          0.45 &            0.1 &              173 &   2.53 &   0.13 &   0.15 &  0.100 &    10.2 &       1.4 &       1.7 \\
 210968143 &     2 &   2.90070 &  0.019 &      0.001 &      0.002 &             12.02 &          86.88 &          0.55 &          0.45 &            0.2 &              133 &   1.31 &   0.12 &   0.13 &  0.035 &    80.8 &      11.1 &      13.2 \\
 211077024 &     1 &   1.41960 &  0.035 &      0.001 &      0.001 &             11.19 &          88.60 &          0.43 &          0.25 &            0.1 &              208 &   1.22 &   0.12 &   0.12 &  0.018 &    56.1 &       6.6 &       7.4 \\
 211305568 &     1 &  11.55059 &  0.041 &      0.004 &      0.553 &             37.16 &          88.70 &          0.48 &          0.26 &            0.2 &              150 &   1.98 &   0.23 &  26.91 &  0.078 &     5.7 &       0.7 &       0.8 \\
 211305568 &     2 &   0.19785 &  0.015 &      0.001 &      0.002 &              2.51 &          79.23 &          0.49 &          0.26 &            0.1 &              157 &   0.72 &   0.08 &   0.10 &  0.005 &  1292.5 &     158.3 &     178.2 \\
 211331236 &     1 &   1.29151 &  0.037 &      0.001 &      0.001 &              8.33 &          88.28 &          0.46 &          0.28 &            0.1 &              208 &   1.96 &   0.12 &   0.12 &  0.019 &   145.8 &      18.9 &      21.5 \\
 211331236 &     2 &   5.44481 &  0.038 &      0.001 &      0.001 &             21.04 &          89.47 &          0.46 &          0.28 &            0.2 &              260 &   2.03 &   0.13 &   0.13 &  0.051 &    21.4 &       2.8 &       3.2 \\
 211336288 &     1 &   0.22181 &  0.018 &      0.002 &      0.002 &              2.28 &          76.90 &          0.48 &          0.33 &            0.1 &              145 &   1.13 &   0.13 &   0.16 &  0.006 &  1140.9 &     151.6 &     172.3 \\
 211357309 &     1 &   0.46395 &  0.017 &      0.001 &      0.001 &              4.29 &          86.70 &          0.47 &          0.27 &            0.1 &              224 &   0.84 &   0.06 &   0.06 &  0.010 &   437.3 &      56.8 &      63.9 \\
 211428897 &     1 &   1.61092 &  0.024 &      0.001 &      0.001 &             12.87 &          89.09 &          0.47 &          0.25 &            0.2 &              254 &   0.75 &   0.08 &   0.08 &  0.021 &    48.0 &       5.9 &       6.8 \\
 211428897 &     2 &   2.17807 &  0.021 &      0.001 &      0.001 &             16.48 &          89.22 &          0.46 &          0.25 &            0.1 &              240 &   0.65 &   0.07 &   0.07 &  0.025 &    32.1 &       3.9 &       4.6 \\
 211428897 &     3 &   4.96883 &  0.021 &      0.001 &      0.001 &             29.34 &          89.53 &          0.47 &          0.25 &            0.1 &              229 &   0.67 &   0.08 &   0.08 &  0.044 &    10.7 &       1.3 &       1.5 \\
 211509553 &     1 &  20.35954 &  0.182 &      0.002 &      0.003 &             47.66 &          89.58 &          0.48 &          0.28 &            0.1 &              142 &  10.85 &   0.59 &   0.59 &  0.119 &     1.8 &       0.3 &       0.4 \\
 211680698 &     1 &  50.92099 &  0.032 &      0.002 &      0.002 &             71.88 &          89.48 &          0.54 &          0.45 &            0.2 &              131 &   2.57 &   0.21 &   0.23 &  0.245 &     4.3 &       0.6 &       0.6 \\
 211694226 &     1 &   1.91828 &  0.021 &      0.002 &      0.002 &             10.53 &          86.20 &          0.49 &          0.26 &            0.1 &              136 &   1.00 &   0.11 &   0.14 &  0.021 &    75.8 &      12.1 &      13.8 \\
 211762841 &     1 &   1.56493 &  0.032 &      0.004 &      0.005 &              7.92 &          83.53 &          0.51 &          0.27 &            0.1 &              151 &   2.22 &   0.30 &   0.38 &  0.023 &   157.6 &      25.9 &      29.7 \\
 211770795 &     1 &   7.72857 &  0.031 &      0.001 &      0.001 &             21.60 &          89.38 &          0.52 &          0.27 &            0.1 &              221 &   2.29 &   0.13 &   0.15 &  0.070 &    54.9 &       8.8 &       9.6 \\
 211791178 &     1 &   9.56276 &  0.028 &      0.001 &      0.001 &             24.17 &          89.55 &          0.49 &          0.38 &            0.1 &              262 &   2.01 &   0.11 &   0.13 &  0.078 &    35.1 &       5.1 &       5.9 \\
 211799258 &     1 &  19.53405 &  0.274 &      0.030 &      0.166 &             80.10 &          89.28 &          0.54 &          0.45 &            0.4 &               92 &   9.82 &   2.13 &   6.29 &  0.087 &     1.0 &       0.6 &       1.4 \\
 211817229 &     1 &   2.17693 &  0.307 &      0.232 &      0.458 &             16.29 &          85.59 &          0.55 &          0.44 &            0.1 &              196 &   7.92 &   6.13 &  11.91 &  0.019 &    15.1 &      11.9 &      56.6 \\
 211818569 &     1 &   5.18575 &  0.110 &      0.006 &      0.005 &             14.72 &          87.21 &          0.50 &          0.27 &            0.2 &              116 &   9.22 &   0.71 &   0.65 &  0.052 &    89.5 &      11.2 &      12.5 \\
 211822797 &     1 &  21.16987 &  0.032 &      0.001 &      0.001 &             48.63 &          89.70 &          0.45 &          0.25 &            0.1 &              207 &   1.99 &   0.11 &   0.11 &  0.131 &     3.5 &       0.5 &       0.5 \\
 211826814 &     1 &   1.53453 &  0.036 &      0.011 &      0.222 &             12.69 &          86.05 &          0.56 &          0.47 &            0.1 &              183 &   1.02 &   0.36 &   6.35 &  0.015 &    26.0 &      18.5 &      63.9 \\
 211831378 &     1 &   3.48928 &  0.015 &      0.001 &      0.001 &             14.05 &          88.87 &          0.49 &          0.37 &            0.1 &              210 &   0.91 &   0.07 &   0.07 &  0.037 &    24.6 &       7.1 &      10.5 \\
 211924657 &     1\tablenotemark{c} &   2.64484 &  0.052 &      0.001 &      0.002 &             14.03 &          88.79 &          0.48 &          0.28 &            0.1 &              221 &   1.83 &   0.21 &   0.25 &  0.026 &    18.8 &       2.6 &       2.9 \\
 211965883 &     1 &  10.55632 &  0.044 &      0.004 &      0.003 &             29.69 &          88.32 &          0.43 &          0.25 &            0.1 &              142 &   2.86 &   0.28 &   0.24 &  0.083 &    11.4 &       1.5 &       1.7 \\
 211969807 &     1 &   1.97424 &  0.037 &      0.001 &      0.002 &              9.97 &          88.02 &          0.50 &          0.26 &            0.1 &              184 &   1.96 &   0.15 &   0.15 &  0.023 &    62.1 &      13.8 &      16.0 \\
 211970234 &     1 &   1.48350 &  0.068 &      0.003 &      0.008 &             13.42 &          88.02 &          0.50 &          0.40 &            0.1 &              175 &   1.41 &   0.30 &   0.32 &  0.015 &    19.2 &       4.3 &       5.0 \\
 211988320 &     1 &  63.84825 &  0.041 &      0.001 &      0.001 &             92.48 &          89.81 &          0.50 &          0.27 &            0.1 &              181 &   2.86 &   0.15 &   0.16 &  0.276 &     0.9 &       0.1 &       0.1 \\
 212006344 &     1 &   2.21940 &  0.020 &      0.001 &      0.001 &             10.45 &          86.36 &          0.47 &          0.25 &            0.2 &              127 &   1.28 &   0.08 &   0.08 &  0.029 &    79.9 &      11.1 &      13.1 \\
 212069861 &     1 &  30.95679 &  0.043 &      0.001 &      0.001 &             61.98 &          89.79 &          0.53 &          0.42 &            0.1 &              219 &   2.65 &   0.15 &   0.15 &  0.167 &     2.9 &       0.4 &       0.5 \\
 212154564 &     1 &   6.41354 &  0.070 &      0.002 &      0.002 &             30.20 &          89.52 &          0.47 &          0.33 &            0.1 &              210 &   2.65 &   0.24 &   0.24 &  0.051 &     8.7 &       1.1 &       1.2 \\
 212398486 &     1 &  21.75025 &  0.050 &      0.002 &      0.002 &             63.41 &          89.71 &          0.45 &          0.25 &            0.1 &              191 &   2.19 &   0.18 &   0.19 &  0.121 &     2.0 &       0.3 &       0.3 \\
 212443973 &     1 &   0.77923 &  0.023 &      0.004 &      0.083 &              7.23 &          83.74 &          0.48 &          0.26 &            0.1 &              166 &   0.87 &   0.17 &   3.11 &  0.011 &    98.7 &      11.5 &      13.0 \\
 212460519 &     1 &   7.38707 &  0.027 &      0.000 &      0.000 &             22.41 &          89.40 &          0.47 &          0.26 &            0.1 &              213 &   1.84 &   0.10 &   0.11 &  0.066 &    35.3 &       5.9 &       6.6 \\
 212554013 &     1 &   3.58816 &  0.118 &      0.002 &      0.002 &             12.11 &          87.24 &          0.54 &          0.27 &            0.1 &              258 &   8.68 &   0.59 &   0.68 &  0.041 &   105.6 &      17.8 &      20.7 \\
 212572452 &     1 &   2.58148 &  0.073 &      0.003 &      0.002 &              9.91 &          86.29 &          0.57 &          0.45 &            0.2 &              262 &   5.38 &   0.38 &   0.38 &  0.033 &   132.0 &      33.7 &      41.8 \\
 212628098 &     1 &   4.35244 &  0.228 &      0.005 &      0.005 &             17.23 &          87.31 &          0.55 &          0.42 &            0.2 &               92 &  14.07 &   0.77 &   0.77 &  0.044 &    81.2 &      11.7 &      13.1 \\
 212634172 &     1 &   2.85177 &  0.088 &      0.023 &      0.430 &             14.76 &          86.39 &          0.55 &          0.44 &            0.1 &              187 &   3.34 &   0.94 &  16.30 &  0.027 &    18.8 &       2.6 &       2.9 \\
 212679181 &     1 &   1.05459 &  0.027 &      0.003 &      0.002 &              7.99 &          83.60 &          0.47 &          0.31 &            0.1 &              140 &   1.29 &   0.17 &   0.14 &  0.016 &   114.7 &      13.9 &      16.5 \\
 212679798 &     1 &   1.83473 &  0.489 &      0.261 &      0.352 &              8.94 &          84.18 &          0.53 &          0.27 &            0.5 &              272 &  29.93 &  16.09 &  21.63 &  0.024 &   164.7 &      28.0 &      35.0 \\
 212686205 &     1 &   5.67581 &  0.017 &      0.001 &      0.001 &             15.57 &          87.37 &          0.48 &          0.26 &            0.2 &              124 &   1.43 &   0.13 &   0.16 &  0.056 &    68.8 &       9.7 &      11.1 \\
 212690867 &     1 &  25.86125 &  0.049 &      0.001 &      0.002 &             64.96 &          89.76 &          0.48 &          0.29 &            0.1 &              213 &   2.20 &   0.18 &   0.19 &  0.133 &     1.4 &       0.2 &       0.3 \\
 212773272 &     1 &   4.68189 &  0.630 &      0.274 &      0.249 &             15.44 &          85.45 &          0.56 &          0.47 &            0.1 &              211 &  29.41 &  12.93 &  11.81 &  0.036 &    13.8 &       2.0 &       2.4 \\
 212773309 &     1 &   4.68199 &  0.754 &      0.316 &      0.187 &             16.87 &          85.63 &          0.48 &          0.26 &            0.1 &              267 &  48.33 &  20.42 &  12.22 &  0.048 &    69.5 &       8.4 &       9.8 \\
 213951550 &     1 &   1.11704 &  0.661 &      0.308 &      0.248 &              6.79 &          80.01 &          0.56 &          0.28 &            0.1 &              234 &  33.94 &  15.98 &  12.94 &  0.016 &   166.0 &      26.5 &      31.7 \\
 214254518 &     1 &   5.05900 &  0.015 &      0.001 &      0.001 &             16.19 &          89.13 &          0.50 &          0.38 &            0.1 &              221 &   1.12 &   0.07 &   0.08 &  0.051 &    56.1 &       7.9 &       9.2 \\
 214522613 &     1 &  10.99044 &  0.024 &      0.002 &      0.002 &             36.32 &          88.86 &          0.49 &          0.26 &            0.2 &              129 &   1.16 &   0.12 &   0.13 &  0.075 &     6.9 &       1.2 &       1.4 \\
 214787262 &     1 &   8.23949 &  0.027 &      0.001 &      0.001 &             34.46 &          89.18 &          0.49 &          0.27 &            0.1 &              147 &   1.05 &   0.09 &   0.10 &  0.057 &     4.5 &       0.5 &       0.6 \\
 216892056 &     1 &   2.78592 &  0.342 &      0.222 &      0.364 &             15.37 &          85.24 &          0.54 &          0.44 &            0.0 &              207 &  14.87 &   9.70 &  15.85 &  0.028 &    25.5 &       3.1 &       3.5 \\
 217941732 &     1 &   2.49413 &  0.015 &      0.001 &      0.001 &              9.38 &          86.49 &          0.49 &          0.27 &            0.1 &              140 &   1.25 &   0.14 &   0.20 &  0.032 &   136.4 &      35.8 &      45.8 \\
\enddata
\tablenotetext{a}{For presentation purposes, this table contains only a subset of the available columns. See the machine-readable version for transit parameters based on fits to k2phot and K2SC photometry as well as errors on all parameters.}
\tablenotetext{b}{See Table~\ref{tab:fpp} for errors on $P$ and estimates of the transit center.}
\tablenotetext{c}{This K2OI displays transit timing variations, but our fits assumed a linear ephemeris.}
\end{deluxetable*}

\section{Revised Planet Properties}
\label{sec:revised_planets}
\begin{figure*}[tbh]
\centering
\includegraphics[width=0.45\textwidth]{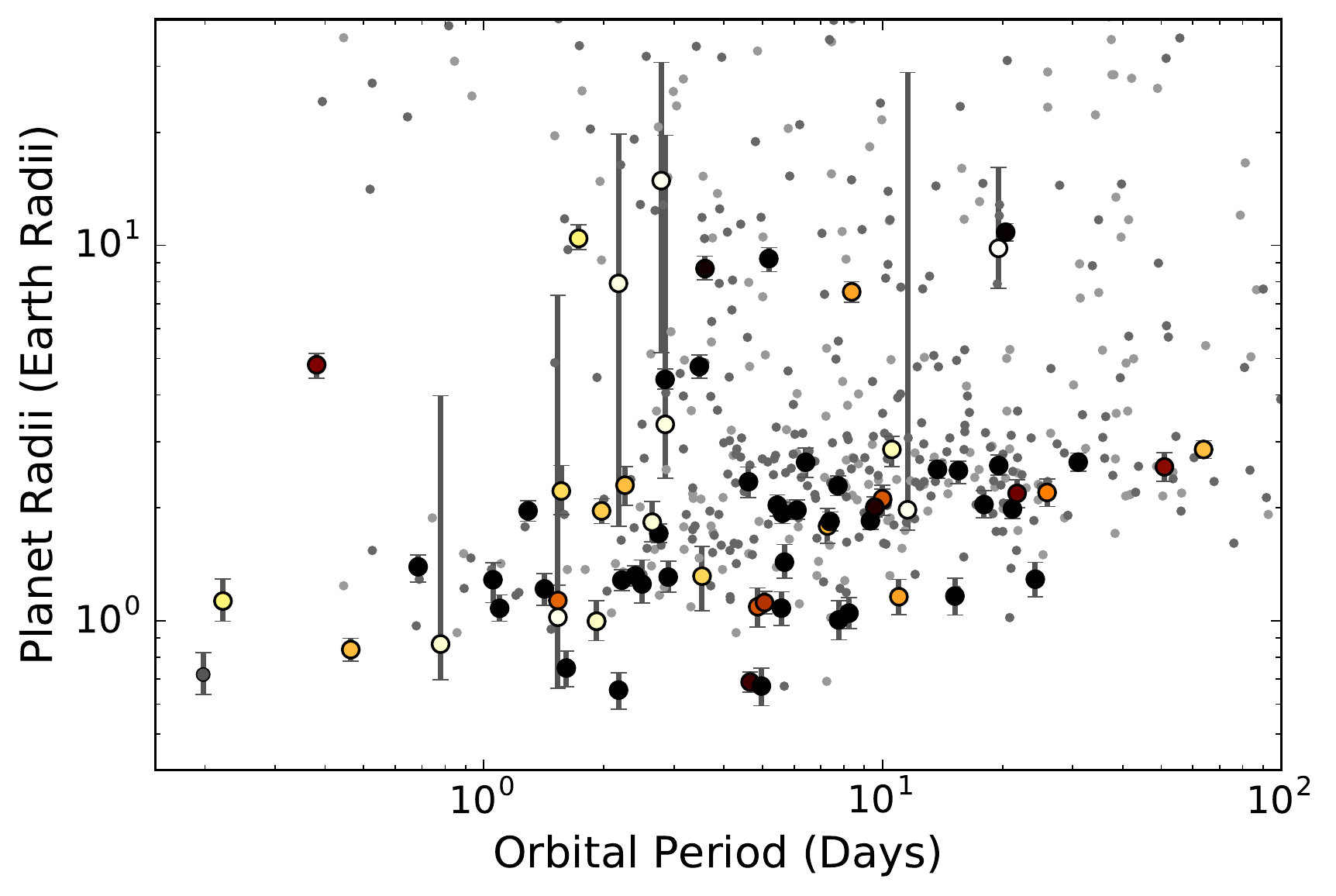}
\includegraphics[width=0.45\textwidth]{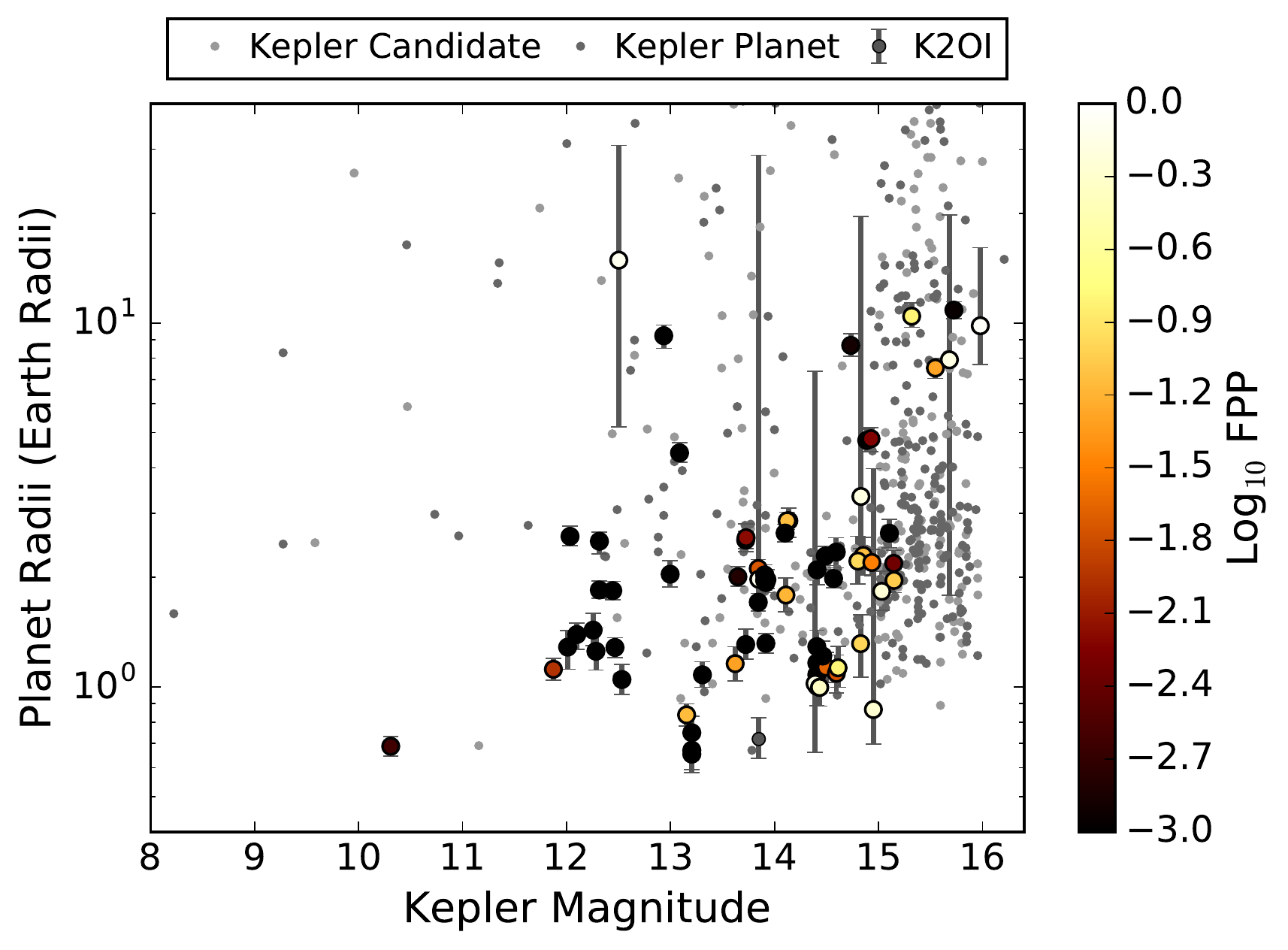}
\caption{Revised radii versus orbital period (left) or $Kp$ (right) for all K2OIs observed in this work that were not classified as false positives. The colors indicate the average of the two false positive probability estimates for each K2OI as shown in the legend. For reference, the smaller circles mark planet candidates (light gray) and validated/confirmed planets (dark gray) detected during the prime \emph{Kepler} mission. We obtained the properties of \emph{Kepler} planets from the NASA Exoplanet Archive \citep{akeson_et_al2013}.}
\label{fig:planets}
\end{figure*}

After refitting the transit photometry and calculating false positive probabilities, we combined our new transit parameters with updated stellar characterizations from D17 to determine the physical properties of each K2OI. We display the revised planet properties in Figure~\ref{fig:planets}. The panels include the full population of K2 planet candidate, validated planets, and false positives as well as planet candidates and confirmed or validated planets identified during the original \emph{Kepler} mission. In general, the left panel demonstrates that the planet size and orbital period distribution of our K2 planet candidates and validated planets is similar to the distribution of short-period Kepler planet candidates. The majority of planets and planet candidates have radii $\lesssim 3\rearth$, but both the \emph{Kepler} and K2 radius distributions have tails extending to larger planet radii. As expected, Figure~\ref{fig:planets} also reveals that K2OIs with high false positive probabilities tend to be larger than K2OIs with lower false positive probabilities. \citet{martinez_et_al2017} noted a similar size difference between the radii of their planet candidates and validated planets (see their Figure~9) and remarked that the radius estimates for candidates tend to be more uncertain.

The primary difference between our K2 planet candidates and the \emph{Kepler} sample is that the K2 sample is biased toward brighter host stars. Considering only stars hosting planet candidates or validated planets, the median $Kp$ of our cool dwarf sample is 14.1.  As shown in the right panel of Figure~\ref{fig:planets}, 77\% of \emph{Kepler} planets and planet candidates orbiting stars cooler than 4800K have fainter host stars. The \emph{Kepler} cool dwarf sample has a median host star brightness of \mbox{$Kp = 15.2$}, fainter than 93\% of our K2 host stars. Interestingly, we note that our false positive sample is biased toward fainter host stars relative to the overall sample: the median $Kp$ of our false positive sample is 14.8. A possible explanation for this trend is that the reduced photon counts for these faint stars cause the S/N of the putative transit events to be below the threshold required to distinguish between bona fide transits and grazing eclipsing binaries. 

In both the \emph{Kepler} and K2 samples, we note a deficit of large planets at short orbital periods. The sole K2 planet with a period shorter than one~day and a plotted radius larger than $1.5\rearth$ is EPIC~201637175.01 (\mbox{K2-22b}), which is reported to be in the process of evaporating \citep{sanchis-ojeda_et_al2015}. Although our best-fit transit model uses a radius of $4.81^{+0.34}_{-0.37}\rearth$, the planet itself is likely sub-Mercury sized and surrounded by large dust clouds. The deep transits are therefore caused by the clouds of debris and do not reflect the underlying radius of the planet \citep{sanchis-ojeda_et_al2015}.

In general, the relative lack of larger short-period planets is consistent with results from the \emph{Kepler} mission: \citet{sanchis-ojeda_et_al2014} observed that all ultra-short-period planets orbiting G,K, or M~dwarfs have radii $\lesssim 2\rearth$. The lack of large planets on ultra-short orbital periods may indicate that the envelopes of highly-irradiated planets are highly vulnerable to photoevaporation \citep[e.g,][]{watson_et_al1981, lammer_et_al2003, baraffe_et_al2004, murray-clay_et_al2009, valencia_et_al2010, sanz-forcada_et_al2011, lopez_et_al2012, lopez+fortney2013, owen+wu2013, kurokawa+kaltenegger2013}. At longer orbital periods, our sample includes planet candidates and validated planets with estimated radii between $0.7\rearth$ and $14.9\rearth$. 

\subsection{Biases in the Planet Sample}
\begin{figure}[tb] 
\centering
\includegraphics[width=0.45\textwidth]{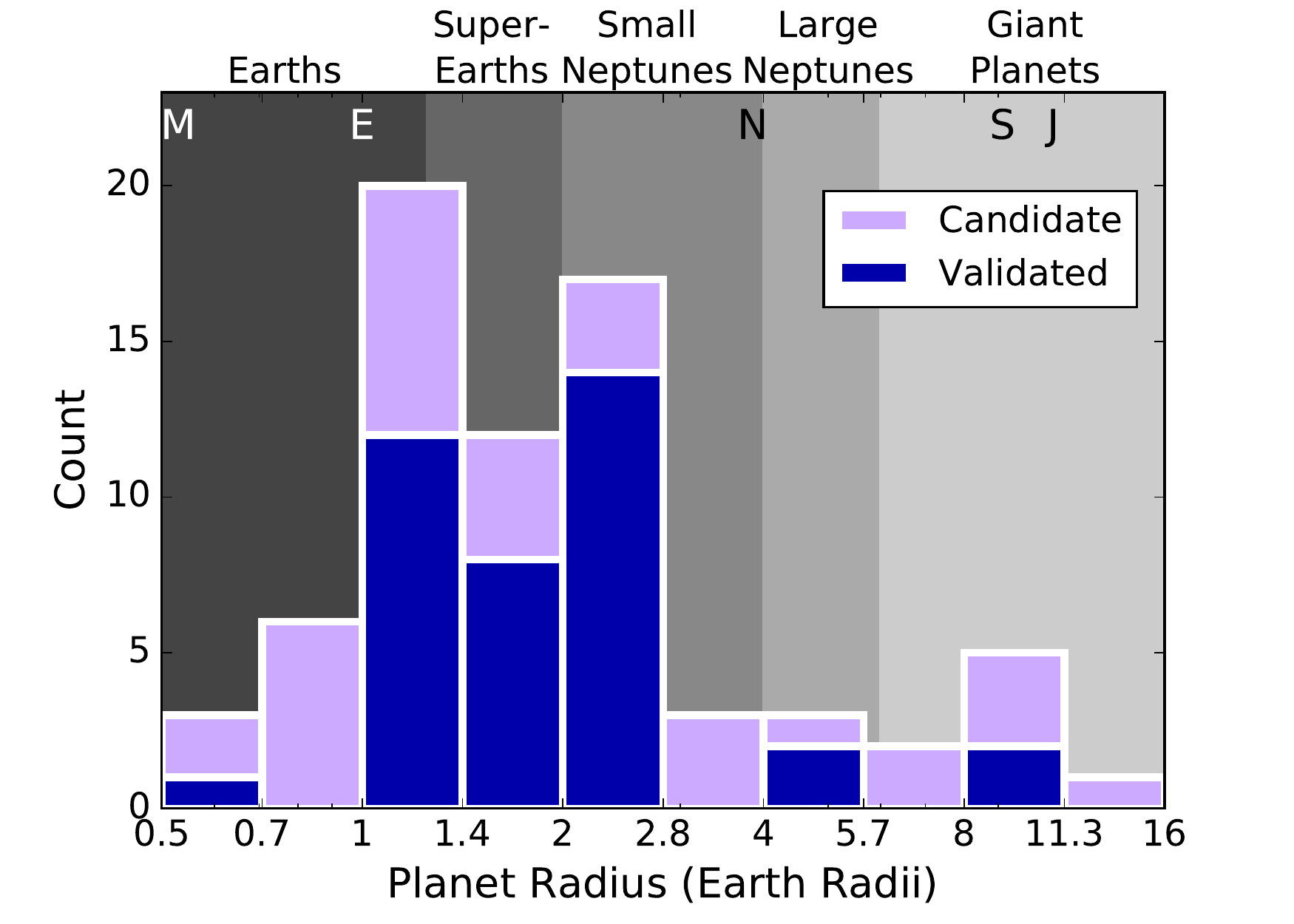}
\caption{Stacked histogram displaying the size distribution for the population of K2 planet candidates (lilac) and validated planets (navy) characterized in this paper. The background shading denotes the planet size ranges defined by \citet{fressin_et_al2013} and adopted in this paper. Letters mark the radii of Mars, Earth, Neptune, Saturn, and Jupiter.}
\label{fig:sizedist}
\end{figure}

Adopting the same planet size categories as \citet{fressin_et_al2013}, our sample contains 20~Earths (\mbox{$<1.25\rearth$}; 7~validated), 20~super-Earths (\mbox{$1.25-2\rearth$}; 14 validated), 20~small Neptunes (\mbox{$2- 4\rearth$}; 14~validated), three~large Neptunes ($4- 6\rearth$; two~validated), and eight~giant planets ($> 6\rearth$; two~validated). The size distribution depicted in Figure~\ref{fig:sizedist} is not representative of the larger population of planets orbiting low-mass stars. Rather, our sample, like most K2 planet catalogs, is shaped by strong selection biases due to the increased detectability of planets on short-period orbits compared to planets with longer periods and our interest in identifying compelling small planets orbiting bright stars for future follow-up observations.

\begin{figure}[tb]
\centering
\includegraphics[width=0.45\textwidth]{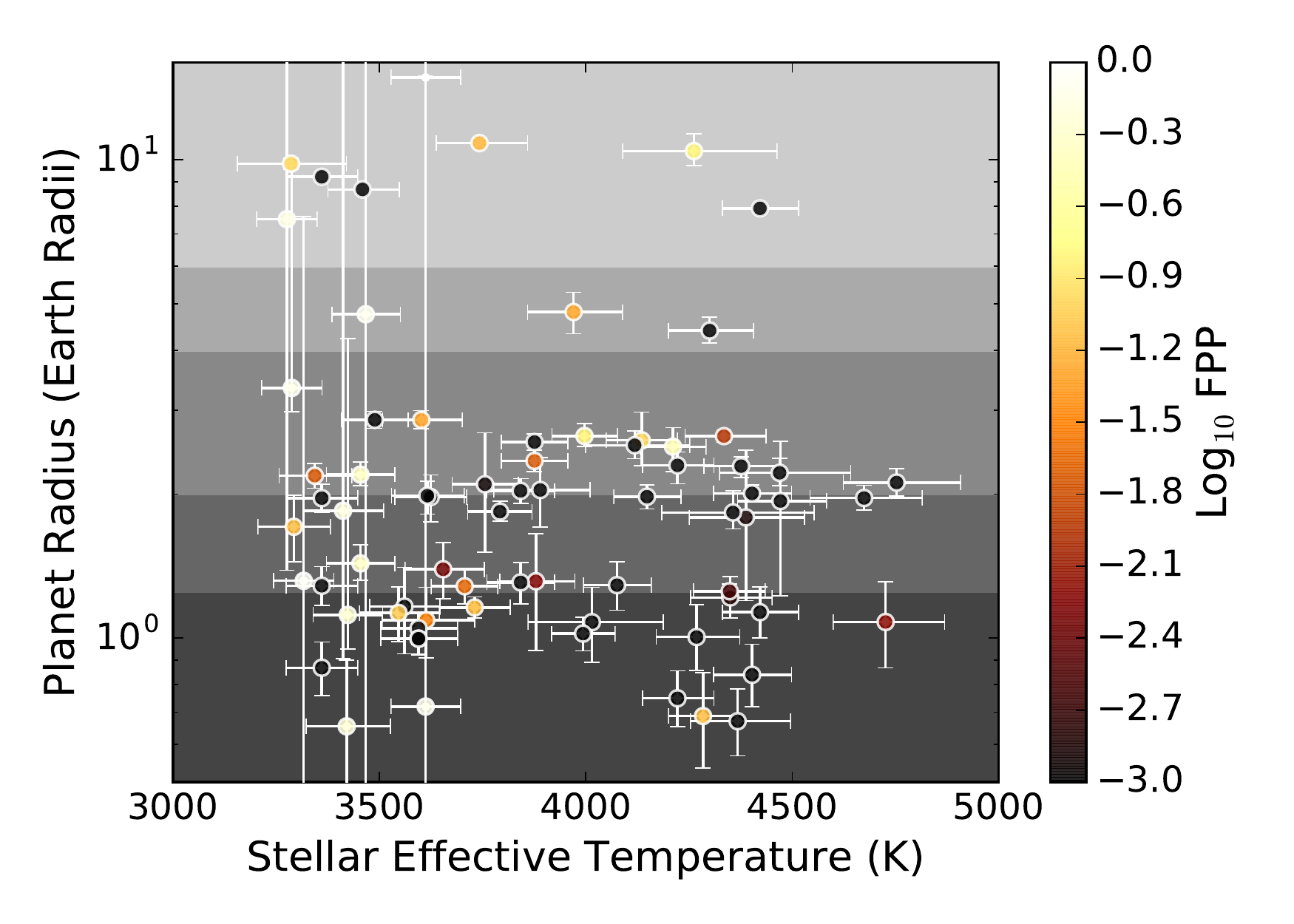}
\caption{Revised radii of K2 planet candidates and validated planets versus the effective temperature of their host stars. As in Figure~\ref{fig:planets}, the planets are color-coded according to their false positive probabilities. The shaded regions are the same as in Figure~\ref{fig:sizedist} and denote Earths (darkest region; at bottom), Super-Earths (second from bottom), Small Neptunes (middle), Large Neptunes (second from top), and Giant Planets (lightest region; at top).}
\label{fig:planets_rp_teff}
\end{figure}

\begin{figure}[tbp]
\centering
\includegraphics[width=.45\textwidth]{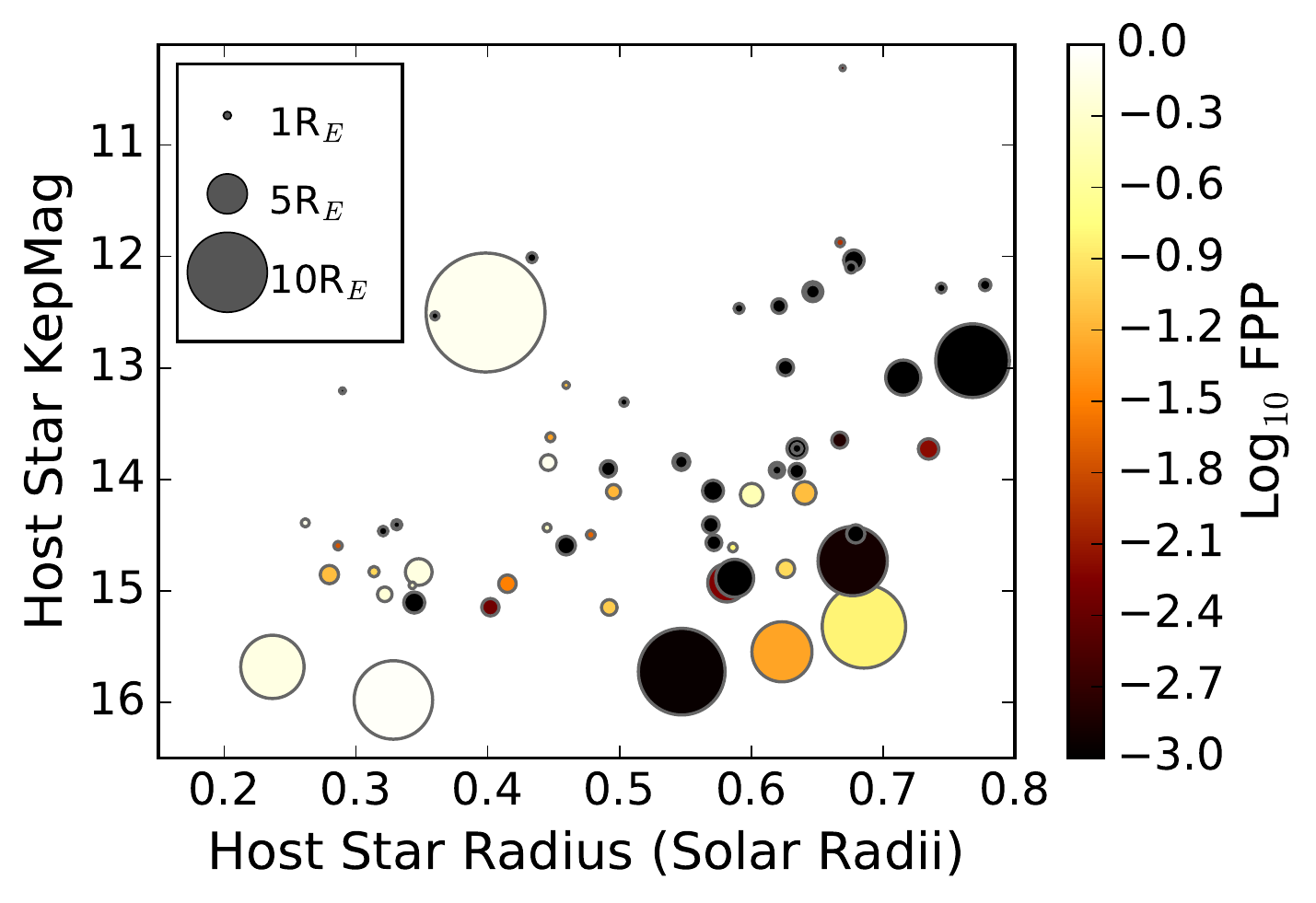}
\caption{Host star magnitude in the \emph{Kepler} bandpass versus estimated stellar radius for K2 targets in our low-mass candidate host star sample. Larger circles indicate larger planets and the planets are colored based on their false positive probabilities using the same scaling as in Figure~\ref{fig:planets}.} 
\label{fig:rs_kepmag}
\end{figure}

Looking at Figure~\ref{fig:planets_rp_teff}, the smallest planets are predominantly detected around the coolest target stars (median $T_{\rm eff} = 3595$~K for $R_p < 1.25\rearth$ compared to $T_{\rm eff} = 3842$~K for the full sample of planet candidates and validated planets), but the observed correlation of planet radius and stellar effective temperature is likely a selection effect due to the $1/(R_*)^2$ scaling of transit depth with stellar radius rather than a reflection of the true underlying occurrence rate of small planets. We further investigate the role of selection biases in Figure~\ref{fig:rs_kepmag} by comparing the host star magnitudes and radii for different sizes of planets. 

As would be expected if the minimum planet radius is a function of search sensitivity, we find that our smallest planet candidates and validated planets are preferentially associated with the brightest and smallest host stars. We also note that the Neptunes and giant planets in the sample fall along the lower right edge of the distribution, indicating that they tend to orbit stars that are fainter and/or larger than the majority of the stars in our target sample. We interpret both of these results as evidence that our planet sample contains significant selection effects and therefore cannot be used to estimate planet occurrence rates unless the detection biases are considered as part of the analysis.

\subsection{Planet Radii versus Insolation Flux: \\ Photoevaporation \& Potentially Habitable Systems}
\begin{figure*}[tbp]
\centering
\includegraphics[width=1\textwidth]{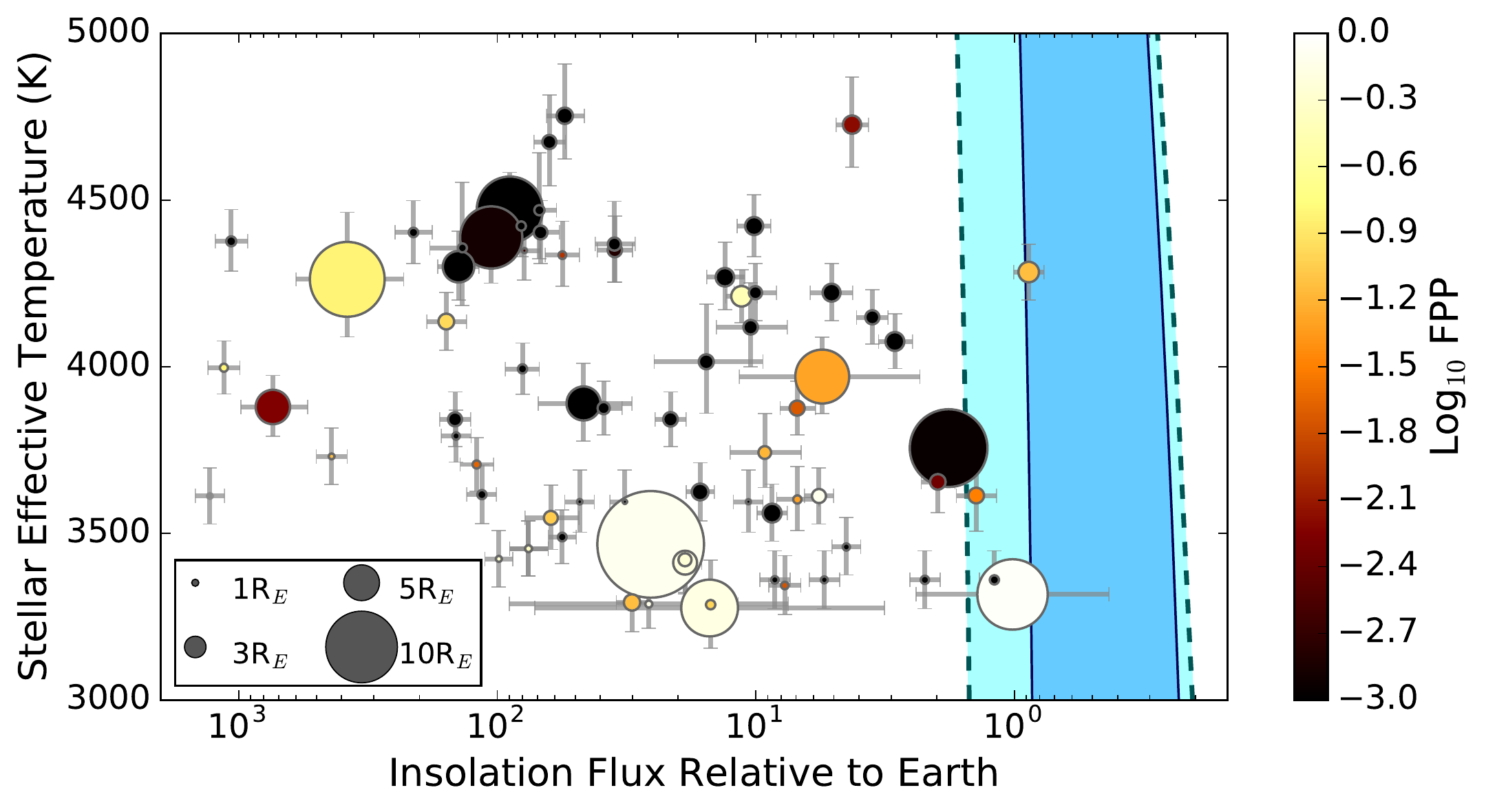}
\caption{Revised stellar effective temperatures versus the insolation flux received by associated planets. As indicated by the legend and colorbar, the data points are scaled by planet radius and colored based on false positive probability. The cyan and blue regions indicate optimistic (early-Venus/recent-Mars) and pessimistic (moist greenhouse/maximum greenhouse) habitable zone boundaries based as calculated by \citet{kopparapu_et_al2013}. The planets plotted within the habitable zone are: EPIC~206209135.04 (small black circle; $1.3 \pm 0.1\rearth$, $F_p = 1.2 \pm 0.2 \fearth$, ${\rm FPP} = 1\times 10^{-5} - 8\times 10^{-5}$), EPIC~211799258.01 (large white circle; $R_p = 9.8_{-2.1}^{+6.3}\rearth$, $F_p = 1.0_{-0.6}^{+1.4} \fearth$, ${\rm FPP} = 90\% - 94\%$), 
EPIC~211988320.01 (medium yellow-orange circle; $R_p = 2.9 \pm 0.2 \rearth$, $F_p = 0.9 \pm 0.1\fearth$, ${\rm FPP} = 7\%$), and EPIC~212690867.01 (medium orange circle; $R_p = 2.2 \pm 0.2 \rearth$, $F_p = 1.4_{-0.2}^{+0.3}\fearth$, ${\rm FPP} = 3-4\%$).}
\label{fig:planets_hz}
\end{figure*}

In Figure~\ref{fig:planets_hz} we plot the insolation flux received by the planet candidates and validated planets in our sample as a function of the effective temperatures of their host stars. We calculate the insolation flux from the stellar luminosities and the orbital semi-major axes, which we derive from the orbital periods and the stellar masses.  

Due to the relatively short 80-day durations of K2 campaigns and the bias of the transit method toward short-period planets, the majority of the planets have very short orbital periods and are therefore highly irradiated. The most heavily irradiated planets in our sample receive over 1000~times the flux received by the Earth ($F_\oplus$) and 21\% of the planets receive at least $100 F_\oplus$

Although our ability to discern relationships between planet radii and insolation flux environment is complicated by selection effects, Figure~\ref{fig:planets_hz} reveals that the population of highly-irradiated planets is dominated by smaller planets. As previously discussed, the shortage of highly-irradiated intermediate-sized planets may be due to photo-evaporation. Of the seven validated planets and planet candidates receiving at least $200F_\oplus$, the only object with a radius between $2\rearth$ and $10\rearth$ is the evaporating planet EPIC~201637175.01 \citep[K2-22b][]{sanchis-ojeda_et_al2015}. 

At the opposite extreme of the insolation flux distribution, our cool dwarf sample contains 21~planets or planet candidates receiving $<10F_\oplus$. In order to assess whether any of these planets might be habitable, we used the polynomial relations from \citet{kopparapu_et_al2013} to determine the insolation flux boundaries corresponding to conservative habitable zone (HZ) limits of the moist greenhouse inner edge and the maximum greenhouse outer edge and to more optimistic limits of recent-Venus inner limit and early-Mars outer limit. A more sophisticated choice would have been to use the planet-mass-dependent relations from \citet{kopparapu_et_al2014}, but we do not know the masses of these planets. 

Given the temperature distribution of the host stars in our cool dwarf sample, the median habitable zone boundaries are $0.26-0.89 F_\oplus$ ($0.22-0.40$~AU) for the conservative case and $0.23 - 1.55 F_\oplus$ ($0.16 - 0.42$~AU) for the more optimistic case. As shown in Figure~\ref{fig:planets_hz}, four of the planets and planet candidates in our sample fall within the optimistic HZ limits (EPIC~206209135.04, 211799258.01, 211988320.01, and 212690867.01). We discuss each of these K2OIs individually in Section~\ref{sec:discussion}.

\section{Discussion}
\label{sec:discussion}
\begin{figure*}[tbp]
\centering
\includegraphics[width=0.32\textwidth]{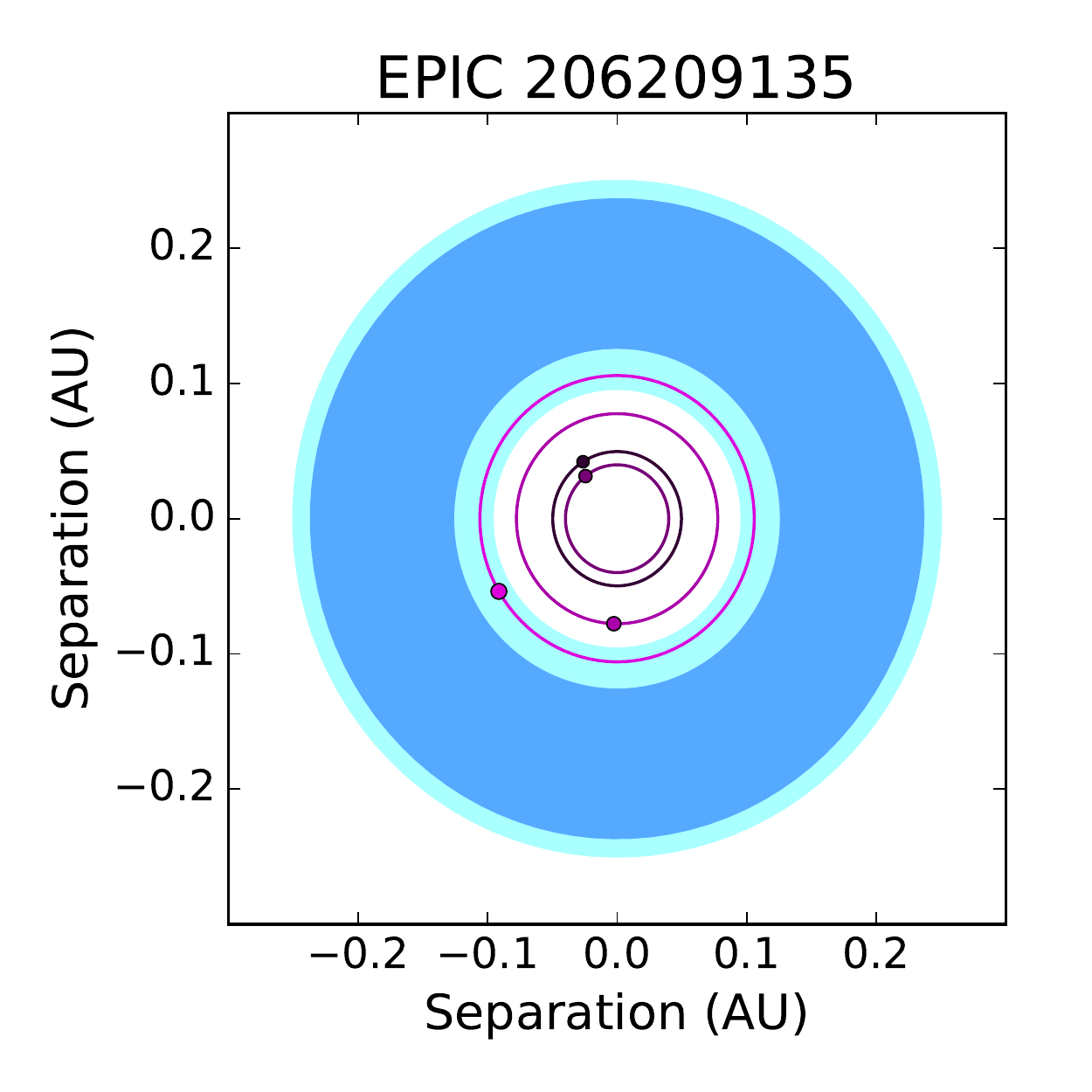} 
\includegraphics[width=0.32\textwidth]{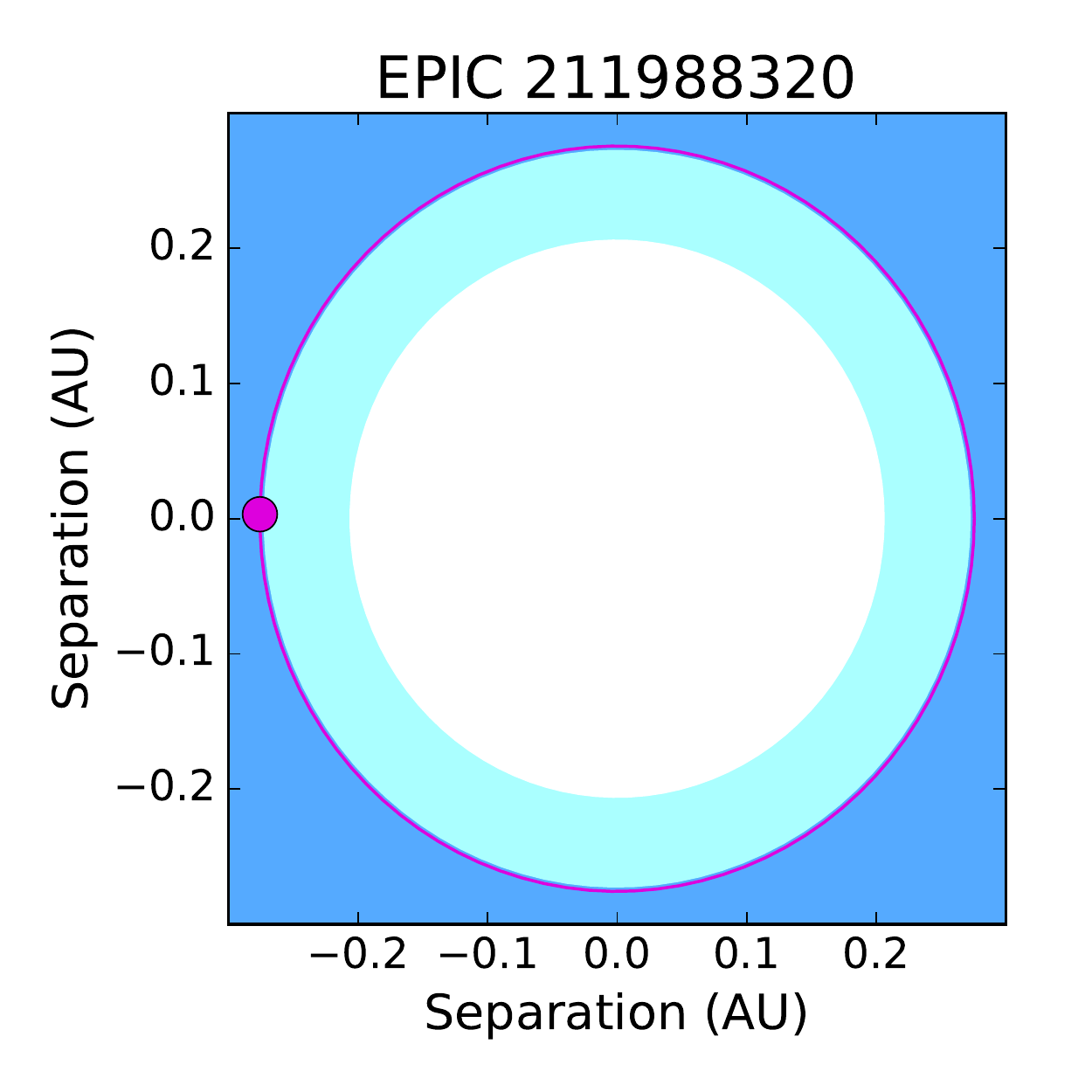} 
\includegraphics[width=0.32\textwidth]{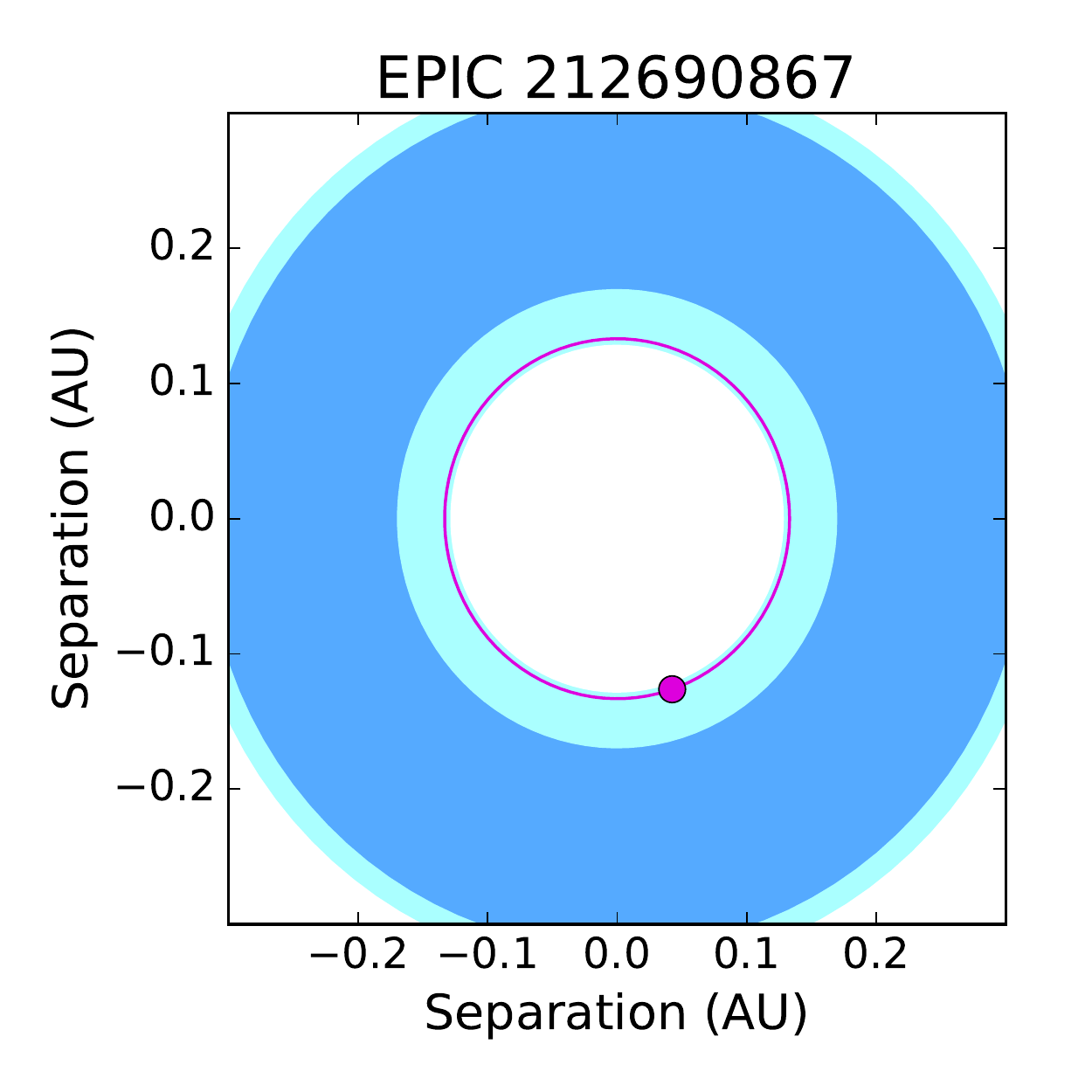} 
\includegraphics[width=0.24\textwidth]{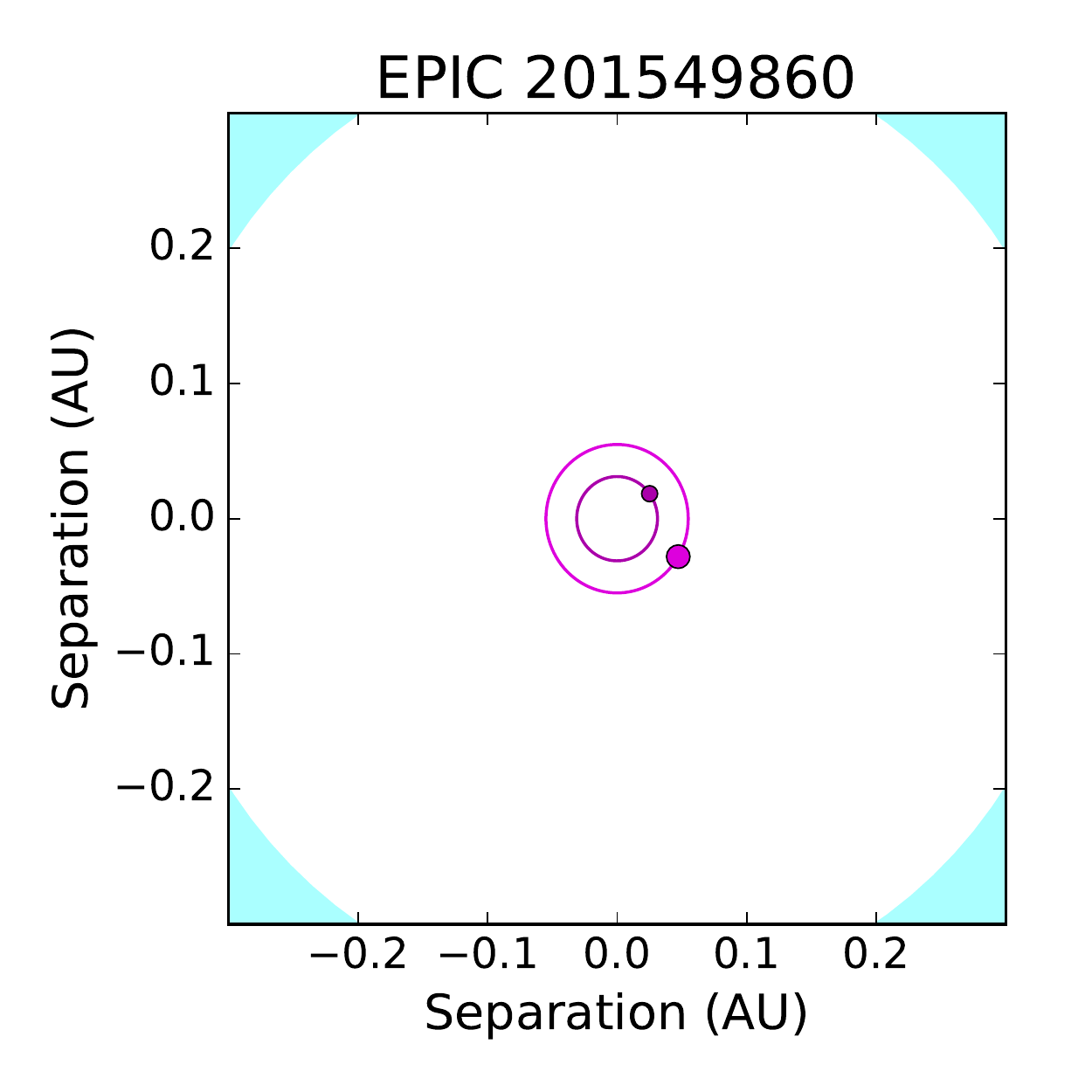} 
\includegraphics[width=0.24\textwidth]{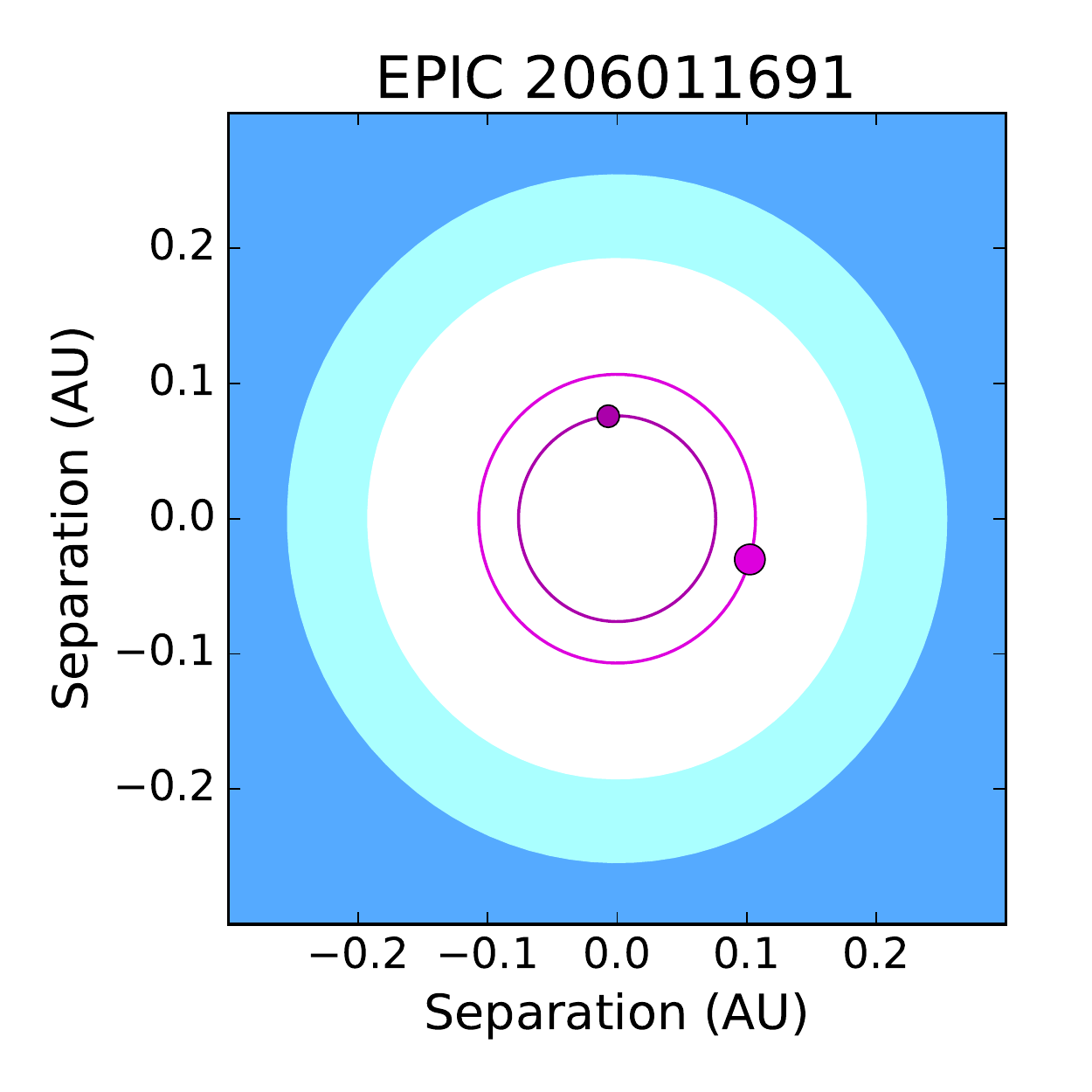} 
\includegraphics[width=0.24\textwidth]{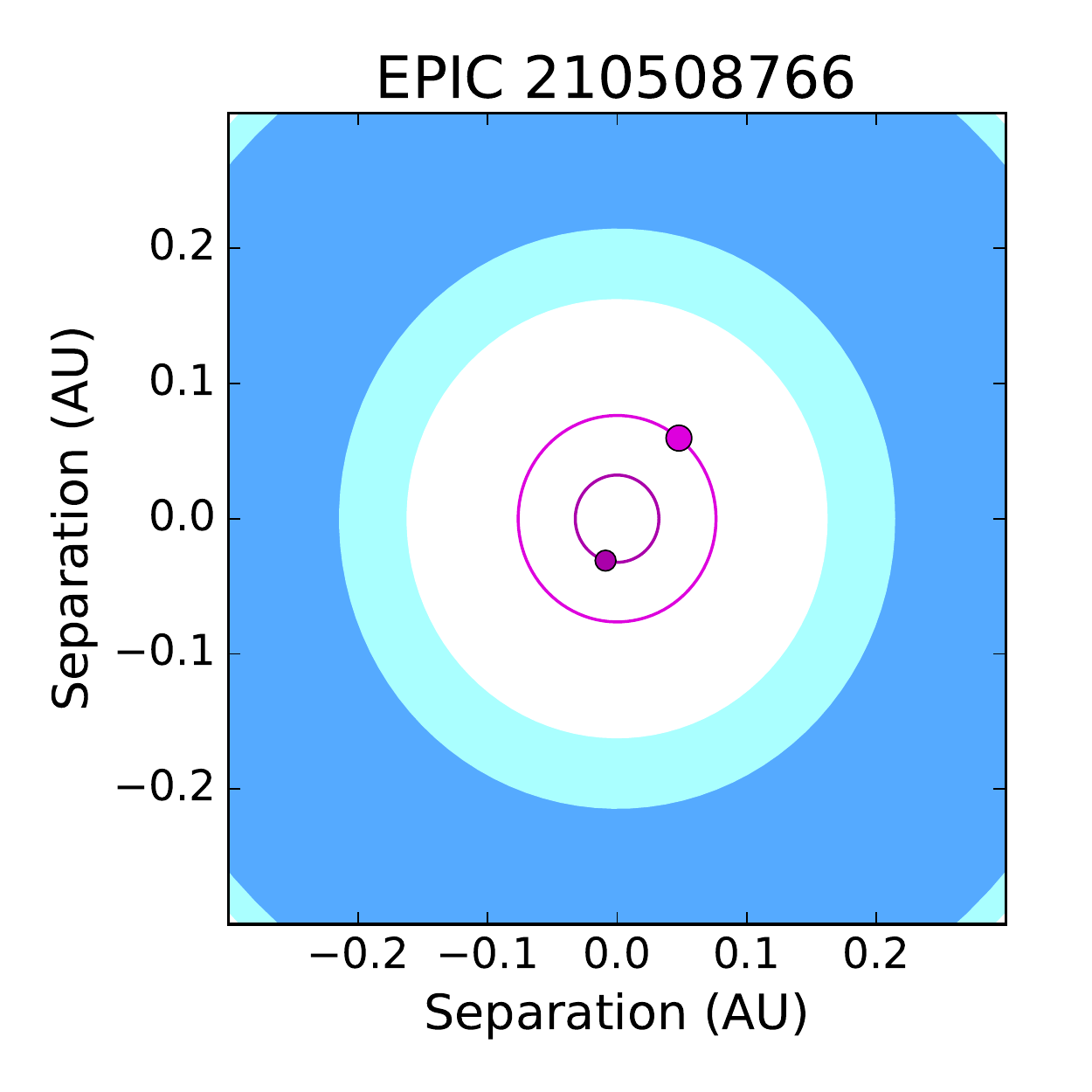} 
\includegraphics[width=0.24\textwidth]{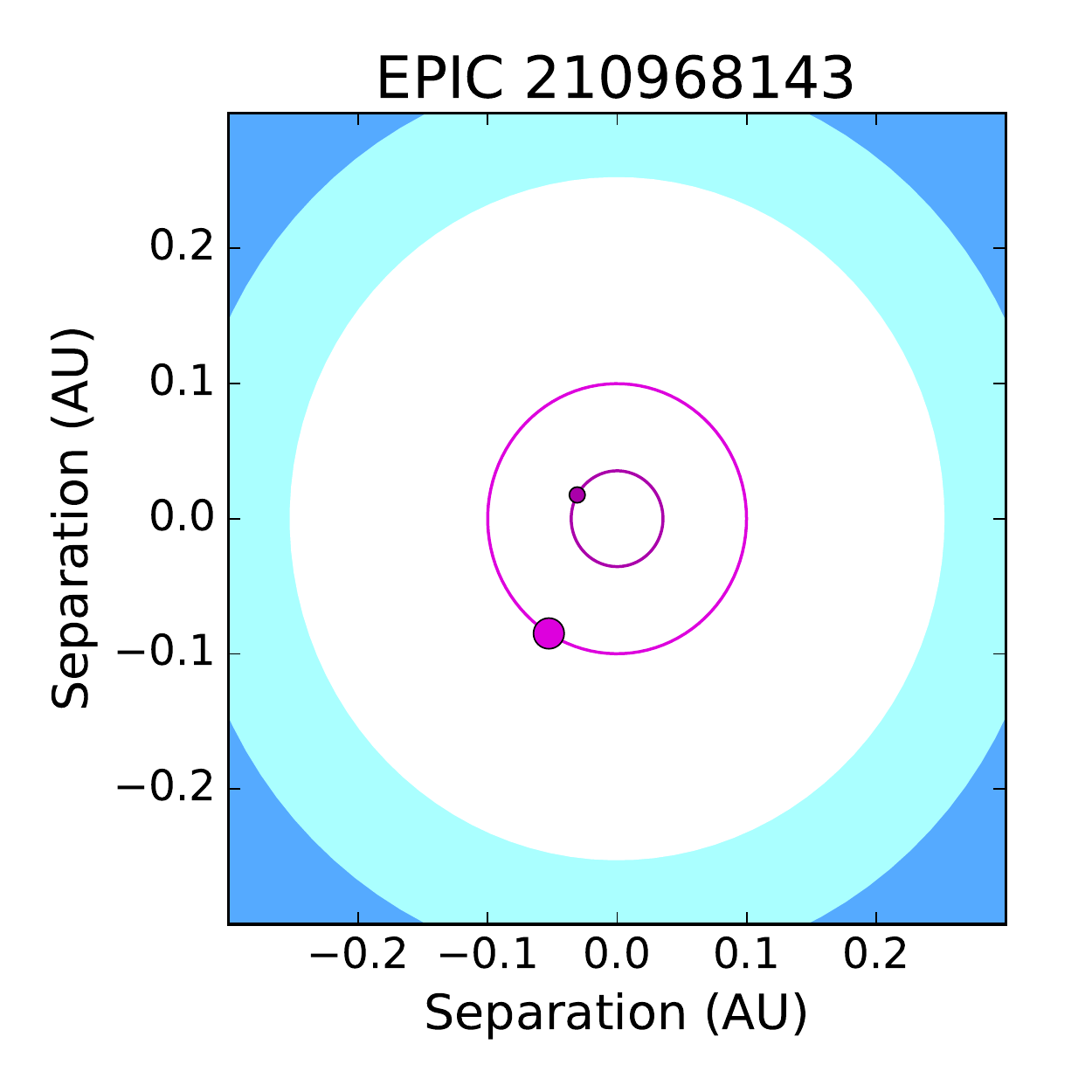} 
\includegraphics[width=0.32\textwidth]{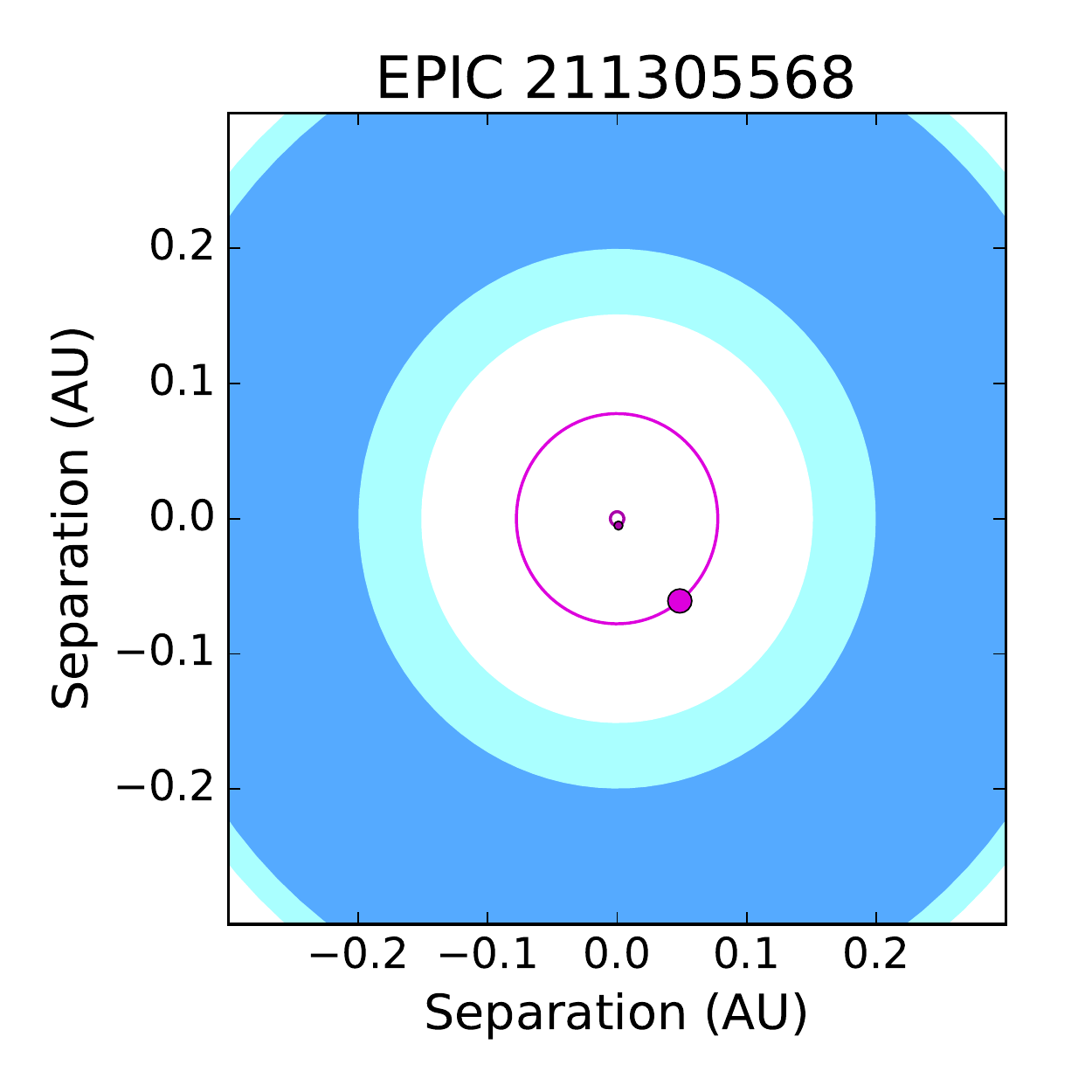} 
\includegraphics[width=0.32\textwidth]{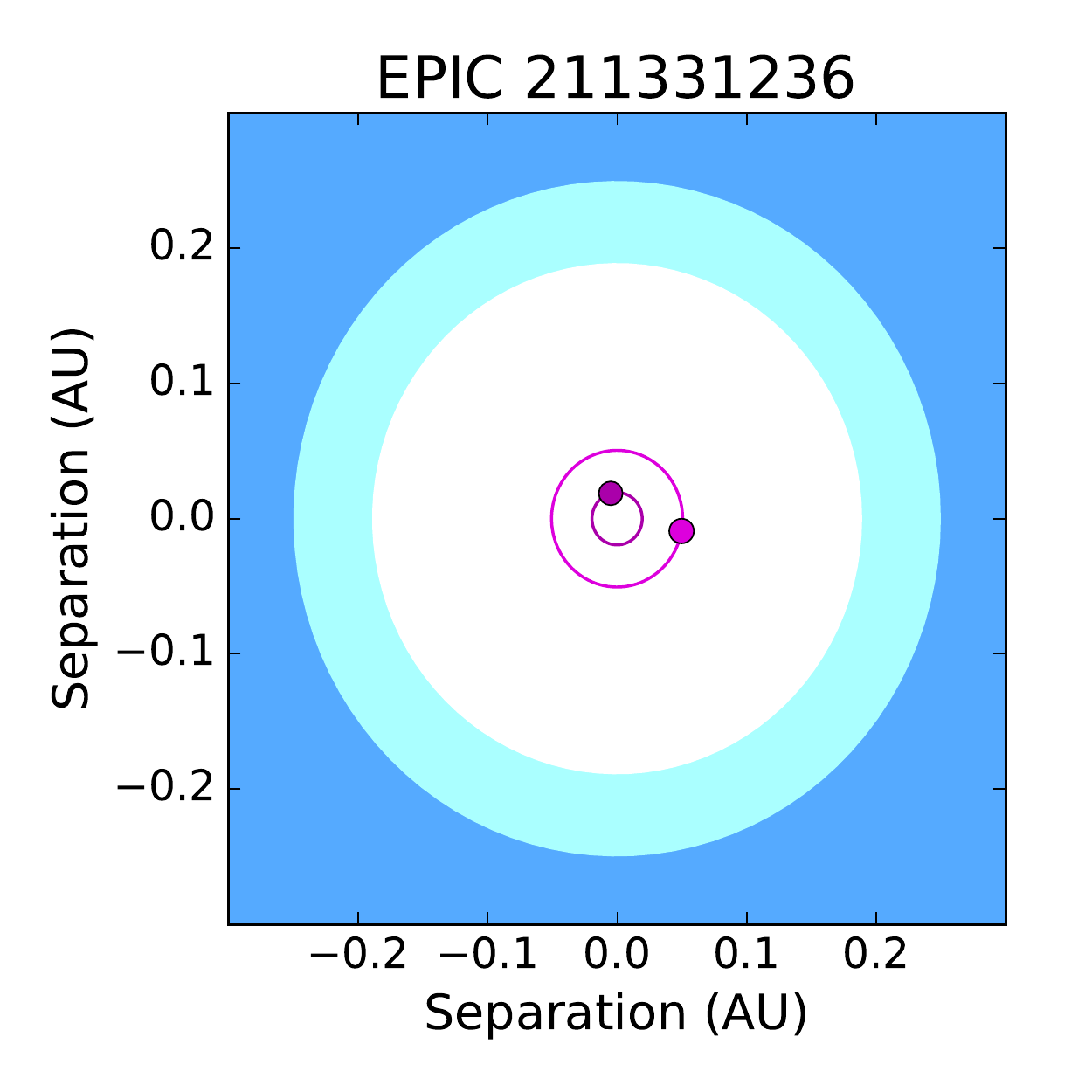}
\includegraphics[width=0.32\textwidth]{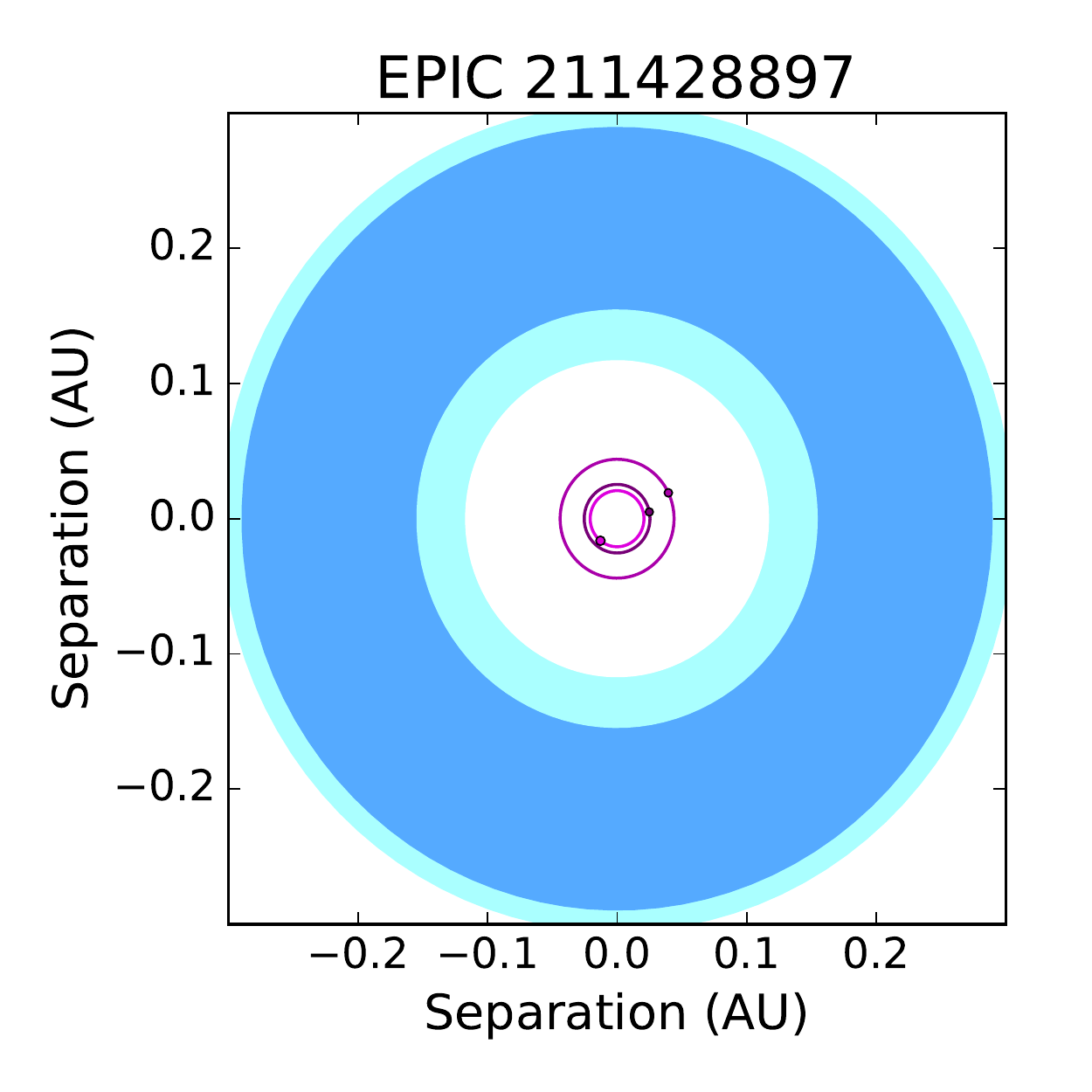} 
\caption{Orbits of potentially habitable planets and planets in multi-planet systems relative to the habitable zones of their host stars. The cyan regions indicate optimistic habitable zones extending from the early-Mars outer boundary to the recent-Venus inner boundary while the blue regions denote more conservative habitable zones set by the maximum greenhouse outer limit and the moist greenhouse inner limit. Planet sizes are to scale relative to other planets, but not relative to the orbits. Each panel covers the same range of semimajor axes. Systems containing planets in the habitable zone are shown in the top row. Note that EPIC~211799258.01 is omitted because it is almost certainly a false positive (see Section~\ref{ssec:211799258}). \label{fig:orbits}}
\end{figure*}

Several of the planets in our sample are particularly interesting because they might be habitable, reside in systems with multiple transiting planets, or orbit particularly bright stars. For reference, we present schematics of all of the multi-planet systems and potentially habitable systems in Figure~\ref{fig:orbits}. We also discuss individual systems in the following sections.

\begin{figure*}[tbp]
\centering
\includegraphics[width=0.49\textwidth]{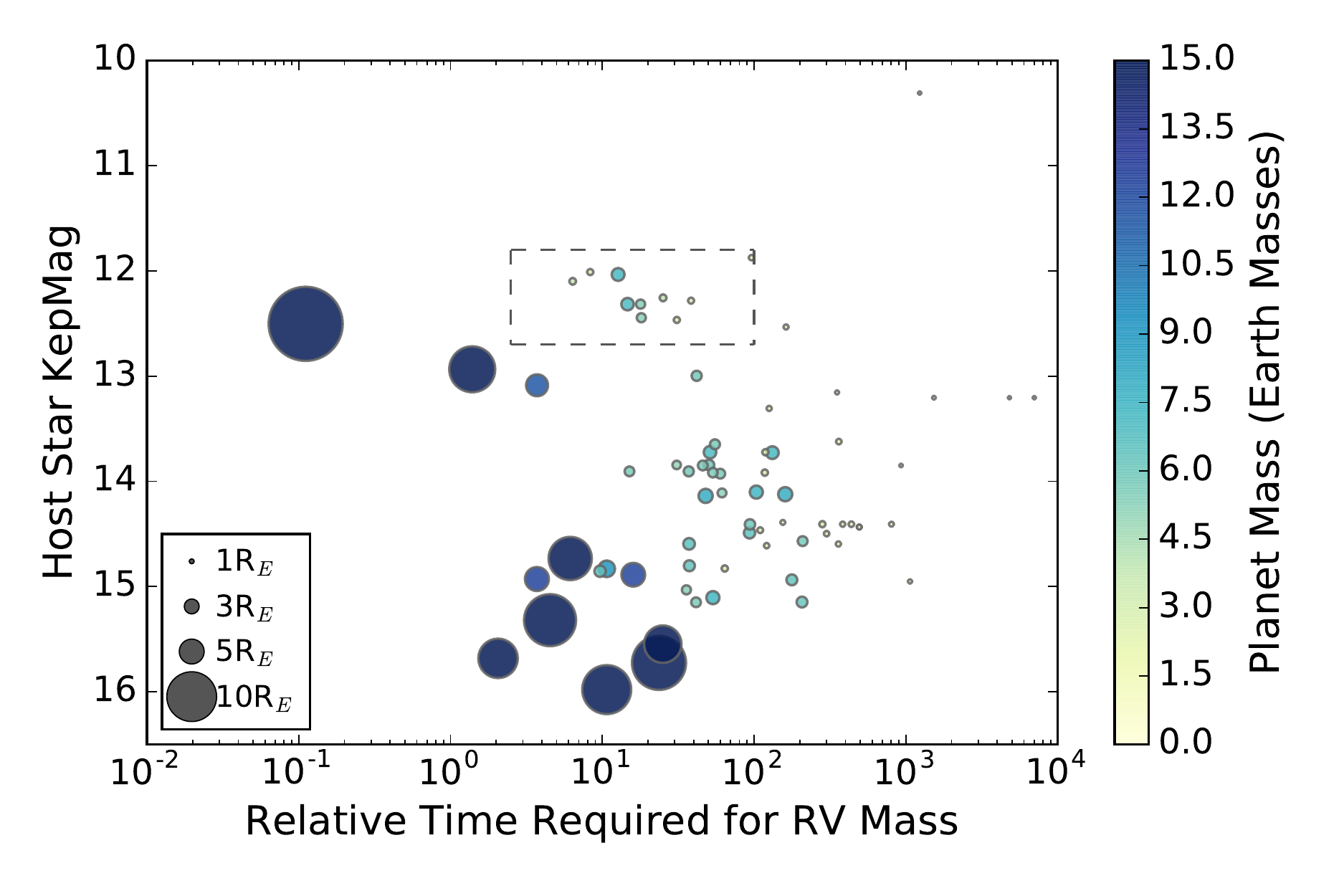}
\includegraphics[width=0.49\textwidth]{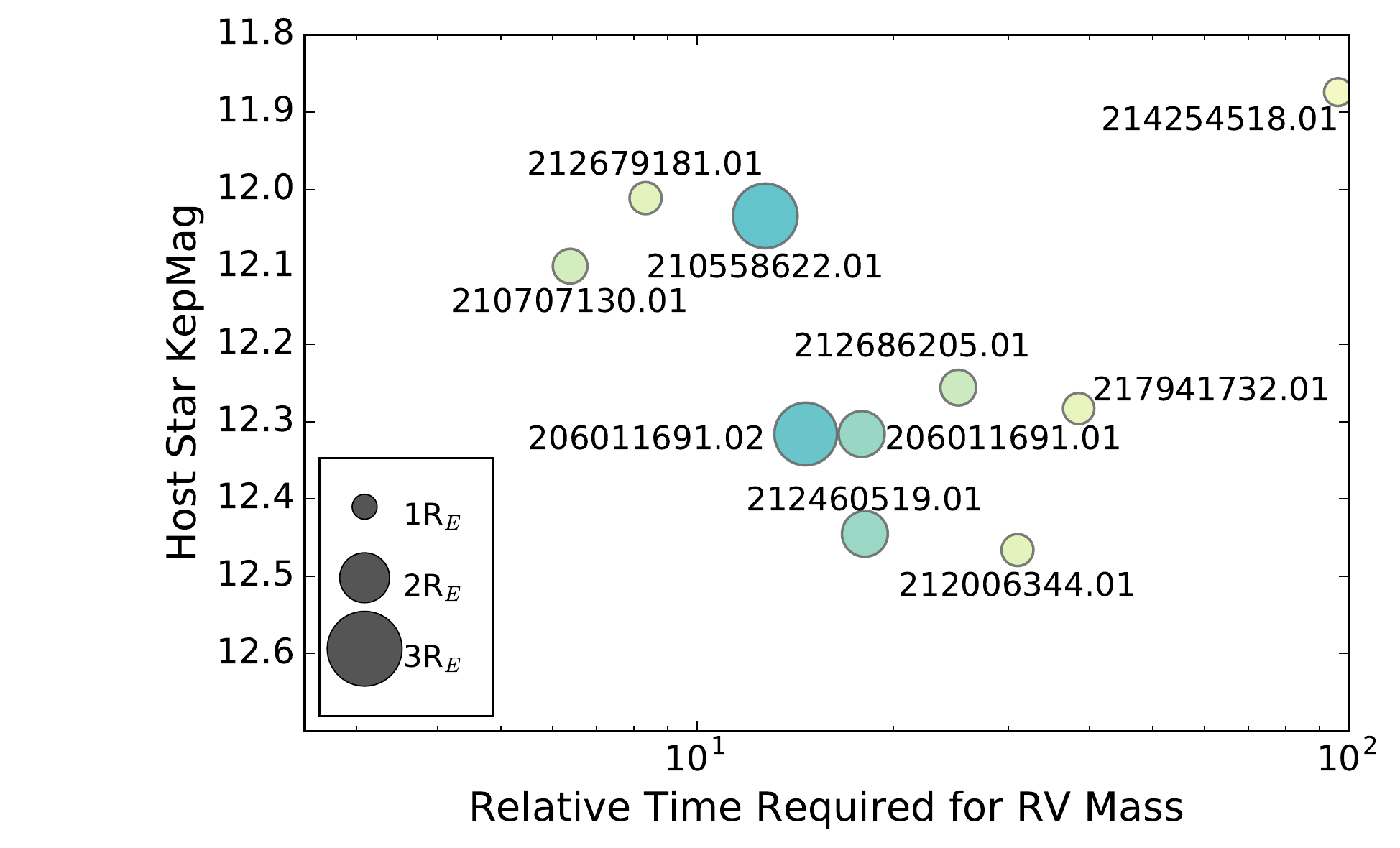}
\caption{Host star \emph{Kepler} magnitude versus the relative time required to measure the mass of the associated planet via radial velocity observations. For reference, obtaining a $6\sigma$ mass measurement of the small planet Kepler-93b ($R_p = 1.5\rearth$, $M_p = 4\mearth$, $P = 4.7$~d, $M_* = 0.91\msun$, $V = 10.2$) would require 9~units of time on this scale. The sizes of data points are scaled so that larger planets have bigger data points and the colors indicate the assumed planet masses. \textbf{Left: } Estimated time investments for all of the planets in our sample. \textbf{Right: } The time required to measure the masses of the small planets most amenable to RV observations. This panel shows a zoom-in of the boxed region in the left panel. We label the planets for ease of reference and discuss the most promising systems individually in Section~\ref{sec:discussion}.}
\label{fig:rvdet}
\end{figure*}

For the purpose of planning potential follow-up observations, we first employed the mass-radius relation presented in \citet{weiss+marcy2014}
to assign masses to the K2OIs and computed the expected radial velocity (RV) signal due to each planet. We then approximated the time investment required to measure the mass of each planet by scaling the expected RV signal by the predicted measurement precision. The absolute time values are meaningless, but the scalings in Figure~\ref{fig:rvdet} allow the targets to be ranked relative to each other. We note that the full range of small planet compositions is not well captured by a one-to-one mass-radius relation \citep{wolfgang+lopez2015, chen+kipping2017}, but these coarse mass estimates are sufficient for assessing the feasibility of future RV investigations. 

In general, we find that these planets would be challenging targets for current RV spectrographs due to the small expected signal (median value~$ =2.3 \,{\rm m }\,{\rm s}^{-1}$) and the faintness of their host stars at optical wavelengths. In addition, several of the easiest candidates in Figure~\ref{fig:rvdet} are unsuitable for RV observations: EPIC~210558622 was identified as an spectroscopic binary \citep{crossfield_et_al2016}, EPIC~212679181 is accompanied by a fainter star $1\farcs5$ away,\footnote{\url{https://exofop.ipac.caltech.edu/k2/edit_obsnotes.php?id=212679181}} and EPIC~216892056.01 is almost certainly a false positive. 

In contrast, the small Neptune EPIC~205924614.01 (K2-55b) is an ideal follow-up target and has already been observed by both Keck/HIRES and \emph{Spitzer}. The validated small planets EPIC~210707130.01 and 212460519.01 are also attractive RV targets. We discuss K2-55b in detail in a separate publication (Dressing et al., \emph{in prep}) and consider the two smaller planets in Section~\ref{ssec:bright}.
 
\begin{figure*}[tbp]
\centering
\includegraphics[width=0.49\textwidth]{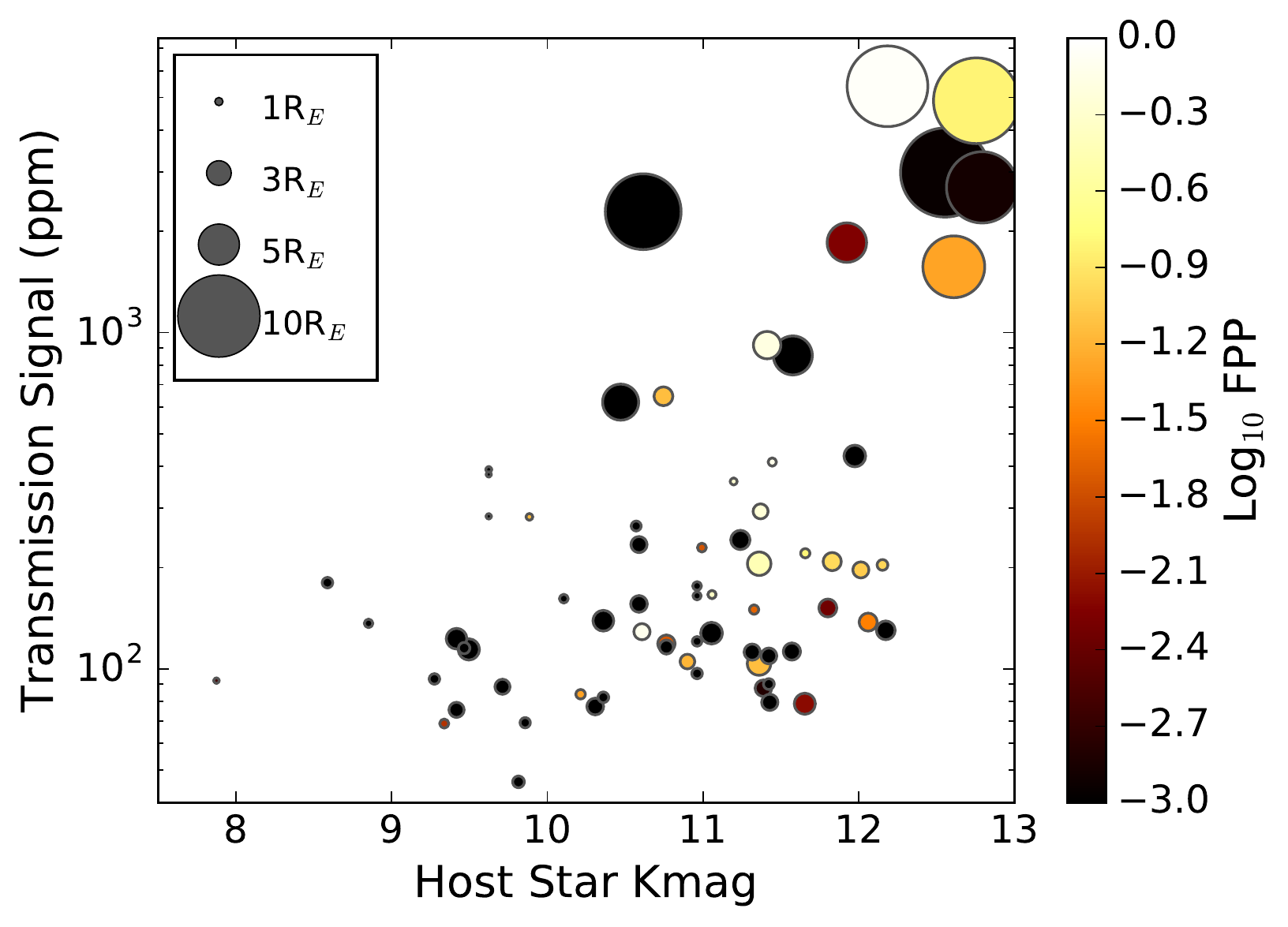}
\includegraphics[width=0.49\textwidth]{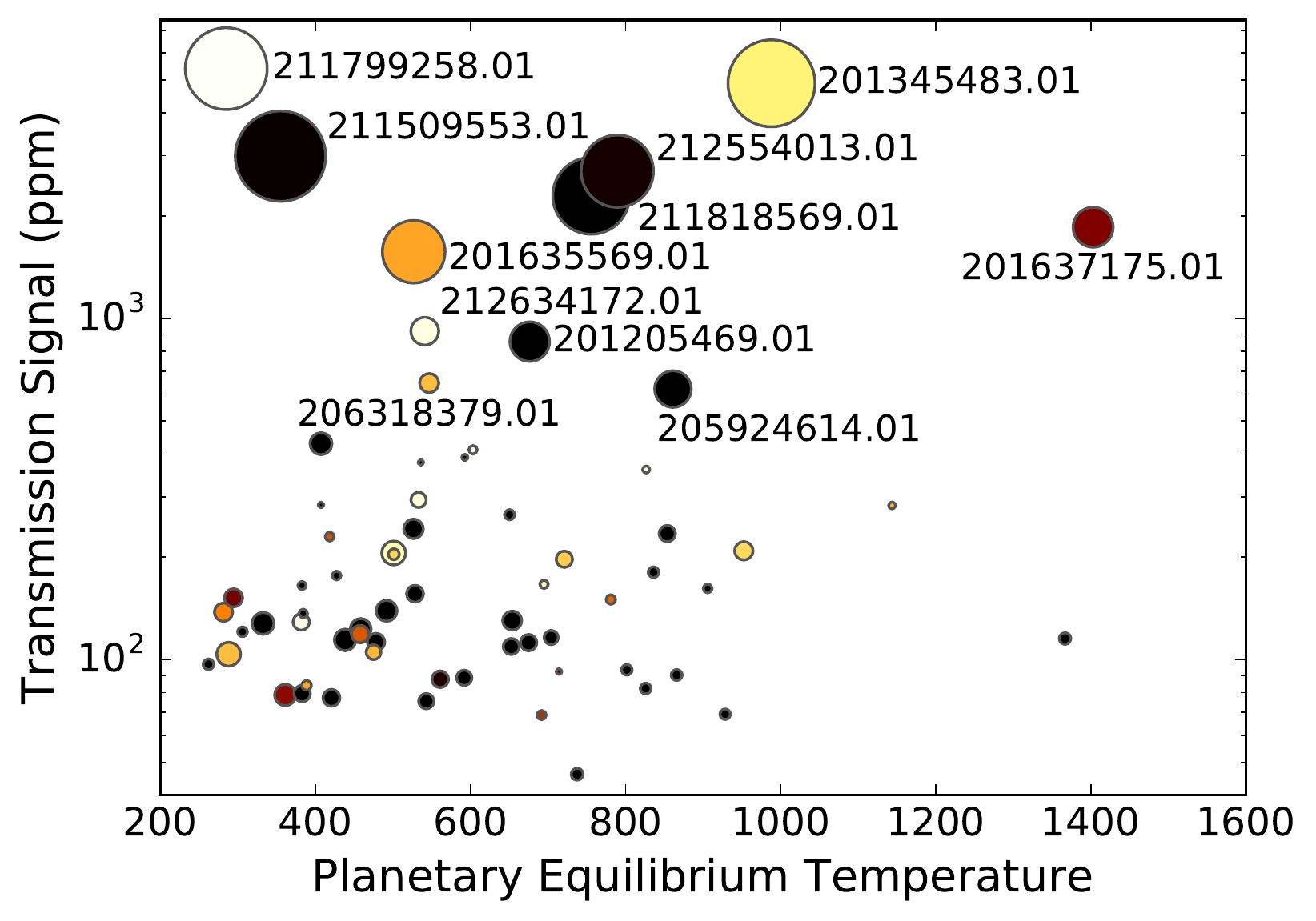}
\caption{Estimated transmission spectroscopy signal versus host star $K$-band magnitude (left) and planetary equilibrium temperature (right). 
As in Figure~\ref{fig:rvdet}, the points are colored based on false positive probability and scaled by planet size. In the right panel, we label the planets with approximated transmission signals larger than 500~ppm. Note that these estimates assume that that the mean molecular weight of the atmosphere is 2.2$m_H$; planets with heavier atmospheres would generate smaller transmission signals.}
\label{fig:transdet}
\end{figure*}

Although our targets are generally poorly suited for RV mass measurements, they are more compelling candidates for transmission spectroscopy. The combination of the small sizes and red colors of their host stars produces larger transmission signals than would be expected for similar planets orbiting similarly bright Sun-like stars. As shown in Figure~\ref{fig:transdet}, several planets should yield detectable transmission signals if our assumptions regarding planet masses and atmospheric compositions are correct. 

We approximated the transmission signals shown in Figure~\ref{fig:transdet} by assuming that the planetary atmospheres extend for five scale heights. When calculating scale heights, we adopted the same planet masses as for the RV predictions and determined planetary equilibrium temperatures using a fixed planetary albedo of $0.3$. We also assumed that each planet possesses a Jupiter-like atmosphere with a mean molecular weight of 2.2$m_H$; planets with water-dominated atmospheres would generate transmission signals an order of magnitude smaller. While exciting from a habitability standpoint, the smallest planets in our sample may be bare rocky cores without substantial atmospheres and would therefore be poor targets for transmission spectroscopy. 

Consulting Figure~\ref{fig:transdet}, the most intriguing targets for atmospheric characterization are EPIC~201205469.01 \citep[K2-43b,][]{crossfield_et_al2016}, 201345483.01 \citep[K2-45b,][]{crossfield_et_al2016}, 201635569.01 \citep[K2-14b,][]{montet_et_al2015, crossfield_et_al2016}, 205924614.01 \citep[K2-55b,][]{crossfield_et_al2016}, 206318379.01 \citep[K2-28b,][]{hirano_et_al2016} (\mbox{K2-28b}), 211509553.01, 211818569.01, and 212554013.01. These planets span a broad range in both size and equilibrium temperature: EPIC~211509553.01 is a cool giant planet ($R_p = 10.8\pm0.6\rearth$, $T_{\rm eq} = 355$~K); EPIC~201345483.01 is a hot  giant planet ($R_p = 10.4_{-0.7}^{+0.9}\rearth$, $T_{\rm eq} = 988$~K); EPIC 201635569.01 , EPIC 211818569.01,  EPIC~212554013.01 are warm giant planets ($R_p = 7.5\pm0.5\rearth$, $T_{\rm eq} = 527$~K, $R_p = 9.2_{-0.7}^{+0.6}\rearth$, $T_{\rm eq} = 756$~K, and $R_p = 8.7_{-0.6}^{+0.7}\rearth$, $T_{\rm eq} = 789$~K, respectively); EPIC~201205469.01 and EPIC~205924614.01 are warm large Neptunes ($R_p = 4.8\pm0.3\rearth$, $T_{\rm eq} = 676$~K and $R_p = 4.4\pm0.3\rearth$, $T_{\rm eq} = 861$~K, respectively); and EPIC~206318379.01  is a warm small Neptune ($R_p = 2.3\pm0.3\rearth$, $T_{\rm eq} = 547$~K). Of these planets, EPIC~205924614.01, 20618379.01, and 211818569.01 have the brightest host stars and are therefore the most favorable targets.

\subsection{Systems with Planets in or near the HZ}

\subsubsection{EPIC~206209135 (K2-72): \\Validated 4-planet System with an Earth-sized Planet in the HZ} 
EPIC~206209135 is an M2 dwarf ($Kp=14.407$, $Ks = 10.962$) orbited by four transiting planets with periods of 5.577, 7.760, 15.189, and 24.167~days. All four planets were previously validated by \citet{crossfield_et_al2016} based on careful scrutiny of the K2 photometry, follow-up imaging with Keck/NIRC2, and low false positive probabilities as computed with VESPA \citep{morton2015}. In their analysis, \citet{crossfield_et_al2016} assumed a stellar radius of $0.232\pm0.056\rsun$, which is  30\% smaller than our revised value of $0.33 \pm 0.03\rsun$ (D17). Despite the significant change in the assumed stellar parameters, our independent {\tt vespa} analysis also returned FPPs low enough to validate all four planets.

The outermost planet (EPIC~206209135.04) has a radius of \mbox{$1.29_{-0.13}^{+0.14} \rearth$}, a semimajor axis of \mbox{$0.106_{-0.013}^{+0.009}$~AU} and an insolation flux of \mbox{$1.2 \pm 0.2\fearth$}, placing it within the \mbox{$0.22-1.52\fearth$} (\mbox{$0.09-0.25$~AU}) boundaries of the optimistic Venus/Mars HZ but outside the \mbox{$0.28-0.91\fearth$} ($0.13-0.24$~AU)  limits of the conservative moist greenhouse/maximum greenhouse HZ. The second outermost planet (EPIC~206209135.02) is slightly smaller \mbox{($R_p = 1.16 \pm 0.13\rearth$)} and lies inside the inner edge of the optimistic habitable zone (\mbox{$F_p = 2.2 \pm 0.3 \fearth$}, \mbox{$a = 0.078_{-0.010}^{+0.007}$}). The two innermost planets (EPIC~206209135.01 and EPIC~206209135.03) have radii of \mbox{$1.08 \pm 0.11\rearth$} and \mbox{$1.01 \pm 0.12 \rearth$} and receive \mbox{$8.5_{-1.1}^{+1.2}\fearth$} and \mbox{$5.4_{-0.7}^{+0.8}\fearth$}, respectively. We note that our radius estimates for all four planets are consistent with the revised values found by \citet{martinez_et_al2017}, who used NTT/SOFI spectroscopy to characterize the host star and then scaled the $R_p/R_\star$ values from \citet{crossfield_et_al2016} accordingly.

The orbital periods of the four planets are in near-resonant configurations, which suggests that the system may have experienced disk-driven migration. Specifically, planets .01 and .03 (5.577~days and 7.760~days) are near the second-order 7:5 mean motion resonance and planets .03 and .02 (7.760~days and 15.189 ~days) are near the first-order 2:1 mean motion resonance. Planets .01 and .04 (5.577~days and 24.167~days) have an orbital period ratio of roughly 3:13. 

Assuming that all four planets have rocky compositions and masses of $1.1 - 2.6\mearth$ as suggested by the \citet{weiss+marcy2014} mass radius relation, the expected radial velocity perturbations from each planet individually would have semi-amplitudes of $0.8-1.4$~m/s. Given that the host star is relatively faint ($Kp = 14.407$, $Ks = 10.962$) the system would be a challenging target for RV surveys. Fortunately, the near-resonant architecture of the system may enable planet mass measurements via TTVs rather than RVs. These TTV-based masses will be useful for interpreting the results of subsequent atmospheric characterization studies.

\subsubsection{EPIC~211799258:\\ New Candidate Cool Jupiter} 
\label{ssec:211799258}
This previously unknown planet candidate has a high false positive probability of $90\%-94\%$, suggesting that the system is most likely a false positive. Our Gemini-N/NIRI imaging of the system was sensitive to companions 6.6 (7.6) magnitudes fainter than EPIC~211799258 at $0\farcs5$ ($1\farcs0$) and did not reveal any companions, but our {\tt vespa} analysis suggested that the most likely false positive configuration for this system was an eclipsing binary that we would not expect to resolve in our AO imaging. If real, the planet candidate is a $9.8_{-2.1}^{+6.3}\rearth$ planet with an orbital period of 19.5 days. The host star is faint ($Kp = 15.979$, $Ks = 12.185$), but the planet should generate a large RV signal. Given the paucity of cool Jupiters known to transit low-mass stars, this system may warrant further study despite the low likelihood that the planetary interpretation is correct. If this transit-like event is indeed caused by eclipses involving the target star, high-resolution spectra acquired at opposite quadratures should reveal significant RV variation. In contrast, the absence of RV variation would rule out the foreground eclipsing binary scenario and possibly allow us to validate EPIC~211799258.01 as a genuine planet. 

\subsubsection{EPIC~211988320:\\ New Candidate Small Neptune in the HZ} 
EPIC~211988320.01 (\mbox{$2.86_{-0.15}^{+0.16}\rearth$}) was originally identified as a transit-like event with an orbital period of 15~days, but our subsequent inspection of the light curve revealed that the orbital period is actually 63.8~days. The longer orbital period places EPIC~211988320.01 just within the boundaries of the conservative moist greenhouse/ maximum greenhouse HZ of its M0V host star. Although this planet is likely too large to be rocky, its host star is bright enough ($Kp = 14.122$, $Ks = 11.36$) to enable future atmospheric characterization studies. Our estimated false positive probability for the candidate is 7\%, but our analysis did not include AO or speckle imaging. While the system is too faint for radial velocity  ($Kp = 14.122$, $Ks = 11.360$), acquiring AO or speckle images would be straightforward. Due to the long orbital period of EPIC~211983320.01, only two transits are visible in the K2 data. We therefore recommend acquiring additional transit observations to refine the orbital ephemeris prior to embarking upon an intensive atmospheric characterization campaign.

\subsubsection{EPIC~212690867:\\ New Candidate Small Neptune in the HZ} 
This small planet (\mbox{$2.20_{-0.18}^{+0.19}\rearth$}) has an orbital period of 25.9~days around a $3614_{-107}^{+118}$~K host star. The planet receives an insolation flux of $1.4_{-0.2}^{+0.3}\fearth$ and would be an intriguing target for atmospheric characterization if its host star were slightly brighter ($Kp = 14.936$, $Ks = 12.061$). Our {\tt vespa} analysis returned a false positive probability too large for validation (FPP = $3-4\%$), but the planet could likely be validated if we acquired an AO or speckle image to rule out the remaining false positive scenarios. 

\subsubsection{EPIC~212398486:\\ Newly Validated Mini-Neptune Near the HZ}
The EPIC~212398486 system consists of a mini-Neptune orbiting an M2~dwarf on a 21.8~day orbit. The $3654_{-92}^{+100}$~K, $0.40 \pm 0.03\rsun$ host star has optimistic habitable zone boundaries of $0.2 - 1.5 \fearth$, just cooler than the $2.0 \pm 0.3 \fearth$ insolation flux received by the planet. At $R_p = 2.19_{-0.18}^{+0.19}\rearth$, the planet is very unlikely to be rocky, but it is still an attractive target for follow-up atmospheric characterization studies. The majority of well-characterized mini-Neptunes are hot; the EPIC~212398486 system provides a complementary example of a cooler small planet. The system is fainter than most of our targets ($Kp = 15.147$, $Ks = 11.802$), but follow-up atmospheric studies should be feasible. Our {\tt vespa} analysis of the K2SFF photometry yielded ${\rm FPP} = 0.2\%-0.7\%$, allowing us to validate this previously unknown planet.

\subsection{Systems with Multiple Transiting Planets}
\subsubsection{EPIC~201549860 (K2-35):\\ Validated 2-Planet System} 
The EPIC~201549860 (K2-35) system was validated by \citet{sinukoff_et_al2016} as a $4680 \pm 60$~K, $0.72\pm0.04$ star hosting a $1.40\pm 0.17\rearth$ inner planet with a period of 2.4~days and $2.09^{+0.43}_{-0.31}\rearth$ outer planet with a period of 5.6~days. They characterized the host star by applying the {\tt SpecMatch} \citep{petigura2015} and {\tt isochrones} \citep{morton2015b} packages to optical spectra acquired with Keck/HIRES. Using IRTF/SpeX NIR spectra, we classified the host star as a K4~dwarf with a cooler temperature of $4402_{-93}^{+96}$~K and a smaller radius of $0.62 \pm 0.03$ (D17). After refitting the transit photometry, we revised the planet radii to $1.32 \pm 0.08 \rearth$ for the inner planet and $1.94_{-0.12}^{+0.13} \rearth$ for the outer planet. Both of these estimates are consistent with the \citet{sinukoff_et_al2016} estimates. 

\subsubsection{EPIC~206011691 (K2-21):\\ Validated 2-Planet System} 
The NASA Exoplanet Archive describes the EPIC~206011691 system as containing a $1.25\pm0.11 \rearth$ inner planet on a 9.324-day orbit and a $1.53\pm0.14\rearth$ outer planet with an orbital period of $15.498$~days. Both planets were announced by \citet{petigura_et_al2015}, independently detected by \citet{vanderburg_et_al2016} and validated by \citet{crossfield_et_al2016}. In addition, \citet{barros_et_al2016} detected the outer planet, but not the inner planet.

The two planets are near the second-order 5:3 mean motion resonance, which may be a residual sign of past convergent migration. Neither planet is habitable and at $V = 12.316$ the host star is a challenging target for precise planet mass measurements via RV, but the system might exhibit transit timing variations.  Accordingly, the system is one of the targets of our ongoing \emph{Spitzer} program to conduct follow-up transit observations of K2 planets \citep[Program 10067, PI: M. Werner;][]{beichman_et_al2016}.

Combining our new stellar characterization in D17 with our revised transit fits, we estimate the radii of the planets as  \mbox{$1.85 \pm 0.1\rearth$} for the inner planet and  \mbox{$2.51_{-0.20}^{+0.15}\rearth$} for the outer planet. Our results are larger than the \citet{petigura_et_al2015} estimates ($1.59 \pm 0.43 \rearth$ and $1.92 \pm 0.53\rearth$, respectively), but they agree within the errors.  

\subsubsection{EPIC~210508766 (K2-83): \\ Validated 2-Planet System} 
The EPIC~210508766 system contains an M1~dwarf orbited by two super-Earths: a $1.71 \pm 0.10 \rearth$ planet on a 2.747~day orbit and a $2.12\pm 0.12\rearth$ planet on a longer 9.997~day orbit. Neither planet is habitable ($F_p = 39 \pm 6\fearth$ and $F_p = 6.9_{-1.0}^{+1.1}\fearth$, respectively) and the system is too faint for RV mass measurement ($Kp = 13.844$, $Ks = 10.765$), but obtaining transmission spectra of both planets would be an interesting exercise in comparative planetology. 

Both planets were previously validated by \citet{crossfield_et_al2016}. As discussed in Section~\ref{sec:id_fps}, we confirm the validation of both planets as long as we account for the fact that planets in multi-planet systems are less likely to be false positives \citep{lissauer_et_al2012, sinukoff_et_al2016, vanderburg_et_al2016b}.

\subsubsection{EPIC~210968143 (K2-90): \\ Validated 2-Planet System} 
EPIC~210968143 is a K5~dwarf hosting two planets validated by \citet{crossfield_et_al2016} and re-validated in this paper. The outer planet ($2.53_{-0.13}^{+0.15}\rearth$) has a 13.734~day orbital period and the inner planet ($1.31_{-0.12}^{+0.13}\rearth$) has a 2.901~day orbital period. The two planets are far from resonance and the host star is too faint at optical wavelengths for precise mass measurement ($Kp = 13.723$), but the brighter infrared magnitude of $Ks=10.36$ renders the system amenable to atmospheric investigations. The inner planet EPIC~210968143.02 is highly irradiated \mbox{($F_p = 81_{-11}^{+13}\fearth$)}, but the outer planet EPIC~210968143.01 receives much less insolation ($F_p = 10.2_{-1.4}^{+1.7}\fearth$). A transmission spectrum of EPIC~210968143.01 might therefore help improve our understanding of less highly-irradiated mini-Neptune atmospheres.

\subsubsection{EPIC~211305568: \\ Candidate 2-Planet System} 
\label{ssec:211305568}
This M1V star was observed by K2 during Campaign~5. There are two K2OIs associated with the target, but neither is particularly compelling. The deeper transit-like events are attributed to 211305568.01, a $2.0_{-0.2}^{+26.9}\rearth$ K2OI with an orbital period of 11.6~days and the shallower events to 211305568.02, a small ($0.7 \pm 0.10\rearth$) K2OI with an ultra-short orbital period of 0.2~days. We calculated a high false positive probability for 211305568.01 (${\rm FPP} = 55\% - 100\%$), but {\tt vespa} could not fit the putative transits of EPIC~211305568.02 and therefore did not return an FPP. We classify both K2OIs as planet candidates in this paper, but we urge readers to obtain additional follow-up observations to verify the veracity of these signals. The star is moderately faint at optical wavelengths ($Kp = 13.849$), but easily observable at redder wavelengths ($Ks = 10.608$). 

\subsubsection{EPIC~211331236: \\ Newly Validated 2-Planet System} 
\label{ssec:211331236}
This system contains two short-period super-Earths ($1.9 \pm 0.12\rearth$ at 1.292~days, $2.03 \pm 0.13\rearth$ at 5.444~days) orbiting an M3~dwarf with a temperature of $3842 \pm 82$~K and a radius of $0.49\pm 0.03\rsun$. The star is too faint for radial velocity observations (\mbox{$Kp = 13.905$}, \mbox{$Ks = 10.589$}), but the system is amenable to atmospheric characterization (see Figure~\ref{fig:transdet}).

\subsubsection{EPIC~211428897: \\ Candidate 3-Planet System} 
The M2 dwarf EPIC~211428897 hosts three transiting Earth-sized planets with orbital periods of 1.611, 2.218, and 4.969 days and radii of $0.75 \pm 0.08\rearth$, $0.65\pm 0.07\rearth$, and $0.67\pm 0.08\rearth$, respectively. The inner two planet candidates are near the first-order 4:3 mean motion resonance and planet candidates .01 and .03 (1.611~days and 4.969~days) are near the second-order 3:1 mean motion resonance. Due to the small planet sizes, the anticipated radial velocity semi-amplitudes of 0.2 - 0.3 m/s are too small to be detected by current spectrographs (and the $Kp = 13.2$, $Ks = 9.624$ host star is too faint at optical wavelengths for high-precision RV observations), but the masses of the planets could be estimated via TTVs. The {\tt vespa} FPPs for these planets would be low enough to merit validation in the absence of nearby stars, but the presence of a nearby star precludes the use of {\tt vespa} for this system and means that the planets are larger than our estimates would suggest. The nearby star is roughly 1~magnitude fainter than and approximately $1\farcs1$ away from the target star. The correction factor for the planet radius estimates depends on which of the targets hosts the planets, but is likely to be on the order of a few \citep{ciardi_et_al2015, ciardi_et_al2017, furlan_et_al2017}. 

\subsection{Systems Orbiting Bright Stars}
\label{ssec:bright}
\subsubsection{EPIC~210707130: \\Relatively Bright Star Hosting a Validated Earth-sized USP Planet} 
\label{ssec:210707130}
As shown in Figure~\ref{fig:rvdet}, EPIC~210707130.01 (K2-85b) is one of the most attractive RV targets in our sample. This ultra-short period $1.39_{-0.12}^{+0.10}\rearth$ planet orbits a moderately bright ($Kp = 12.099$, $Ks = 9.466$) K5~dwarf every 0.685~days and was validated by \citet{crossfield_et_al2016}. Using the mass-radius relation from \citet{weiss+marcy2014}, we estimate a planet mass of $3.5\mearth$, which is consistent with the expectation that highly irradiated small planets tend to have rocky compositions \citep{dressing_et_al2015}. Given the estimated host star mass of $0.70\msun$, the anticipated RV semi-amplitude is 3.2~m/s.

\subsubsection{212460519.01:\\Moderately Bright Star Hosting a Newly Validated Super-Earth}
\label{ssec:212460519}
EPIC~212460519 is a moderately bright star ($Kp = 12.445$, $Ks = 9.712$) hosting a $1.89_{-0.10}^{+0.11}\rearth$ planet with a 7.4~day orbital period. The planet is too hot to be habitable ($F_p = 35_{-6}^{+7}\fearth$), but might be an interesting target for atmospheric characterization. Assuming $m_p = 5\mearth$ as predicted by the \citet{weiss+marcy2014} mass-radius relation, EPIC~212460519.01 is expected to induce an RV semi-amplitude of 2.2~m/s. 

\subsection{Comparison of K2 Photometric Pipelines}
\label{ssec:photcomp}
In order to investigate the influence of systematic effects in the reduced K2 photometry on the derived planet properties, we repeated the transit fits using photometry processed by the K2 Systematics Correction Pipeline\footnote{\url{https://archive.stsci.edu/prepds/k2sc/}}$^{,}$\footnote{\url{https://github.com/OxES/k2sc}} \citep{aigrain_et_al2015, aigrain_et_al2016} and the k2phot pipeline \citep{petigura_et_al2013a}. We note that lightcurves produced using the k2varcat \citep{armstrong_et_al2014b,armstrong_et_al2015,armstrong_et_al2016} and EVEREST \citep{luger_et_al2016} pipelines are also available on the MAST, but we did not perform fits using those lightcurves.

\begin{figure*}[tbh]
\centering
\includegraphics[width=0.3\textwidth]{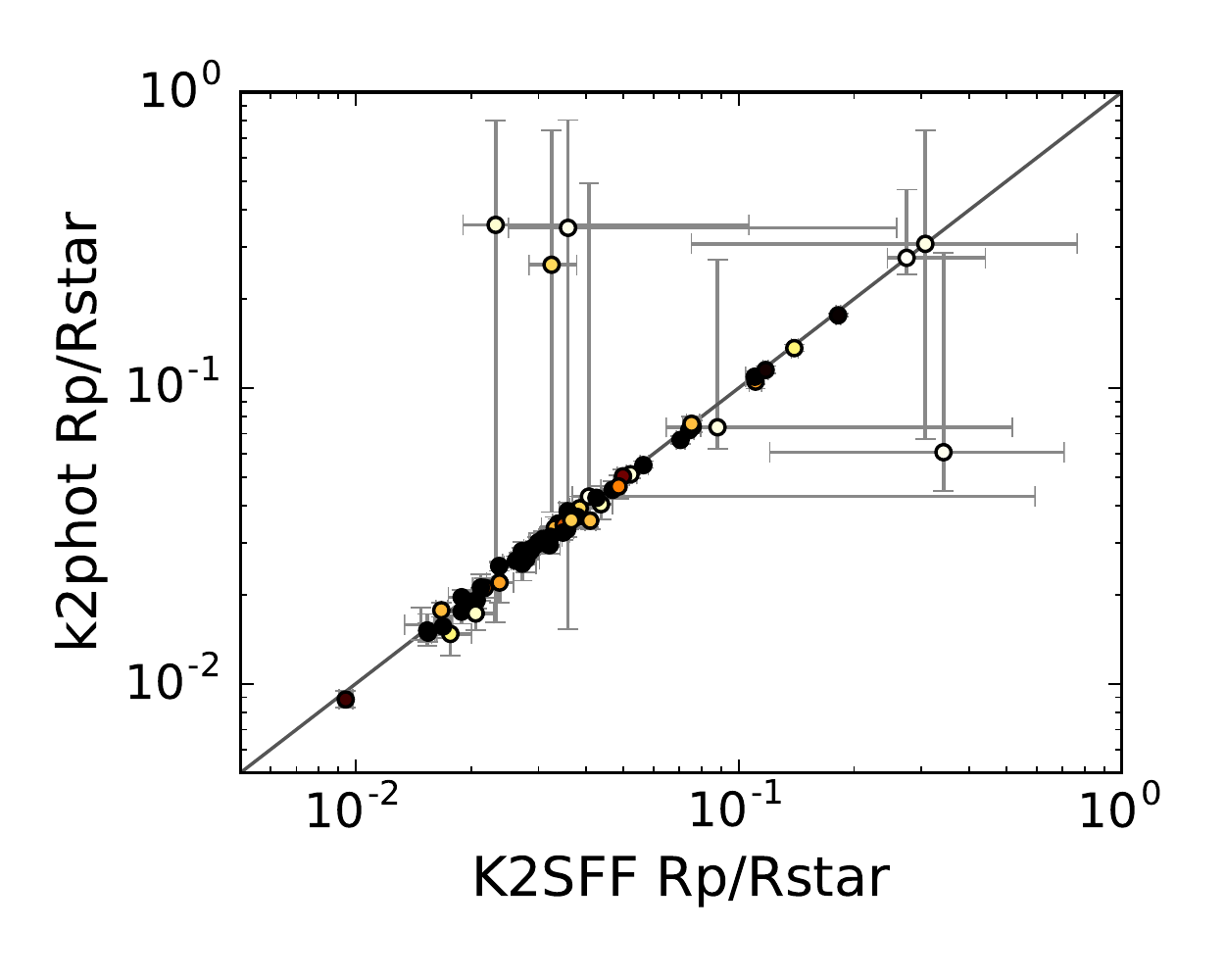}
\includegraphics[width=0.3\textwidth]{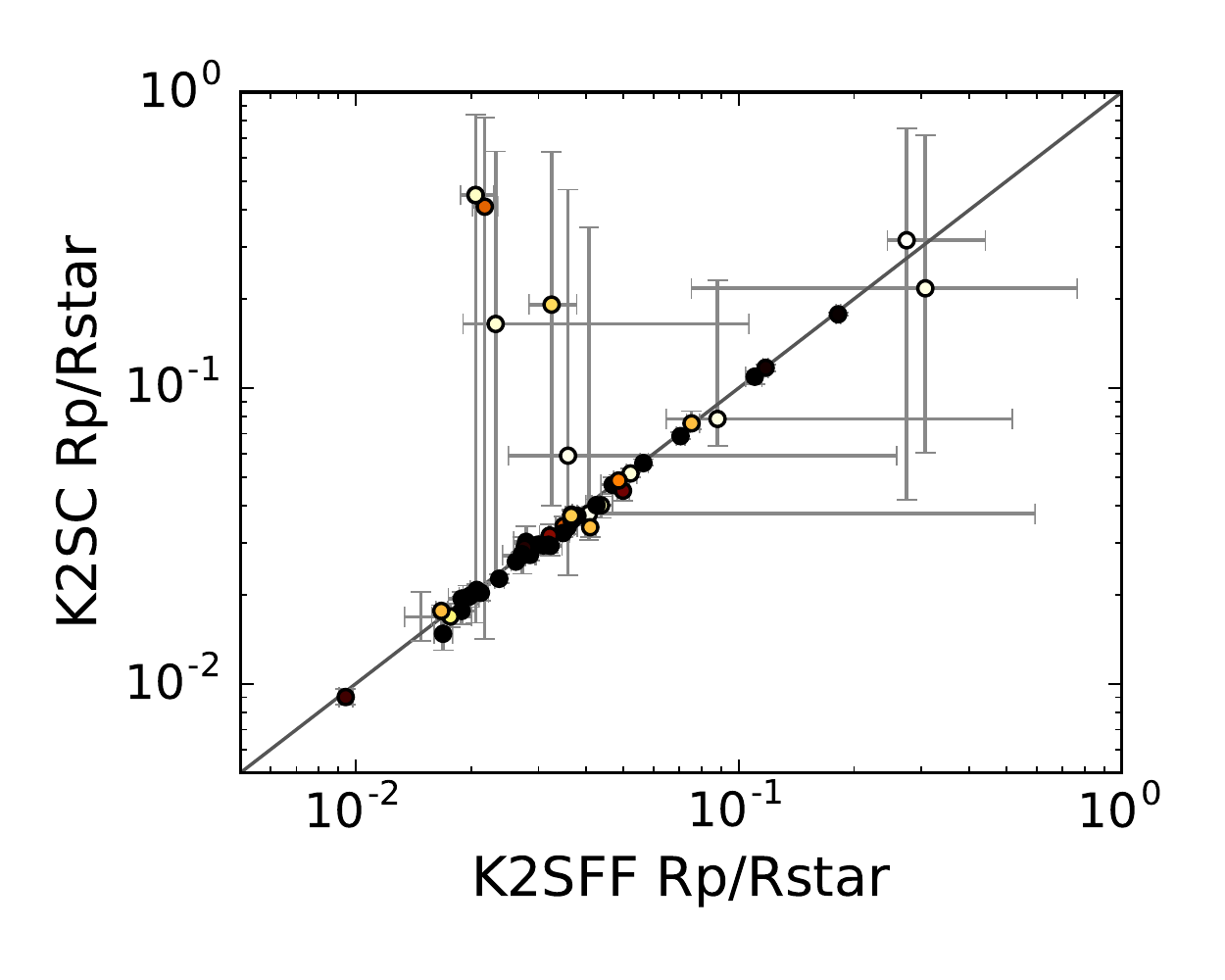}
\includegraphics[width=0.3\textwidth]{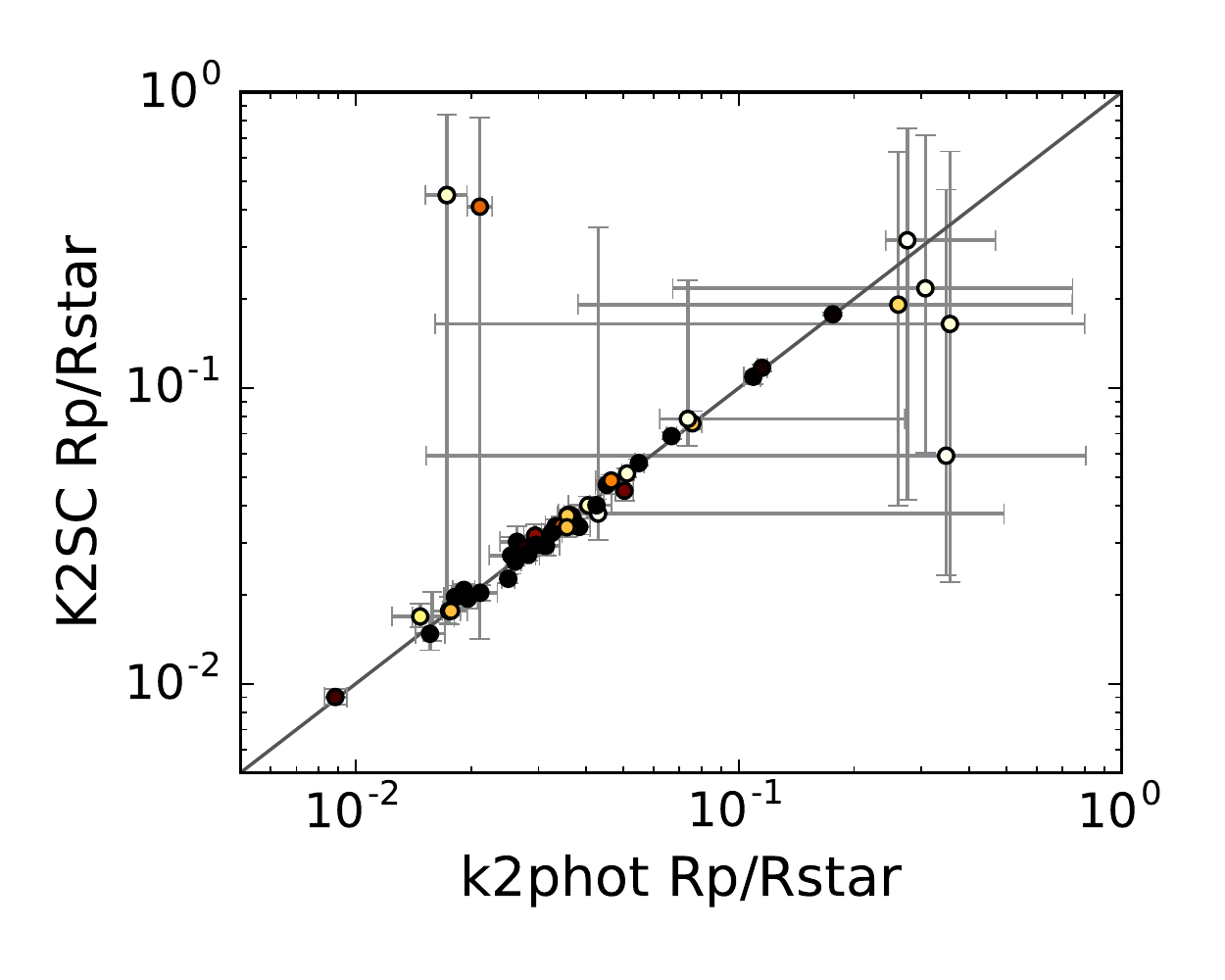}\\
\includegraphics[width=0.3\textwidth]{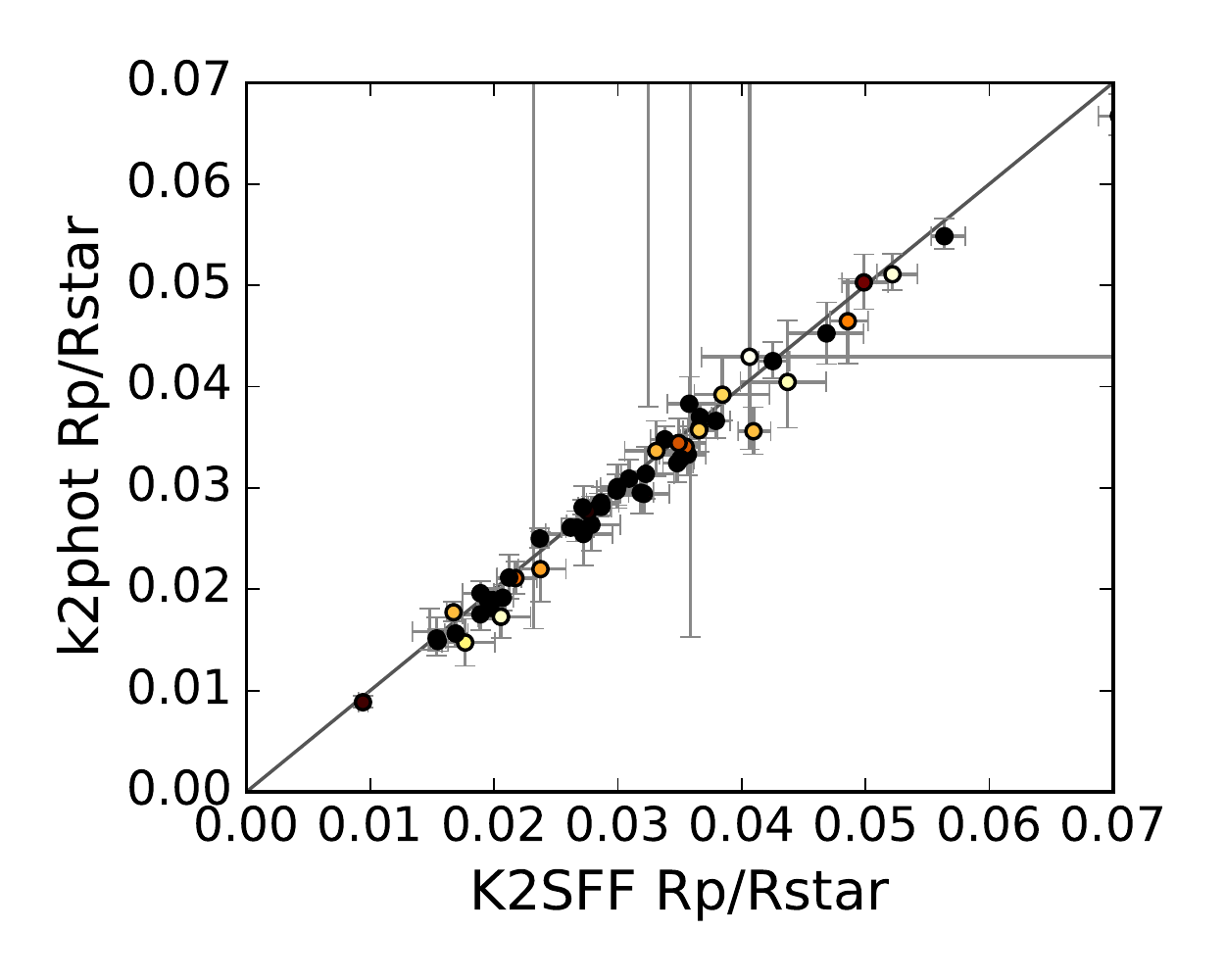}
\includegraphics[width=0.3\textwidth]{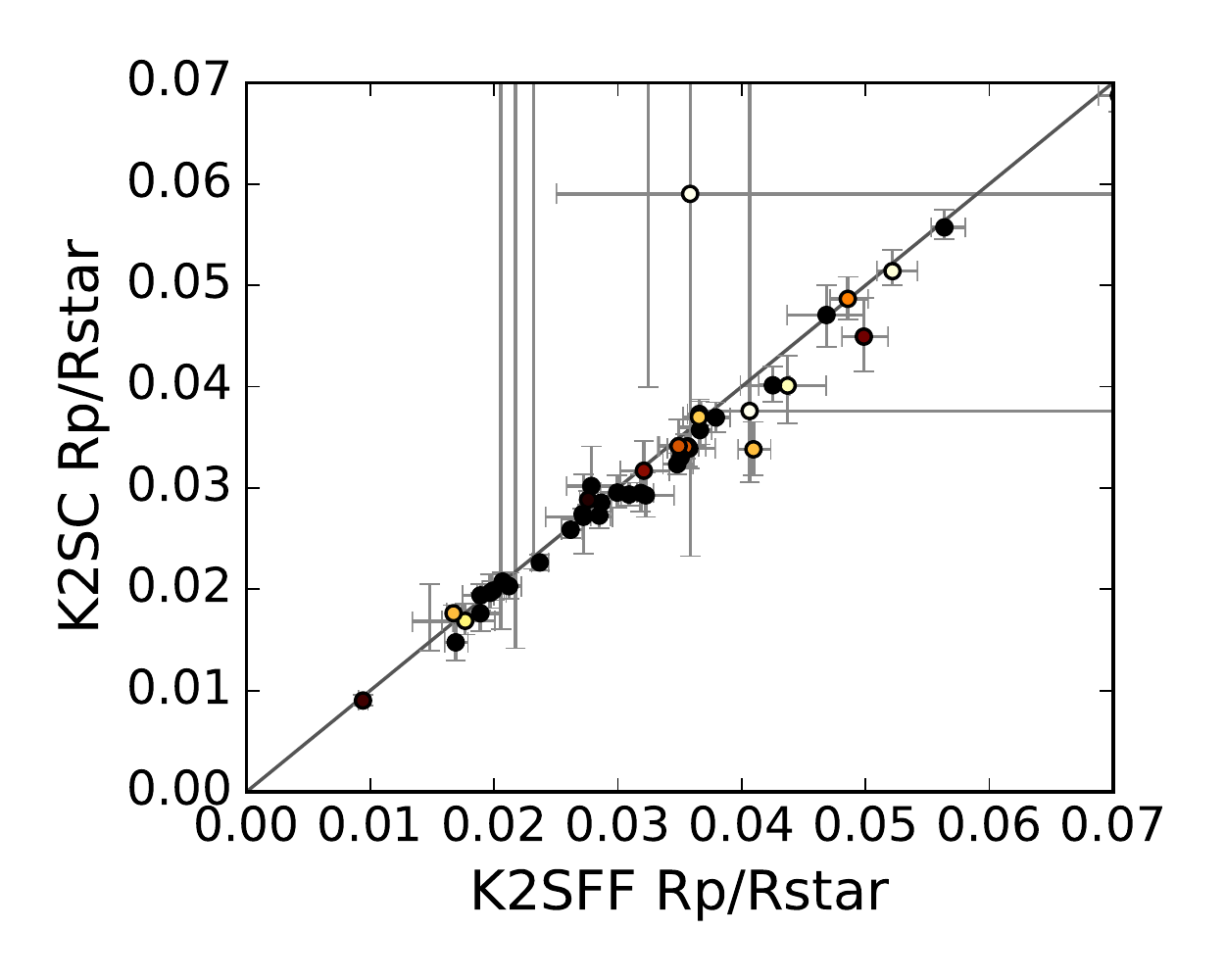}
\includegraphics[width=0.3\textwidth]{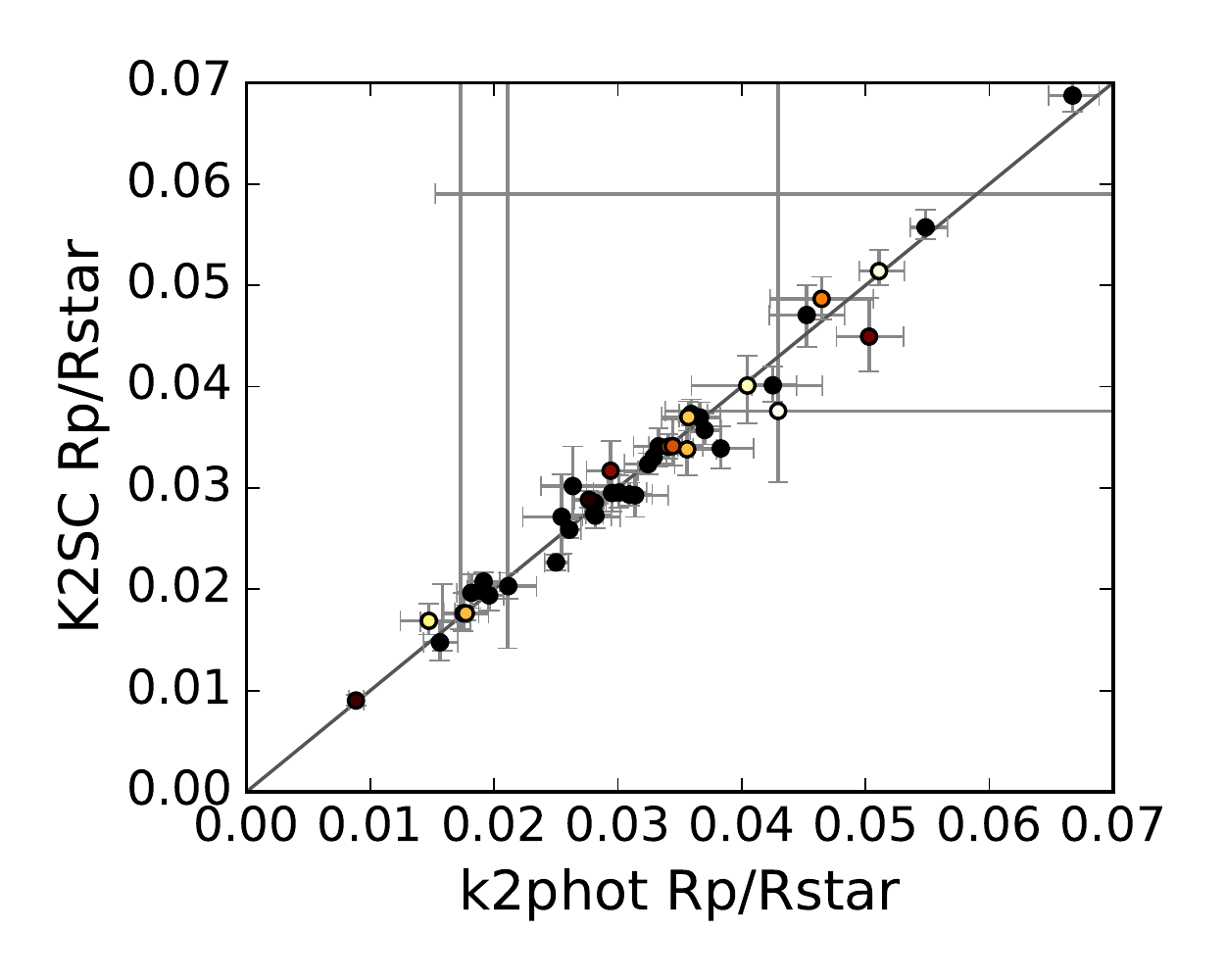}
\caption{Comparison of planet/star radius ratios found by fitting photometry from different pipelines. The symbols are colored based on false positive probability using the same color scaling as in Figure~\ref{fig:transdet}; K2OIs with lower FPPs have darker colors. The top panels display all K2OIs while the bottom panels zoom in to highlight the K2OIs with small $R_p/R_\star$ ratios. Only planet candidates and validated planets are shown. \emph{Left: } k2phot fits versus K2SFF fits.  \emph{Middle: } K2SC fits versus K2SFF fits.  \emph{Right: } K2SC fits versus k2phot fits. }
\label{fig:phot_comp}
\end{figure*}

Figure~\ref{fig:phot_comp} compares the resulting planet/star radius ratios found by fitting photometry from the K2SFF, K2SC, and k2phot pipelines. For K2OIs with small $R_p/R_\star$, we find generally consistent planet properties regardless of our choice of photometric pipeline, but the estimates for K2OIs with $R_p/R_\star > 0.05$ can be quite discrepant. In general, the K2OIs with the largest radius disagreement are false positives for which the transit model provides a poor fit to the data. In contrast, nearly all of the K2OIs with $R_p/R_\star > 0.05$ appear well-fit by a transit model and many were classified as confirmed planets in Section~\ref{sec:vespa} based on their K2SFF lightcurves. 

Specifically, the median absolute difference in $R_p/R_{\star,{\rm KSFF}}$ and $R_p/R_{\star,{\rm k2phot}}$ is $\Delta R_p/R_{\star,{\rm  K2SFF - k2phot}}= 0.001$ for validated planets, $\Delta R_p/R_{\star,{\rm  K2SFF - k2phot}} = 0.002$ for planet candidates, and $\Delta R_p/R_{\star,{\rm  K2SFF - k2phot}} = 0.04$ for false positives. Comparing the KSFF and K2SC fits, we find similar MAD of 0.001 for validated planets and 0.003 for planet candidates, but a higher MAD of 0.12 for false positives. These values are nearly identical to the MAD of 0.001, 0.002, and 0.10 found when comparing the $R_p/R_\star$ fits from the k2phot and K2SC pipelines for validated planets, planet candidates, and false positives, respectively. 

Despite the overall agreement between the transit fits produced using distinct sets of photometry, there are a few K2OIs with large $R_p/R_\star$ differences across pipelines. For instance, we noticed that the K2SC pipeline appeared to remove the transits of EPIC~211799258.01. In addition, our k2phot fits for EPIC~211826814.01 and EPIC~212443973.01 and our K2SC fits for EPIC~206312951.01 and EPIC~211694226.01 failed to converge on the period and transit midpoint. Finally, our K2SFF fit for EPIC~211988320.01 found a larger planet radius than our k2phot and K2SC fits: \mbox{$R_p/R_\star = 0.041 \pm 0.001$} for K2SFF versus \mbox{$R_p/R_\star =0.036 \pm 0.002$} for k2phot and \mbox{$R_p/R_\star = 0.034 \pm 0.003$} for K2SC. In general, we found that comparing fits from different pipelines was a convenient way to bolster confidence in borderline detections and reject astrophysical false positives due to blended photometry.

\section{Conclusions}
\label{sec:conc}
Due to the dependence of planet radius estimates on host star characterization, our recent work improving the classification of cool dwarfs observed by K2 significantly affected the assumed properties of the associated planet candidates. In general, we found that the host star radii were underestimated by $8-40$\% (D17), implying that the initial planet radius estimates were also undersized. In this paper, we investigated 79~candidate transit signals in the lightcurves of 74~low-mass K2 target stars and provided an updated catalog of planet properties. Our revisions to the system properties include both improved classifications of the host stars from D17 and our new transit fits. As part of the analysis, we also assessed the credibility of each planet candidate by using the {\tt vespa} framework developed by \citet{morton2012, morton2015} to calculate the probability that each transit-like event was caused by an astrophysical false positive rather than a genuine planetary transit.
 
In total, we considered 79~putative transit events. We rejected six as false positives, validated~18 as bona fide planets, and classified 17~new planet candidates. We also upheld the previous classifications of two false positives and provided updated planet radius estimates for 15~planet candidates and 21~validated planets announced in earlier publications.

Our cool dwarf planet sample is dominated by small worlds: 56\% (40) are smaller than $2\rearth$ and 85\% (60) are smaller than Neptune.  Compared to the planets detected during the prime \emph{Kepler} mission, our candidates tend to orbit brighter stars that are more amenable to follow-up observations. In particular, the thirteen small planets with radii of $1.5-3\rearth$ and host stars brighter than $Ks=11$ are likely targets for atmospheric characterization studies with \emph{Spitzer}, HST, and JWST. Furthermore, radial velocity observations may be feasible for the brightest systems, particularly given the advent of red spectrographs like CARMENES \citep{quirrenbach_et_al2010}, the Habitable Zone Planet Finder \citep{mahadevan_et_al2010}, the Infrared Doppler instrument \citep{tamura_et_al2012, kotani_et_al2014}, iSHELL \citep{rayner_et_al2012}, and SpiROU \citep{delfosse_et_al2013, artigau_et_al2014}.  

As the K2 Mission continues, additional planets will be detected around stars across the ecliptic. We will continue to conduct follow-up observations for future targets and eagerly await the opportunity to characterize the best systems with JWST. Beginning in 2018, we will expand our follow-up program to include planet candidates detected by the Transiting Exoplanet Survey Satellite \citep{ricker_et_al2014, sullivan_et_al2015}, which is scheduled to launch in March~2018.

\begin{acknowledgments}
Many of our targets were provided by the K2 California Consortium (K2C2). We thank K2C2 for sharing their candidate lists and vetting products. We are grateful to Tim Morton for making {\tt vespa} publicly available and to Jennifer Winters, Nic Scott, and Lea Hirsch for assisting with speckle imaging. We also acknowledge helpful conversations with Chas Beichman, Eric Gaidos, Jessie Christiansen, Michael Werner, and Arturo Martinez. 

This work was performed under contract with the Jet Propulsion Laboratory (JPL) funded by NASA through the Sagan Fellowship Program executed by the NASA Exoplanet Science Institute. A.V. is supported by the NSF Graduate Research Fellowship, grant No. DGE 1144152. This publication was made possible through the support of a grant from the John Templeton Foundation. The opinions expressed here are those of the authors and do not necessarily reflect the views of the John Templeton Foundation. This paper includes data collected by the K2 mission, which is funded by the NASA Science Mission directorate. This research has made use of the NASA Exoplanet Archive, and the Exoplanet Follow-up Observation Program website, which are operated by the California Institute of Technology, under contract with the National Aeronautics and Space Administration under the Exoplanet Exploration Program. We appreciate the willingness of ExoFOP users to share their follow-up observations with the broader K2 community.

Our stellar characterization follow-up observations were obtained at the Infrared Telescope Facility, which is operated by the University of Hawaii under contract NNH14CK55B with the National Aeronautics and Space Administration and at Palomar Observatory. We thank the staff at both observatories and the Caltech Remote Observing Facilities staff for supporting us during our many observing runs. We are grateful to the IRTF and Caltech TACs for awarding us telescope time. 

Our contrast curves were based on observations obtained at Gemini Observatory, Kitt Peak National Observatory via the NN-EXPLORE program time allocation at the WIYN telescope, and the W.M. Keck Observatory. Kitt Peak National Observatory, National Optical Astronomy Observatory, is operated by the Association of Universities for Research in Astronomy (AURA) under a cooperative agreement with the National Science Foundation. Gemini Observatory is operated by the Association of Universities for Research in Astronomy, Inc., under a cooperative agreement with the NSF on behalf of the Gemini partnership: the National Science Foundation (United States), the National Research Council (Canada), CONICYT (Chile), Ministerio de Ciencia, Tecnolog\'{i}a e Innovaci\'{o}n Productiva (Argentina), and Minist\'{e}rio da Ci\^{e}ncia, Tecnologia e Inova\c{c}\~{a}o (Brazil). The W.M. Keck Observatory is operated as a scientific partnership among the California Institute of Technology, the University of California and the National Aeronautics and Space Administration. The Observatory was made possible by the generous financial support of the W.M. Keck Foundation. 

The authors wish to recognize and acknowledge the very significant cultural role and reverence that the summit of Mauna Kea has always had within the indigenous Hawaiian community.  We are most fortunate to have the opportunity to conduct observations from this mountain. 
\end{acknowledgments}

\facilities{Kepler, K2, IRTF (SpeX), Palomar:Hale (TripleSpec, PALM-3000/PHARO), Gemini-N (DSSI, NIRI), Gemini-S (DSSI), Keck:II (NIRC2), WIYN (DSSI)}

\bibliography{mdwarf_biblio.bib}

\appendix
\section{Vetting Plots}
\label{sec:appendix}
\begin{figure*}[hbp]
\centering
\includegraphics[width=0.3\textwidth]{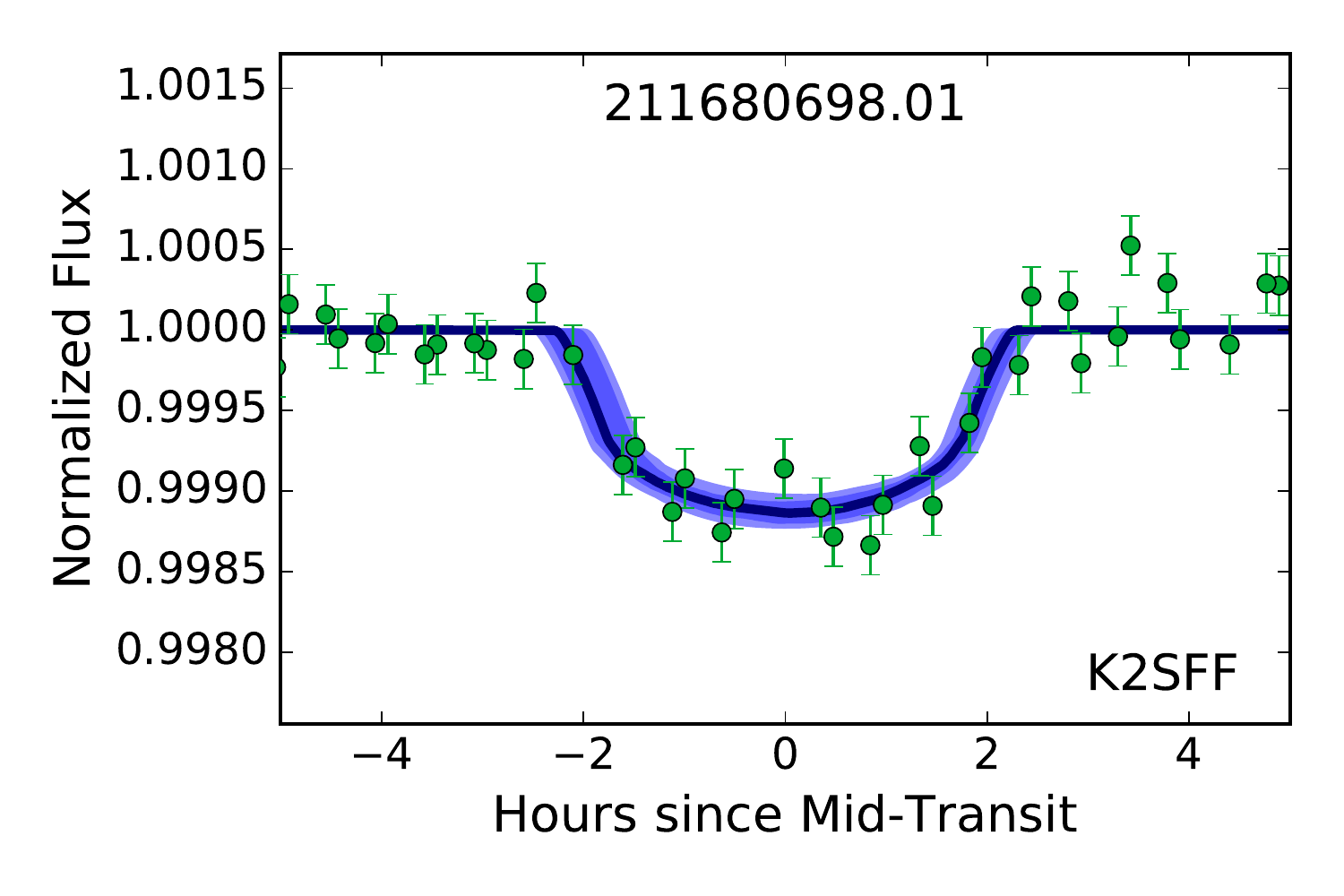} 
\includegraphics[width=0.3\textwidth]{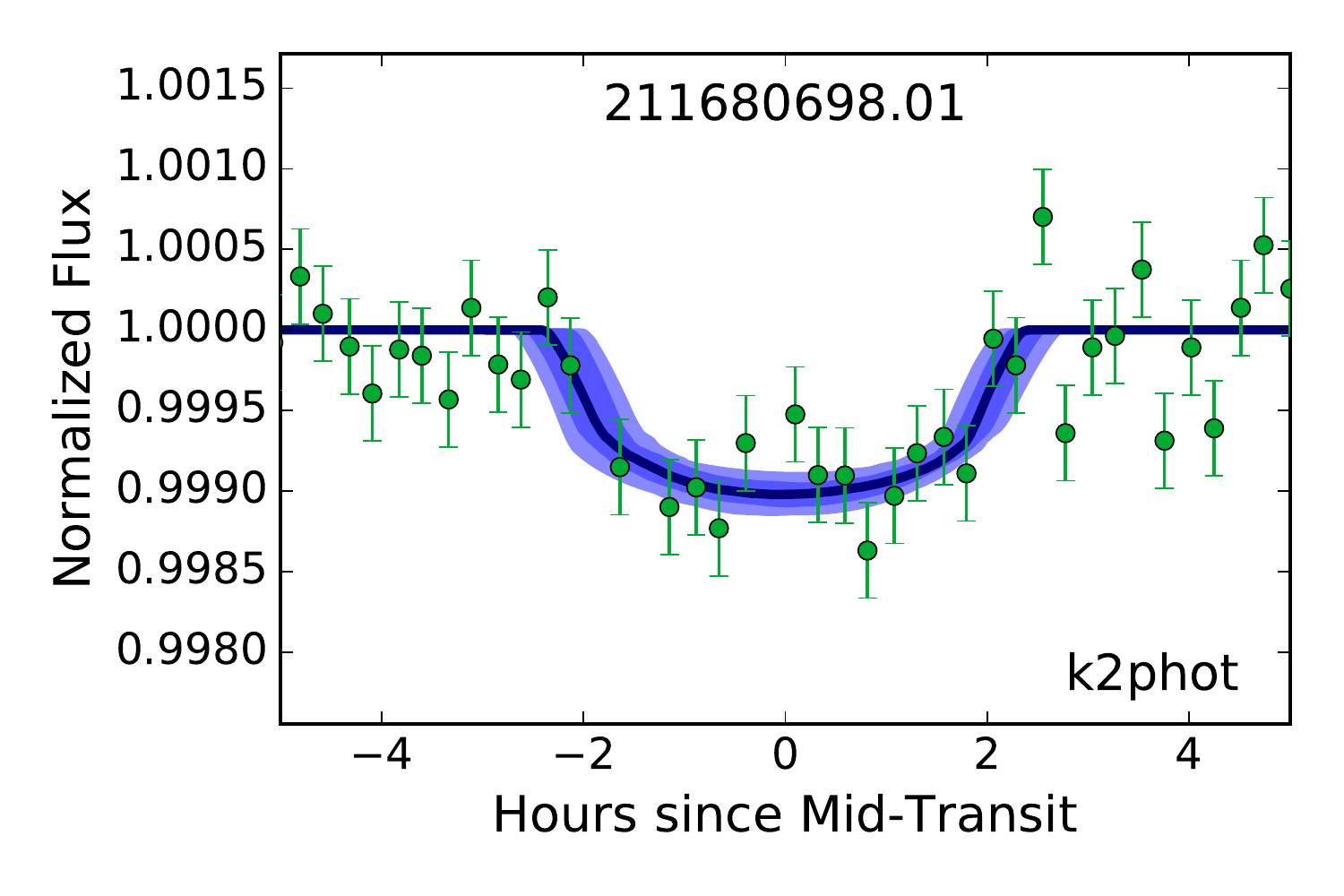} 
\includegraphics[width=0.3\textwidth]{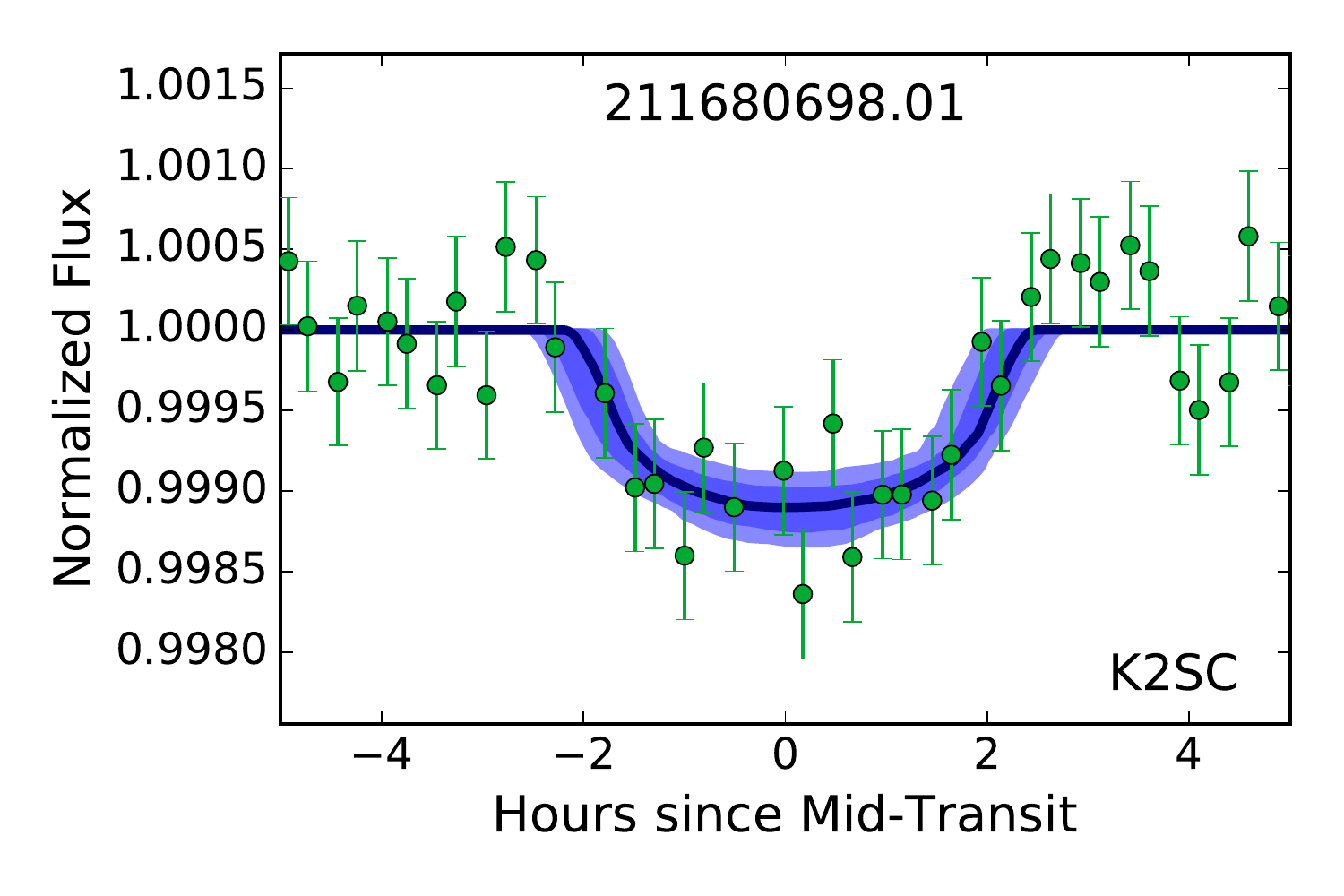} 
\includegraphics[width=0.3\textwidth]{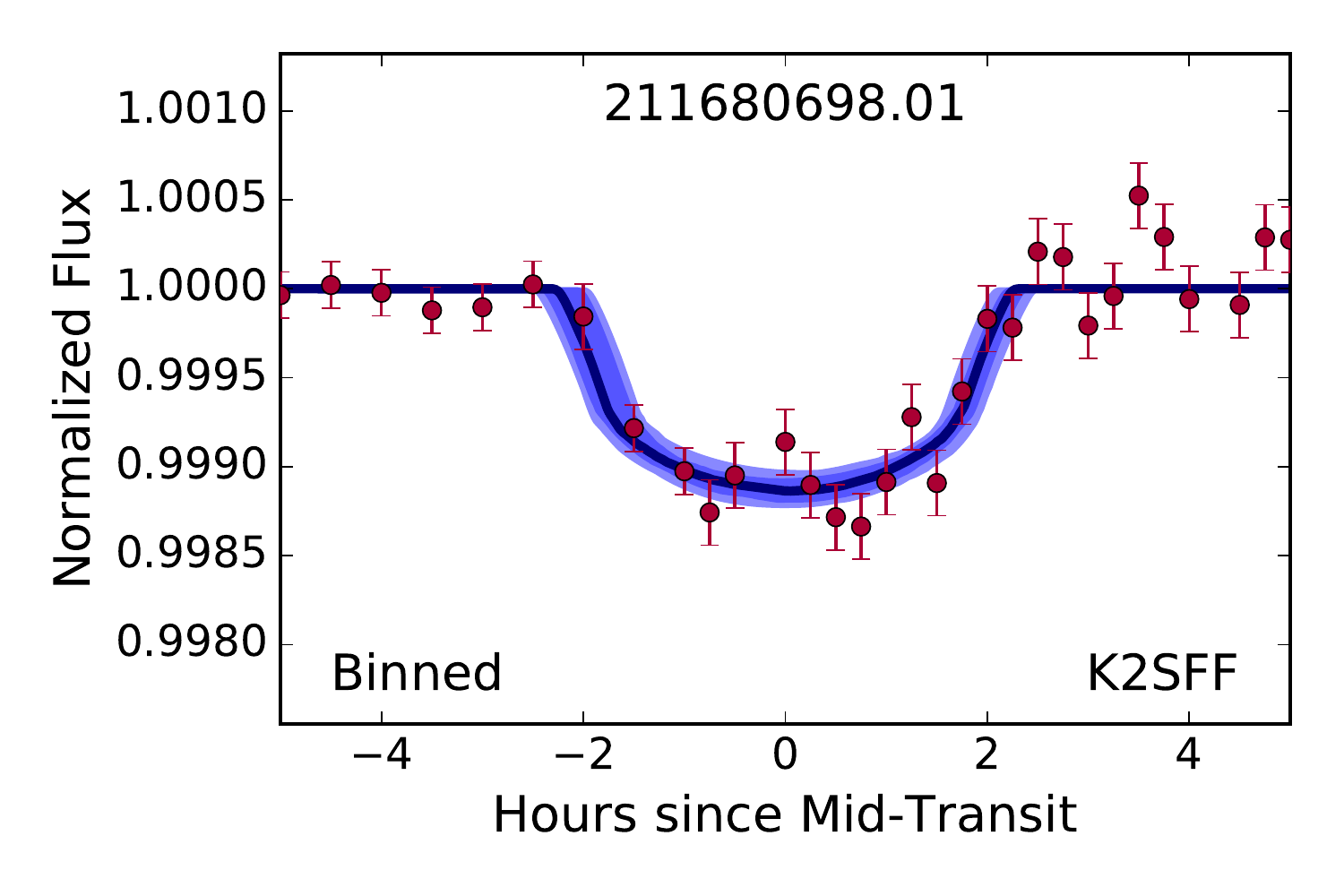} 
\includegraphics[width=0.3\textwidth]{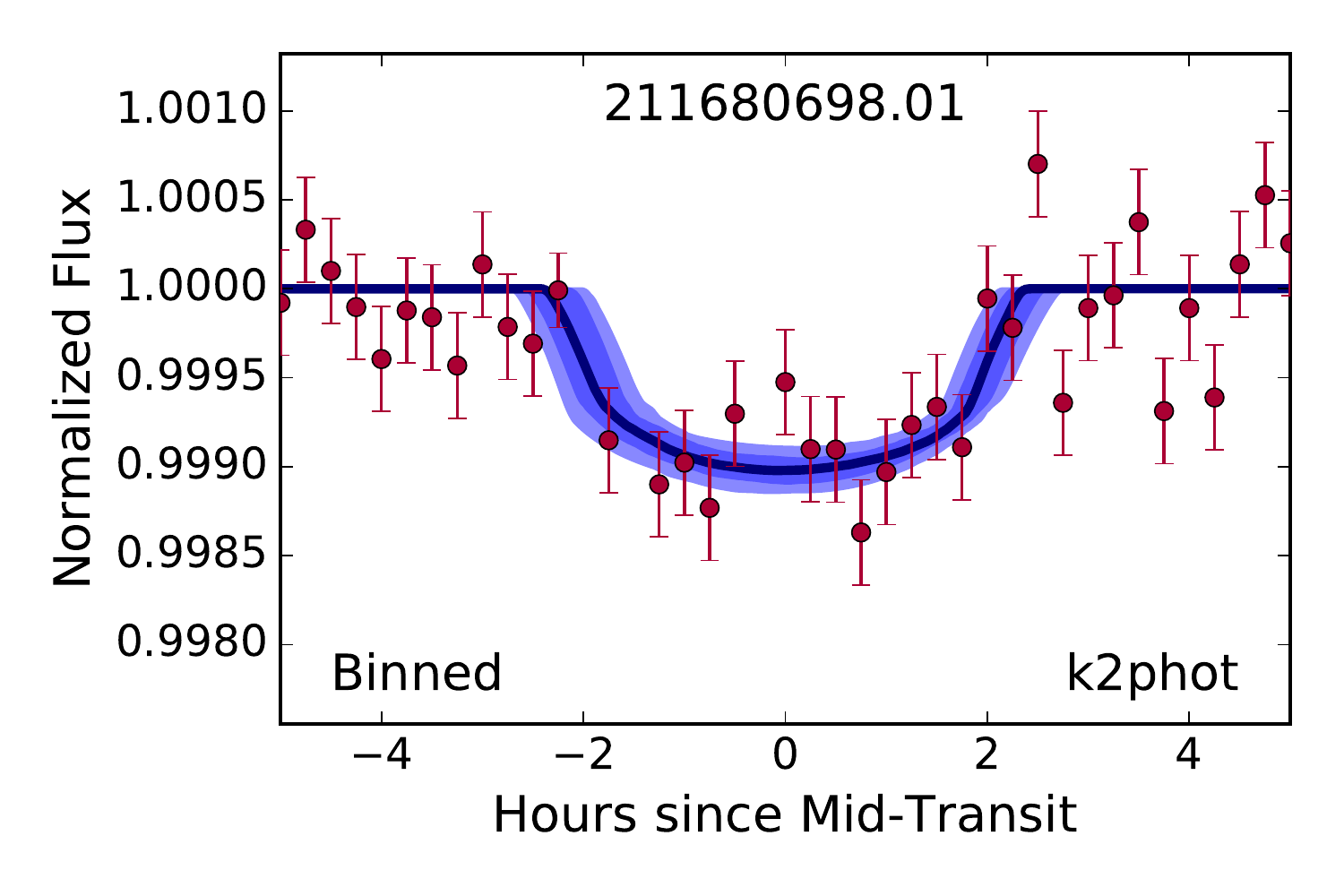} 
\includegraphics[width=0.3\textwidth]{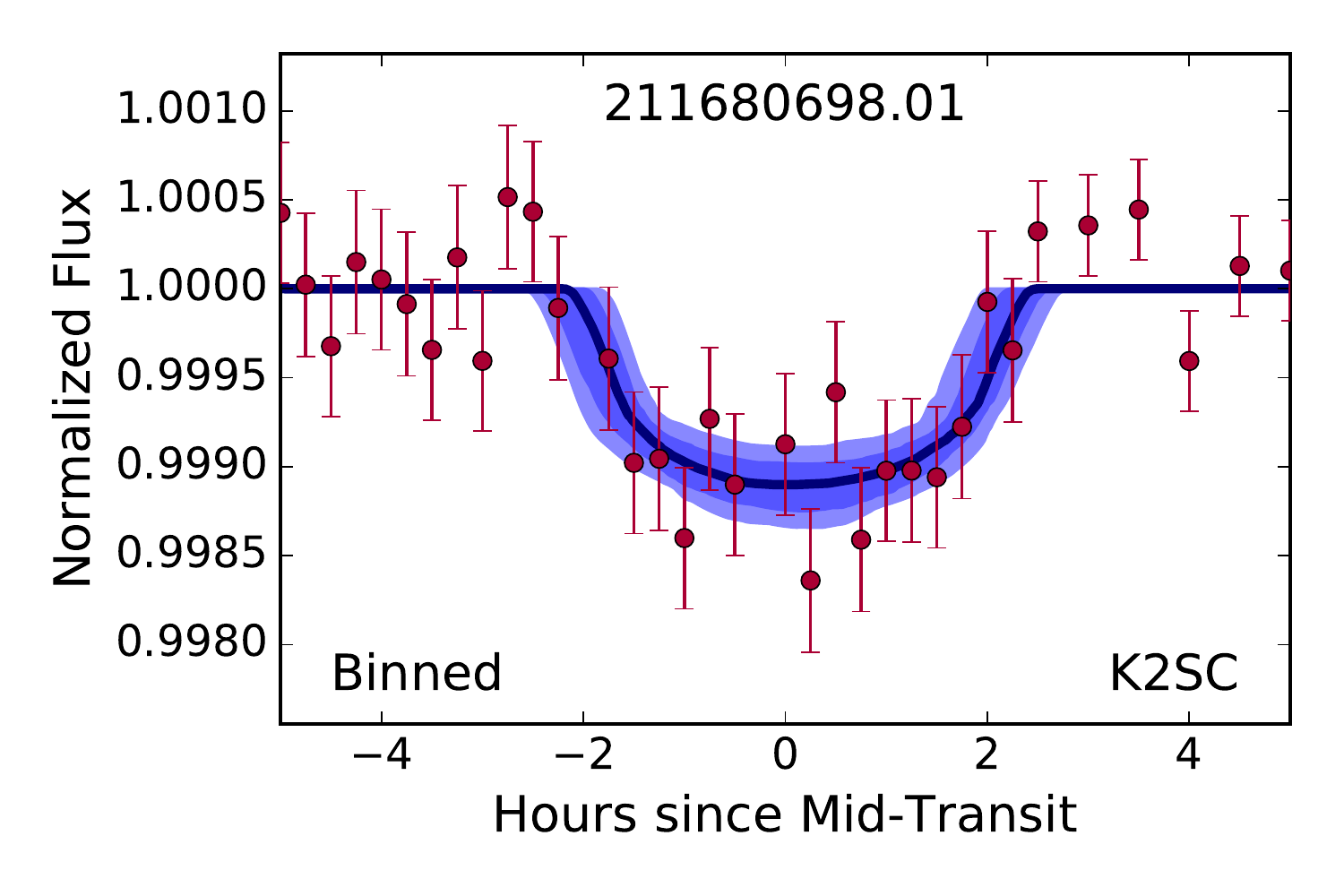} 
\includegraphics[width=0.3\textwidth]{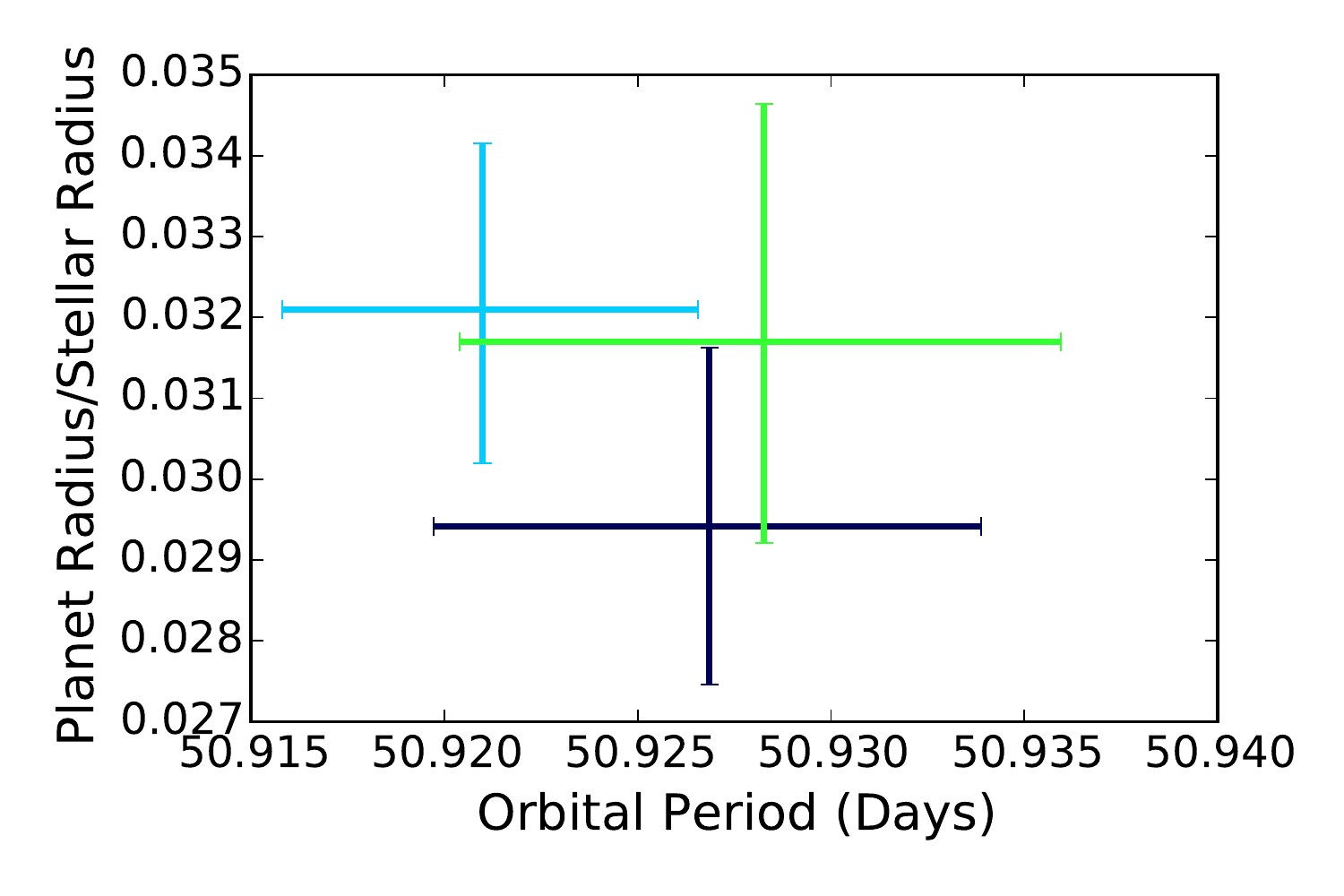} 
\includegraphics[width=0.3\textwidth]{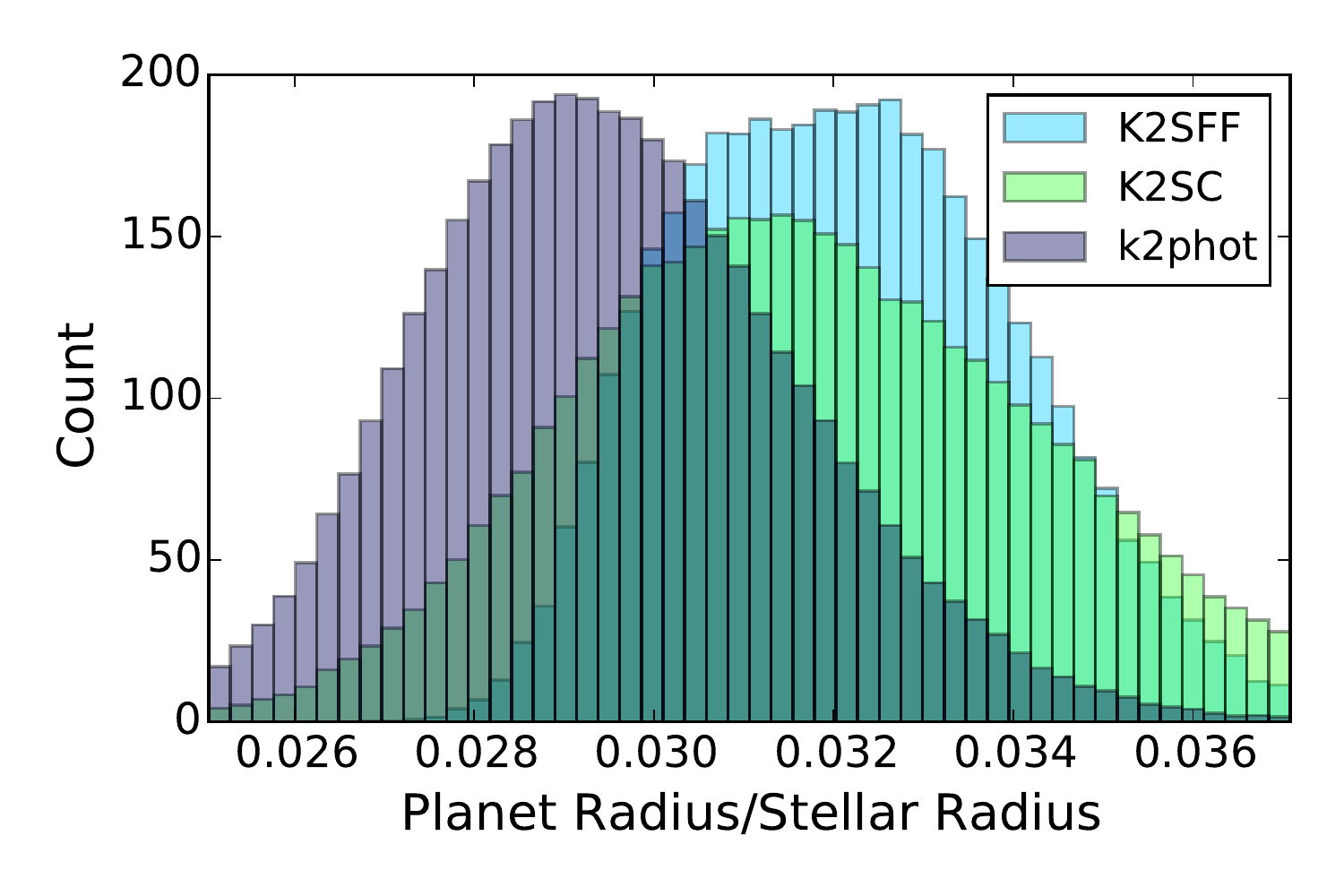} 
\includegraphics[width=0.3\textwidth]{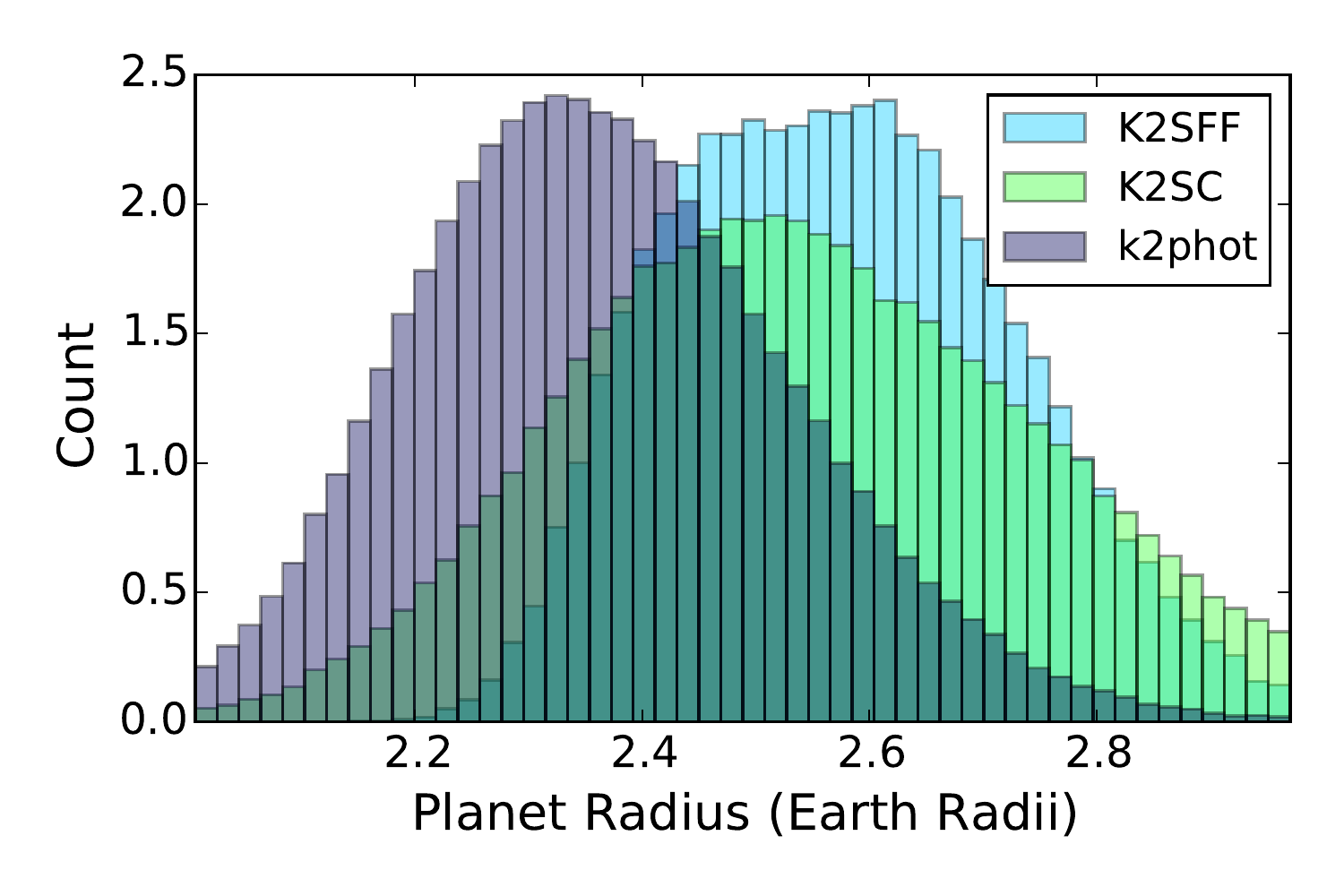} 
\includegraphics[width=0.3\textwidth]{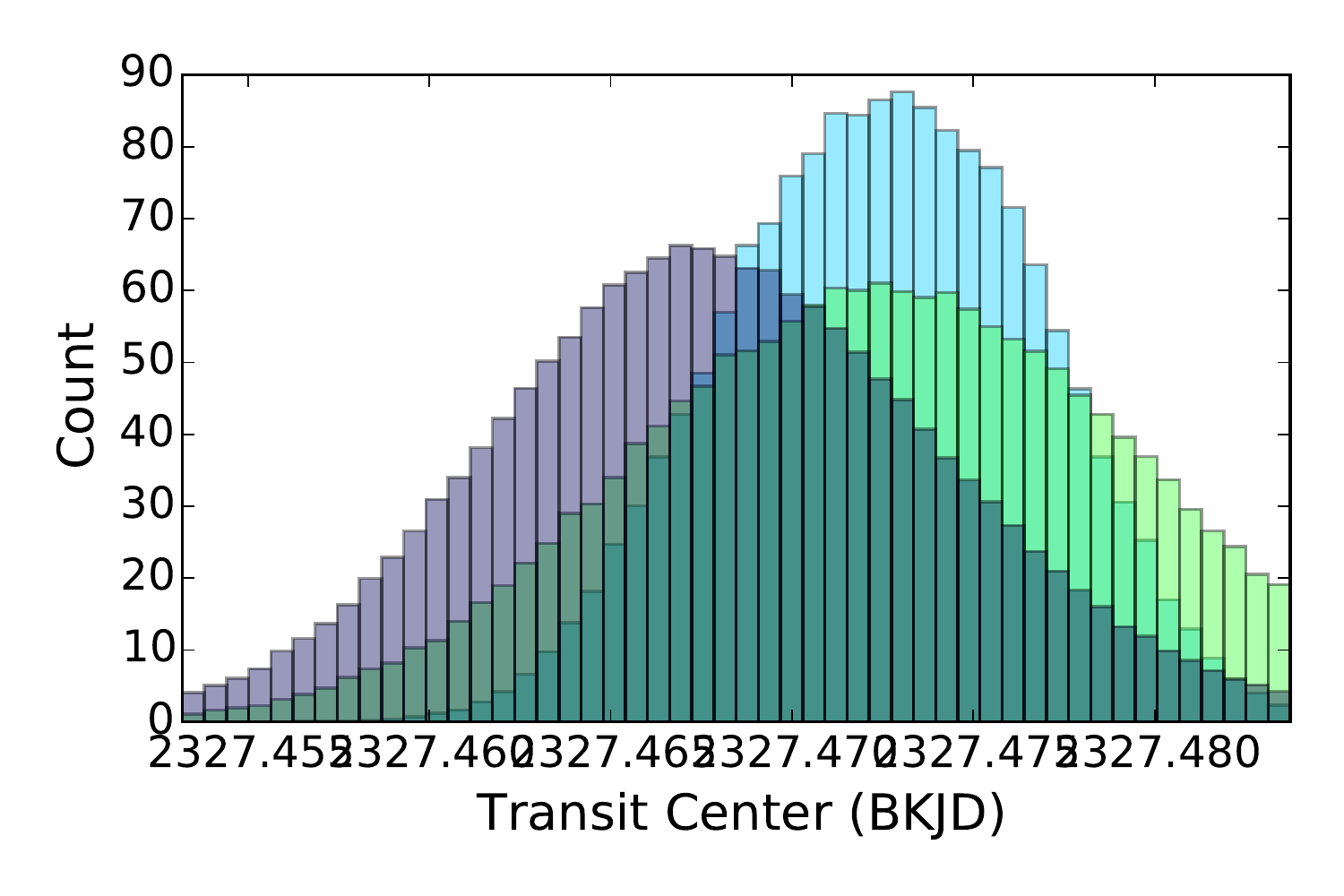} 
\includegraphics[width=0.3\textwidth]{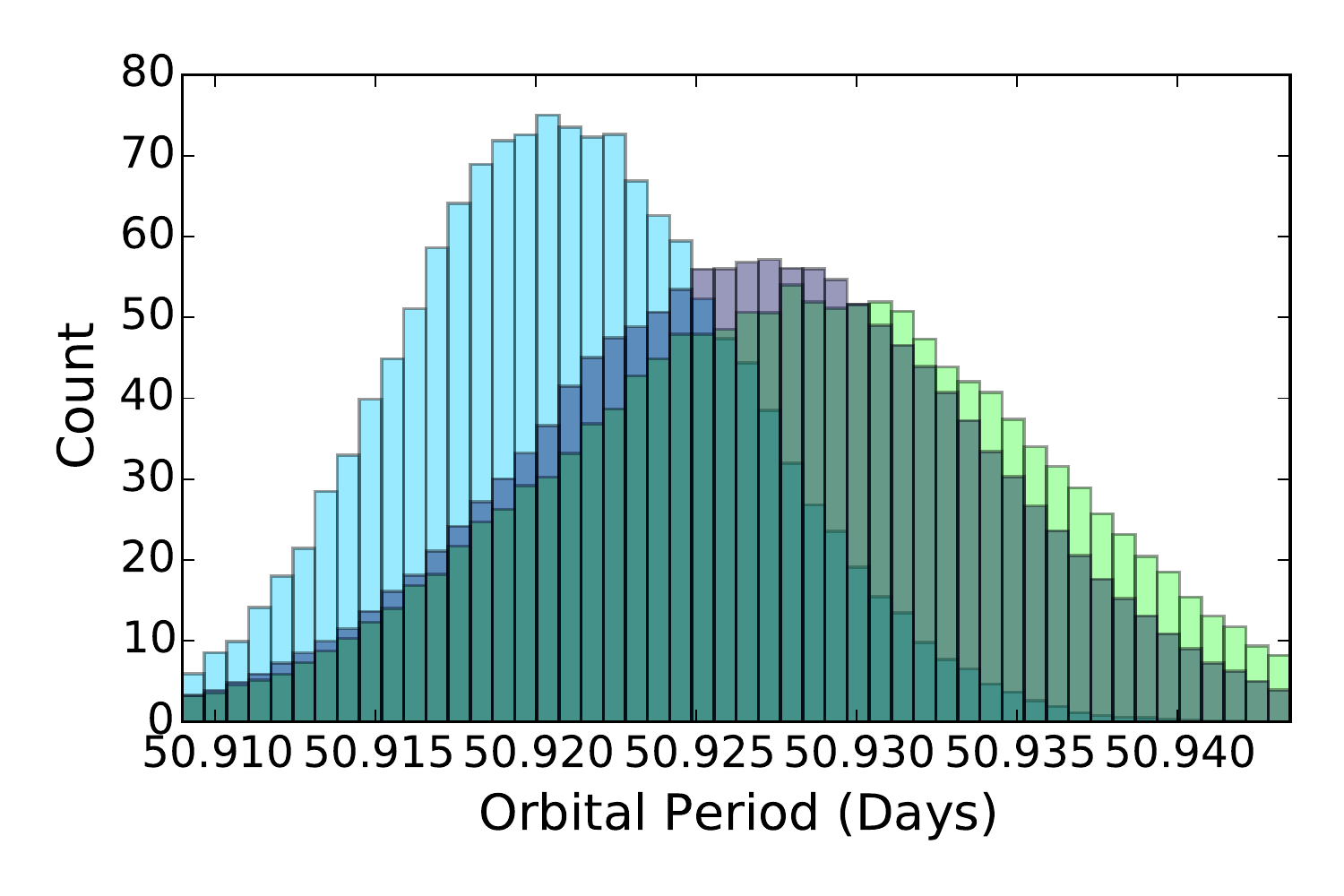} 
\includegraphics[width=0.3\textwidth]{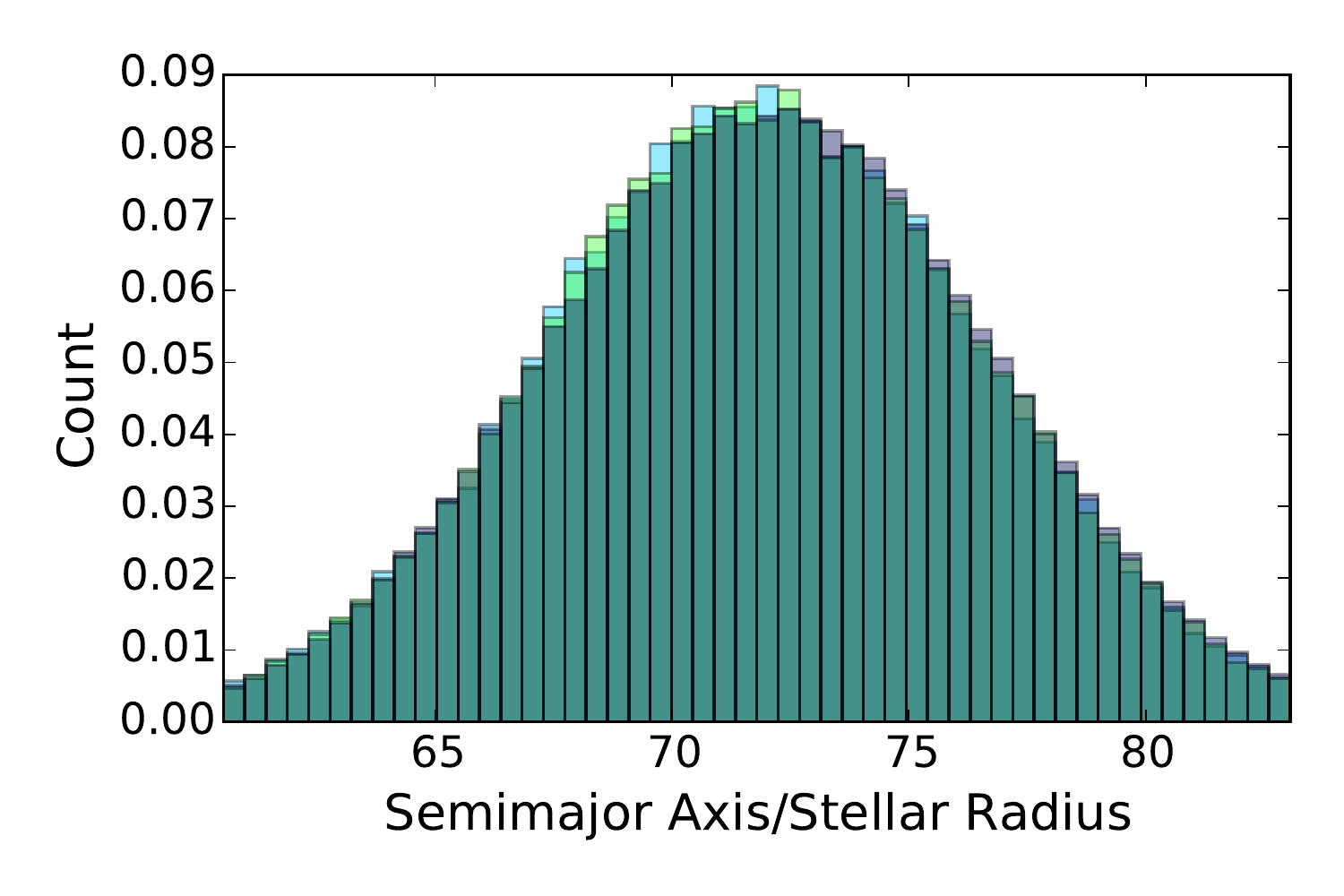} 
\includegraphics[width=0.3\textwidth]{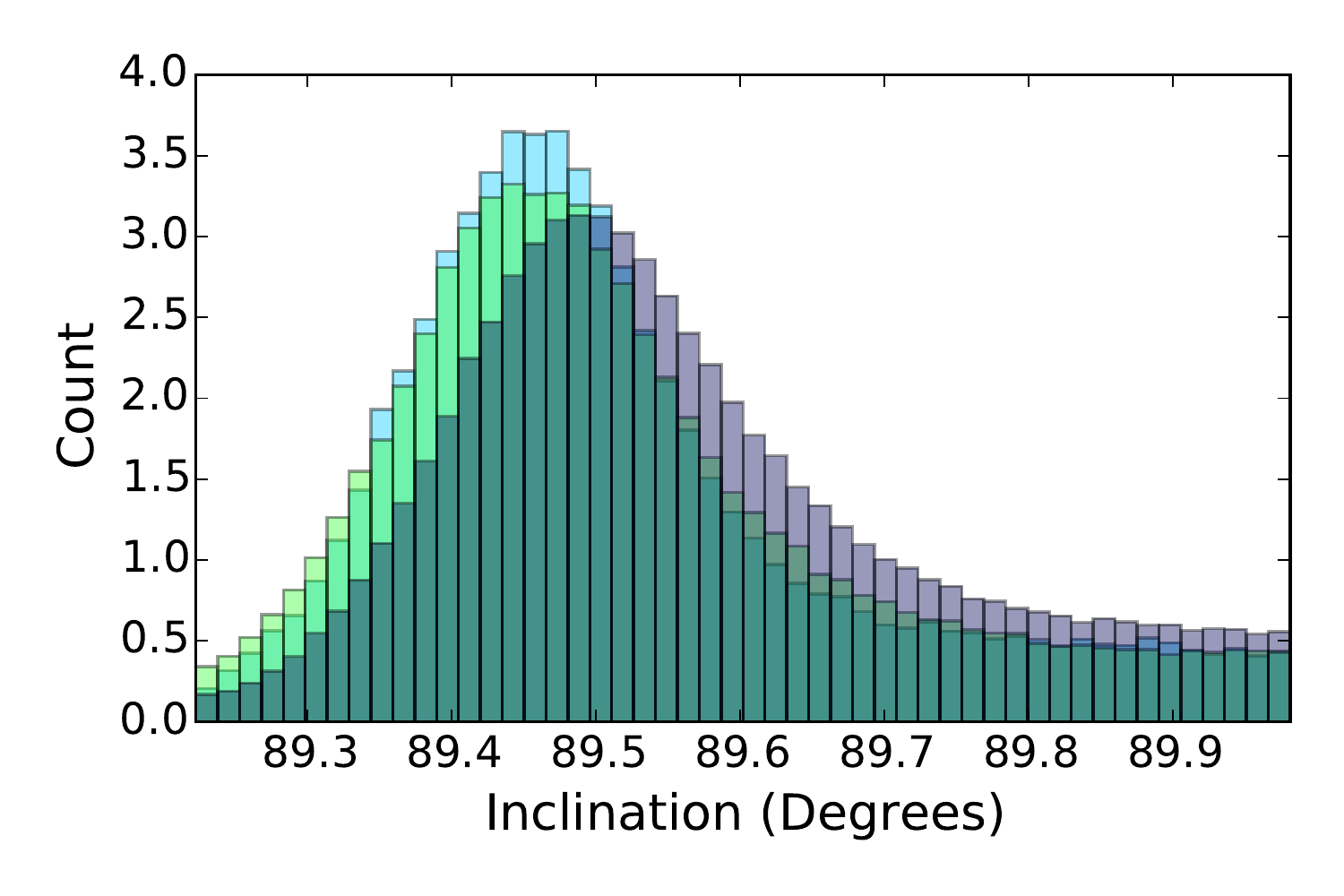} 
\includegraphics[width=0.3\textwidth]{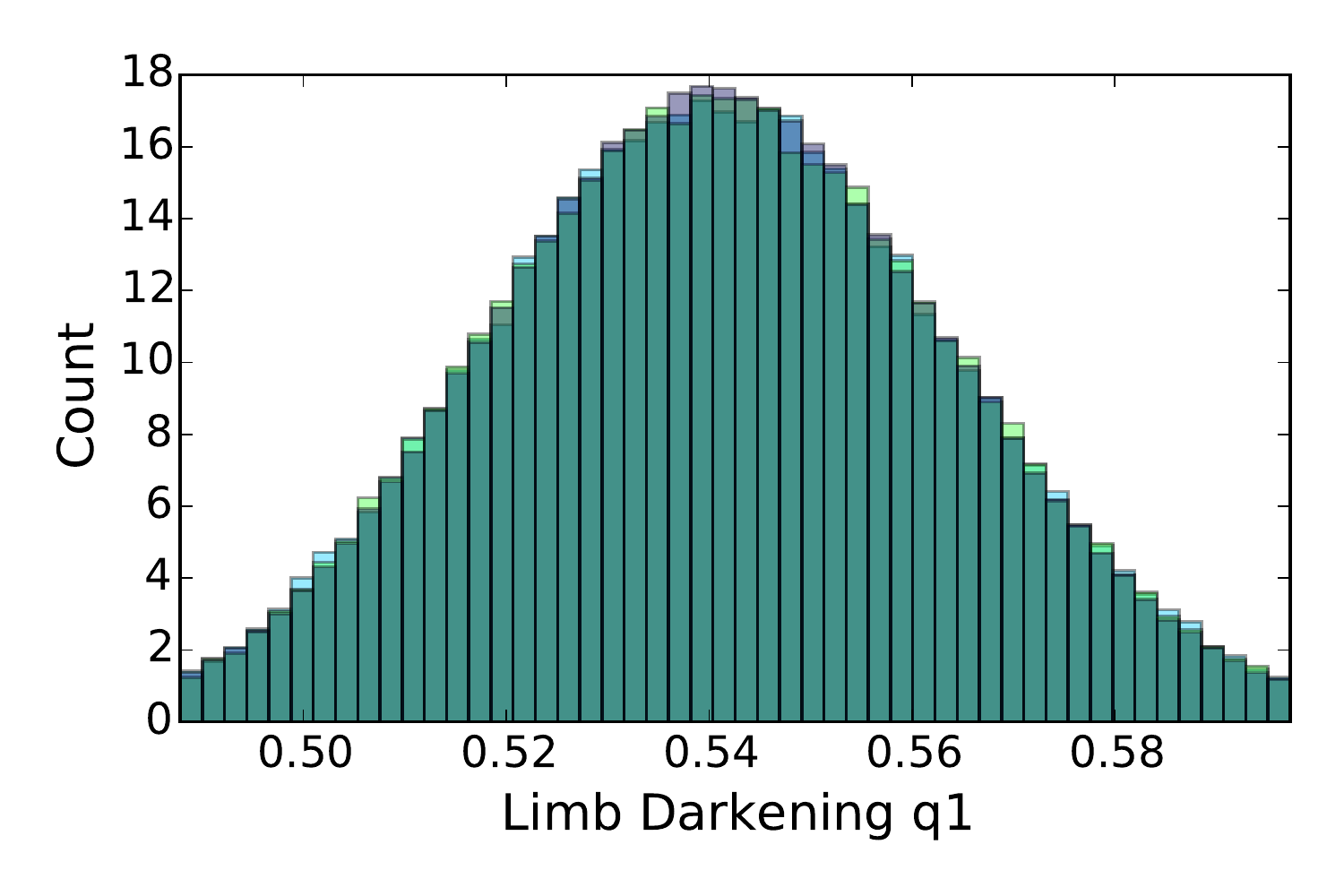} 
\includegraphics[width=0.3\textwidth]{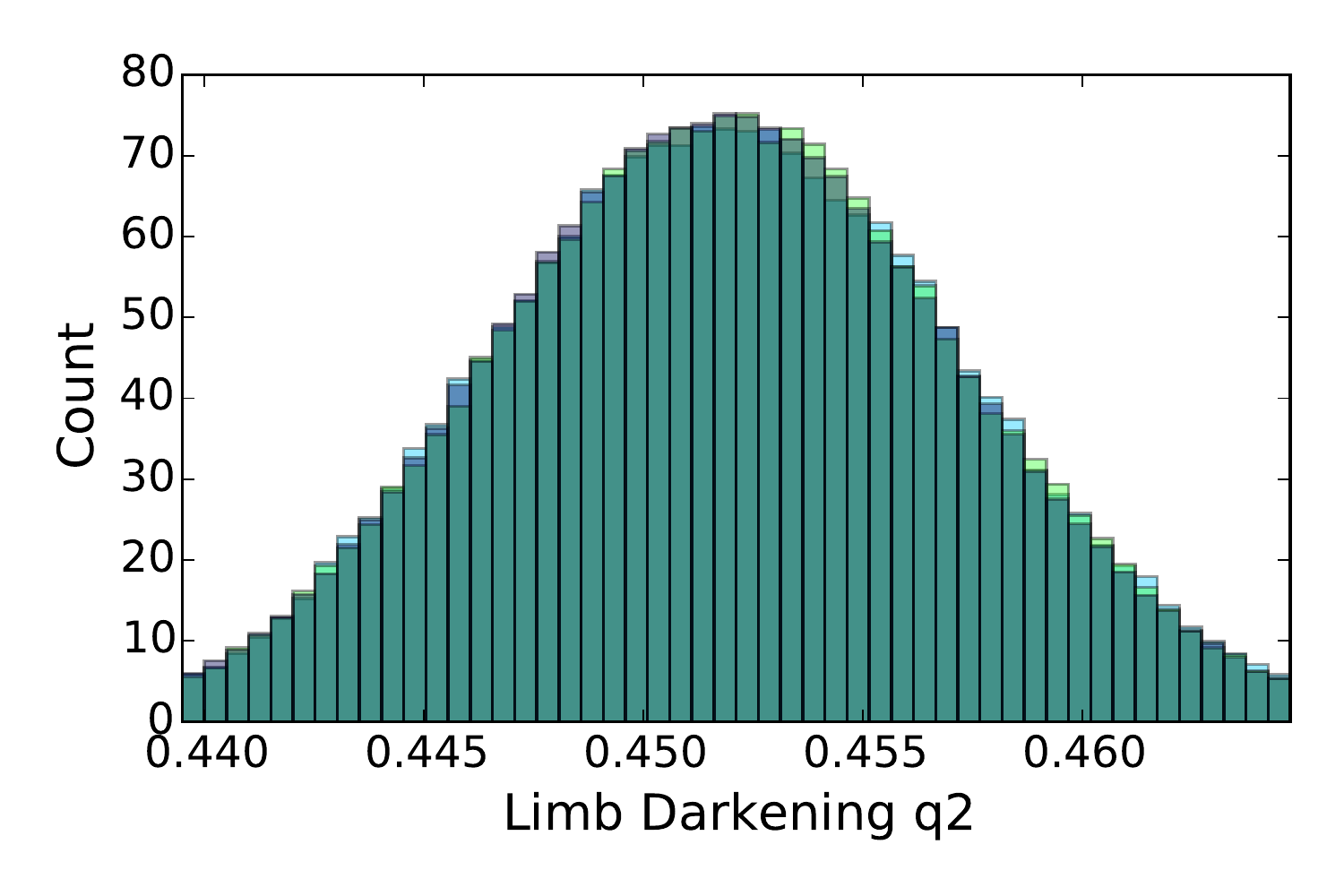} 
\includegraphics[width=0.3\textwidth]{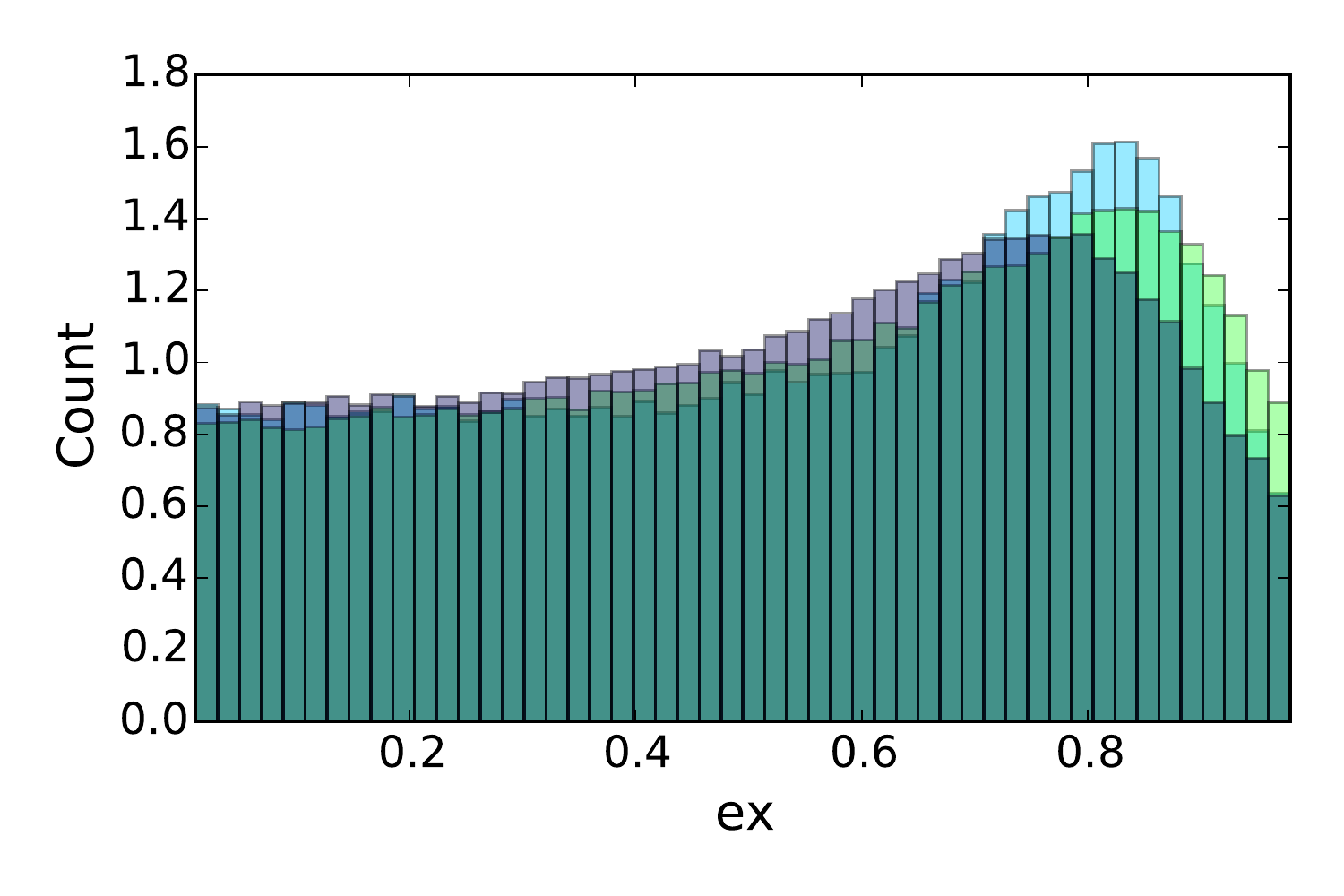} 
\includegraphics[width=0.3\textwidth]{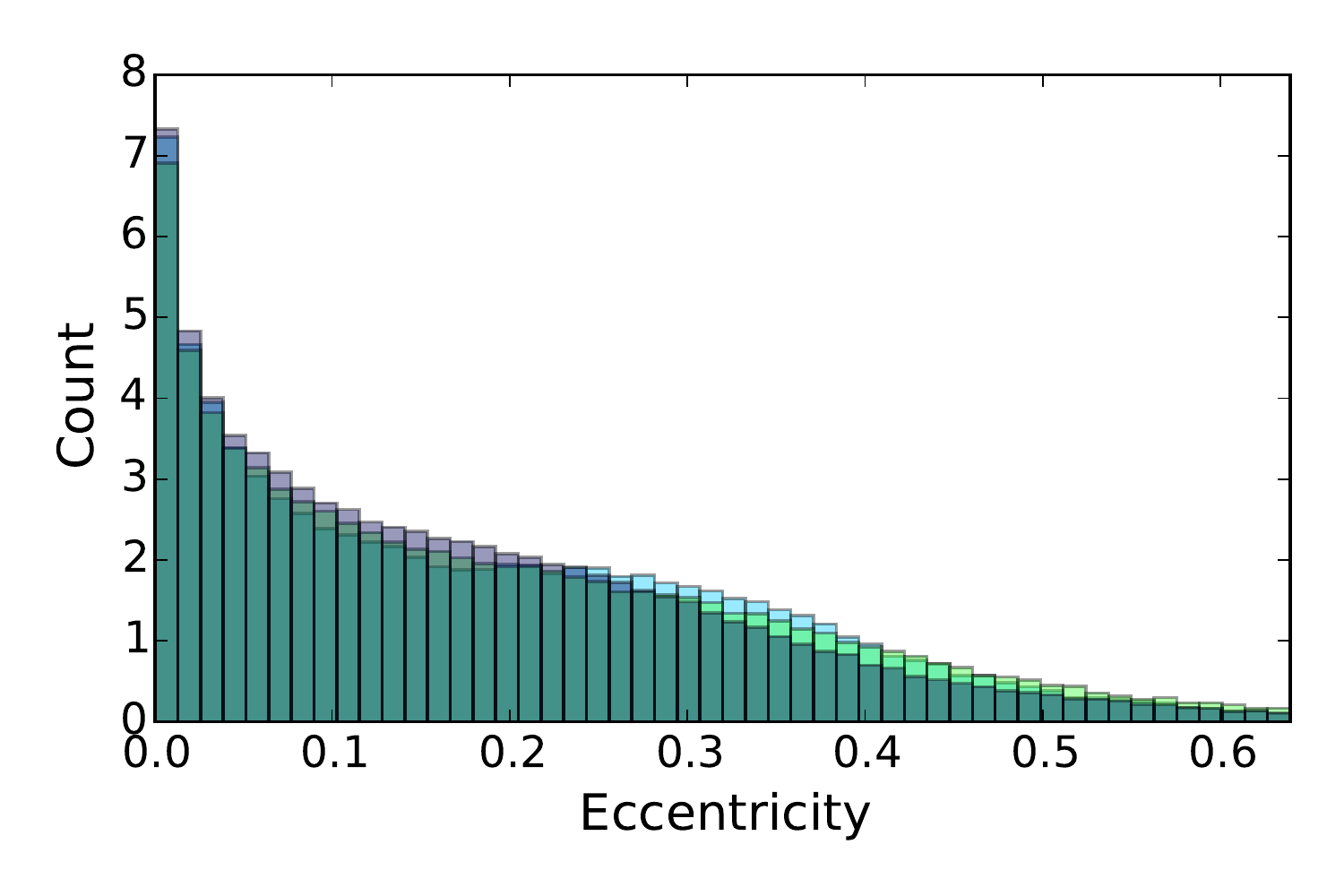} 
\includegraphics[width=0.3\textwidth]{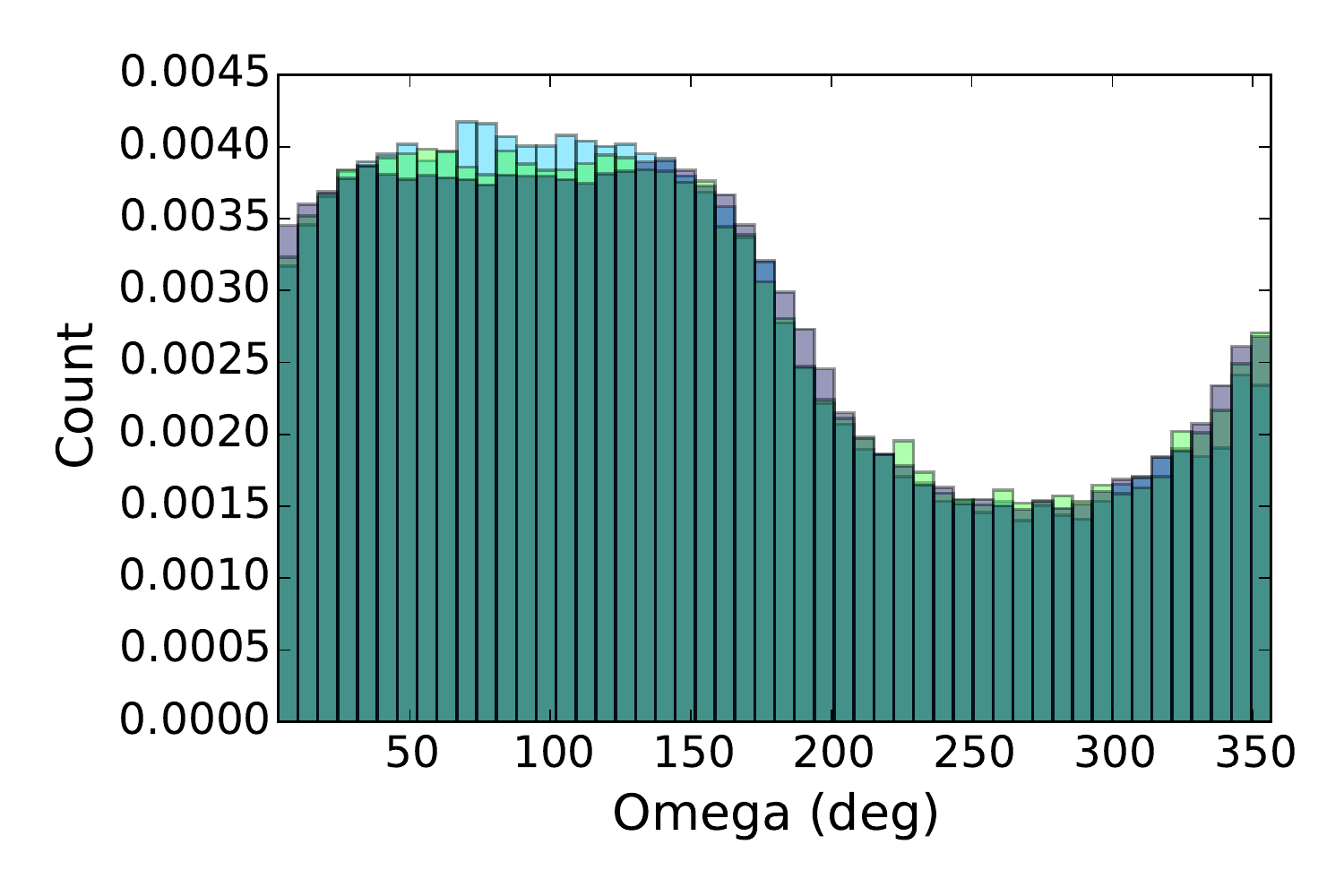} 
\caption{Example set of vetting plots showing transit fits and parameter posterior distributions for EPIC~211680698.01, a previously unknown planet validated in this paper. \emph{Top Row: } K2 data processed by the K2SFF (left), k2phot (middle), and K2SC (right) pipelines and phase-folded to the orbital period of EPIC~211680698.01. The dark blue lines indicate the median transit model fit to each version of the photometry and the medium (light) blue mark $1\sigma$ ($2\sigma$) contours.  \emph{Second Row: } Same as top row, but the K2 data points are binned to better reveal the quality of the transit fits. \emph{Third Row, Left Column: } Comparison of the planet/star radius ratios and orbital periods found when fitting K2SFF (light blue), K2SC (green), and k2phot (dark purple) photometry. \emph{All Remaining Panels: } Posterior probability distributions for the indicated transit parameter. The translucent histograms compare the results produced by fitting K2 photometry processed using K2SFF (light blue), K2SC (light green), and k2phot (gray).}
\label{fig:vet1} 
\end{figure*}

\begin{figure*}[hbp]
\centering
\includegraphics[width=0.3\textwidth]{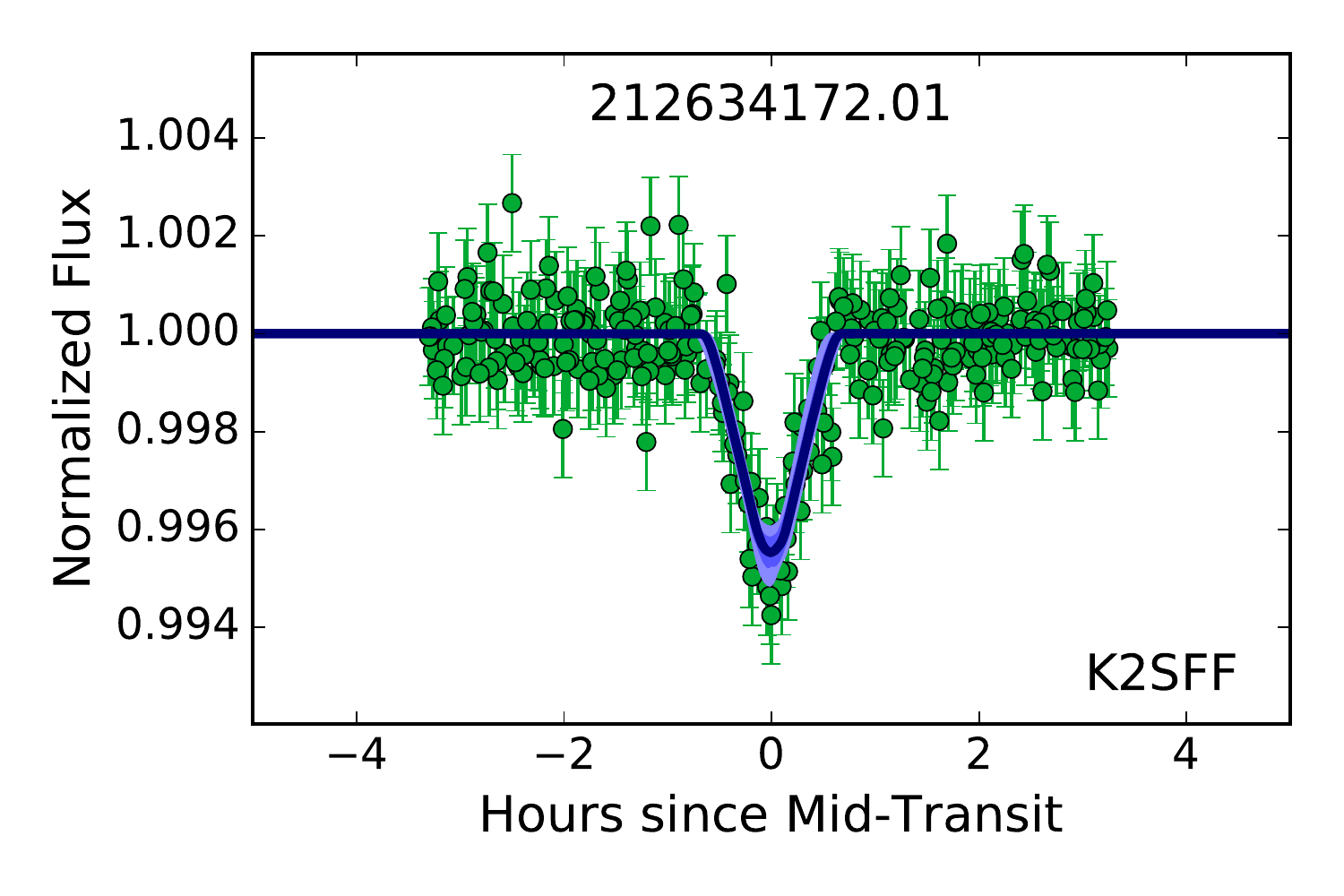} 
\includegraphics[width=0.3\textwidth]{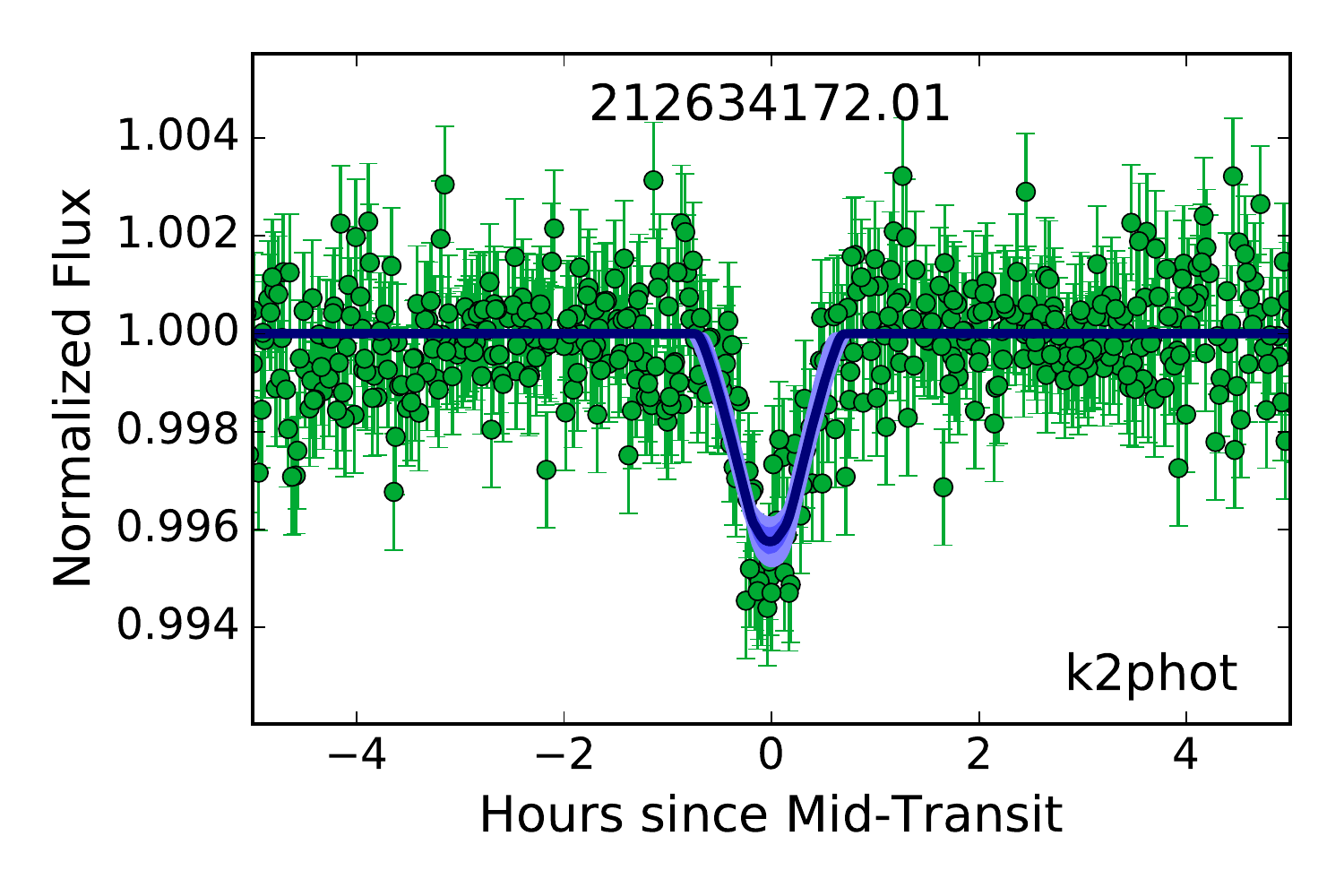} 
\includegraphics[width=0.3\textwidth]{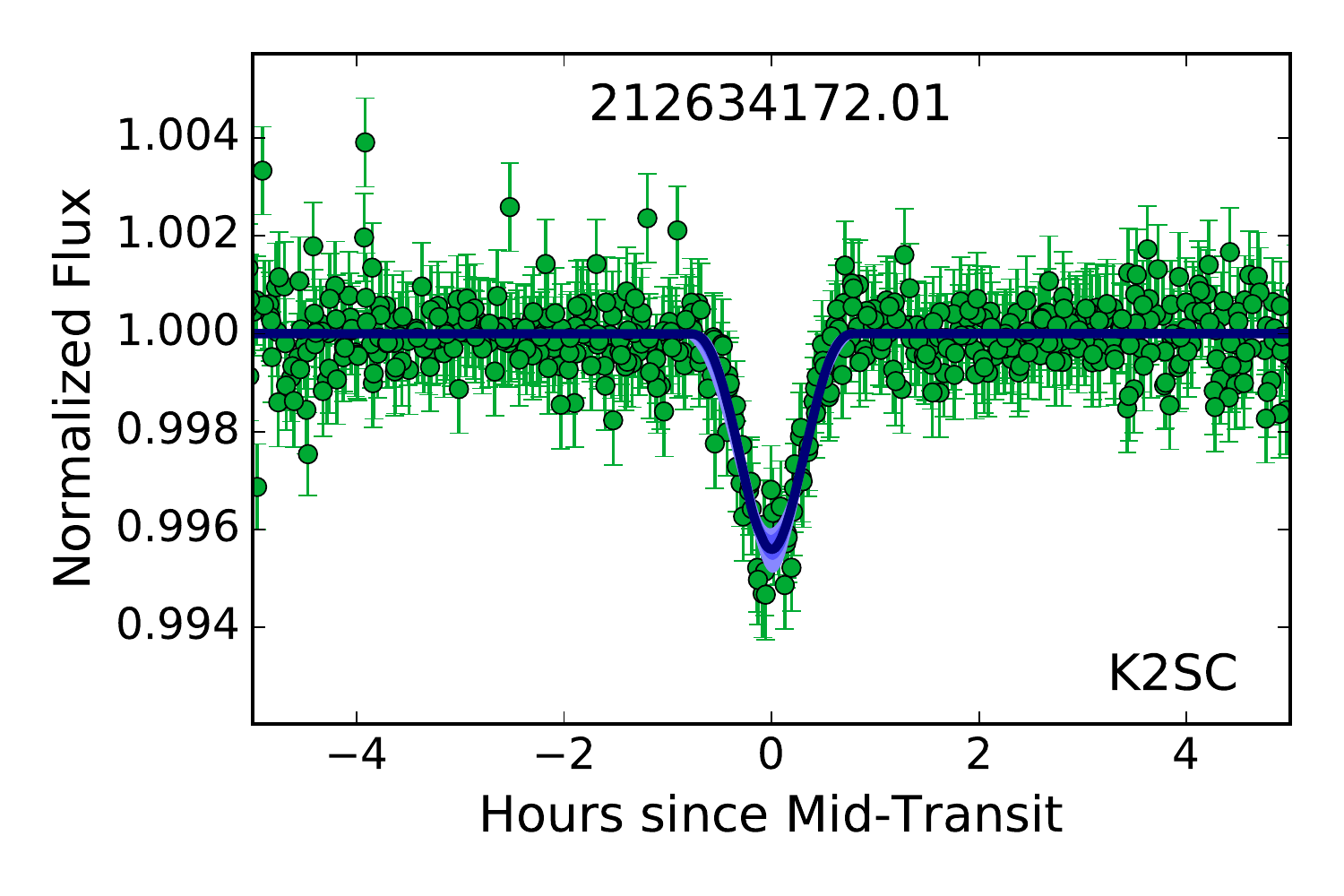} 
\includegraphics[width=0.3\textwidth]{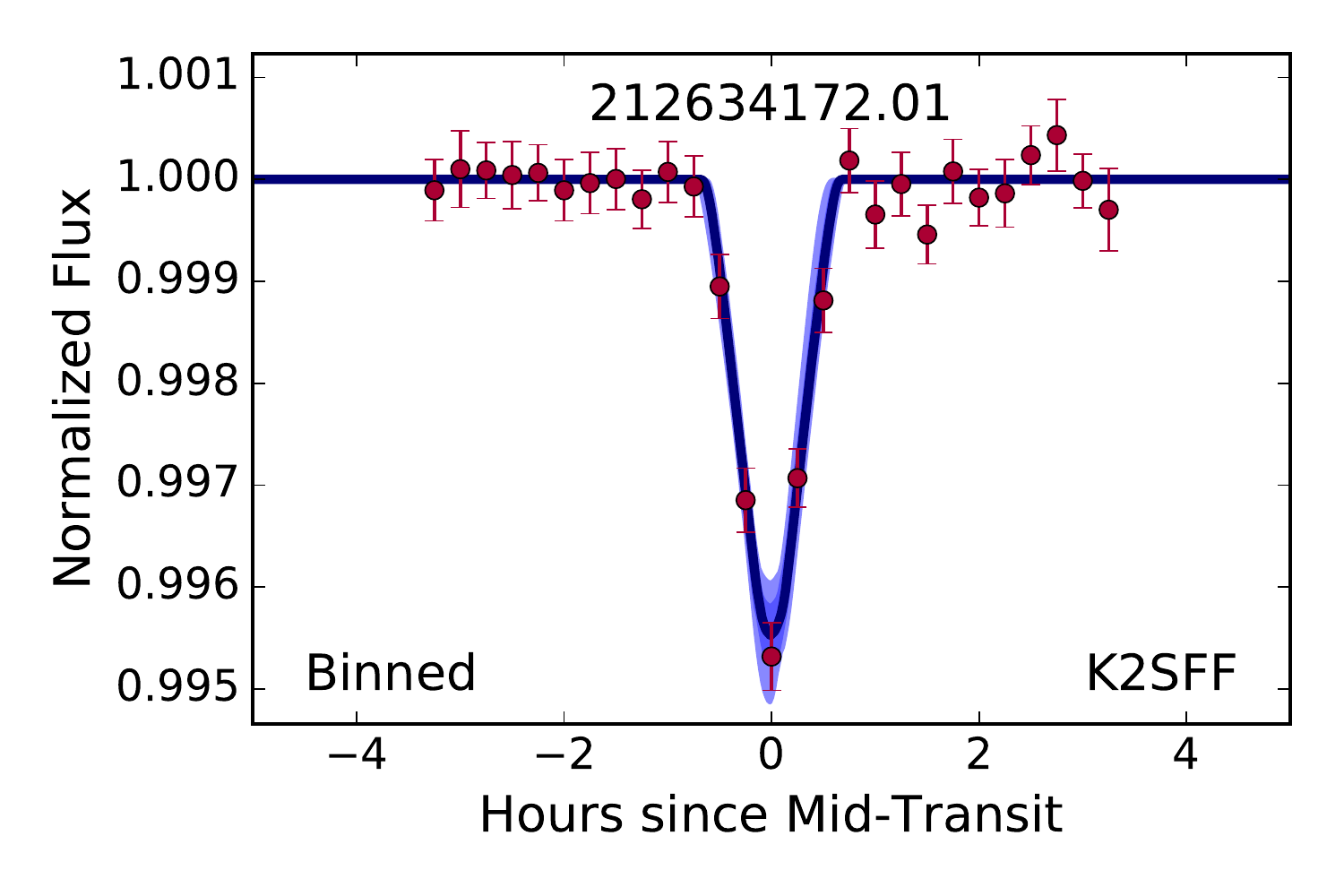} 
\includegraphics[width=0.3\textwidth]{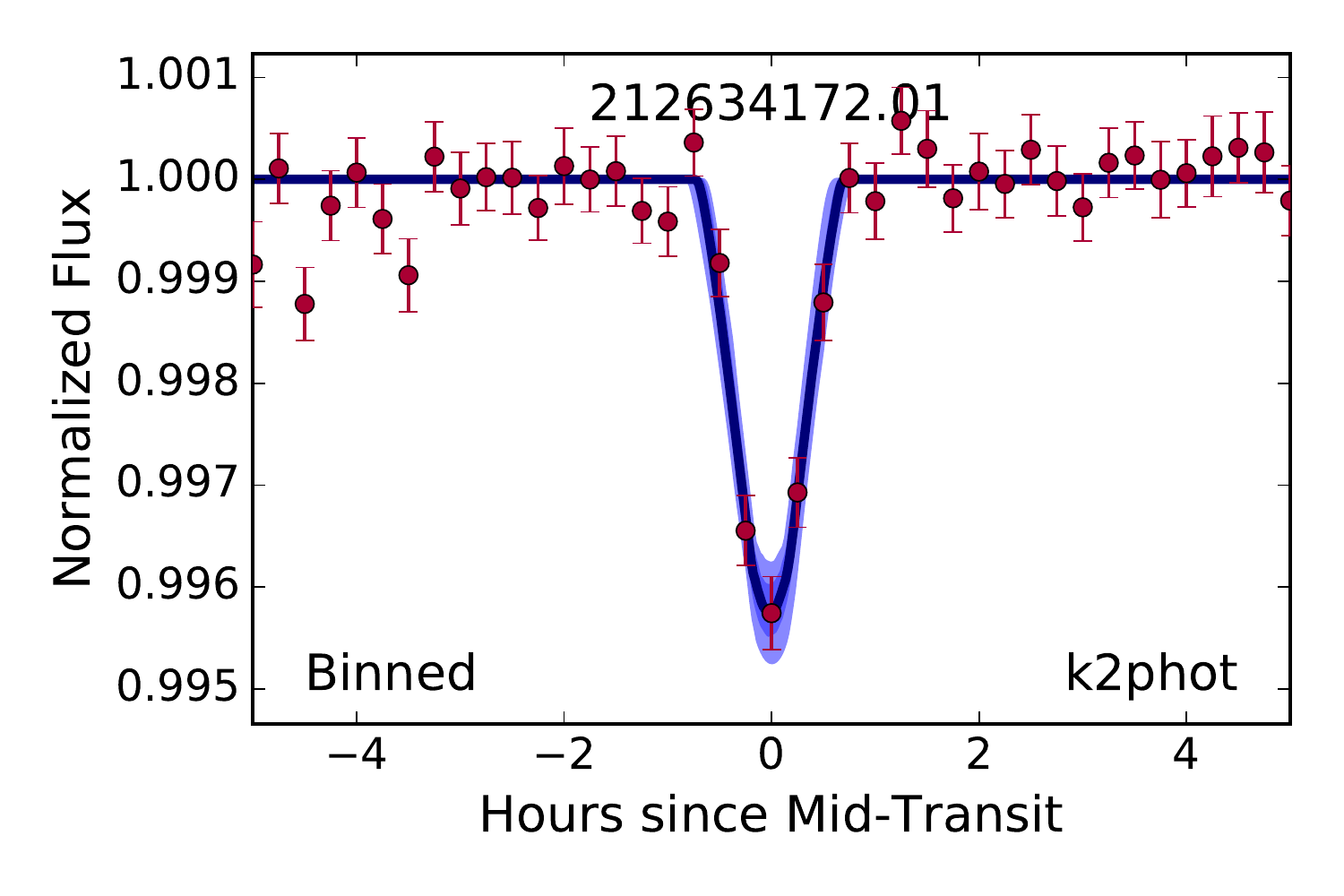} 
\includegraphics[width=0.3\textwidth]{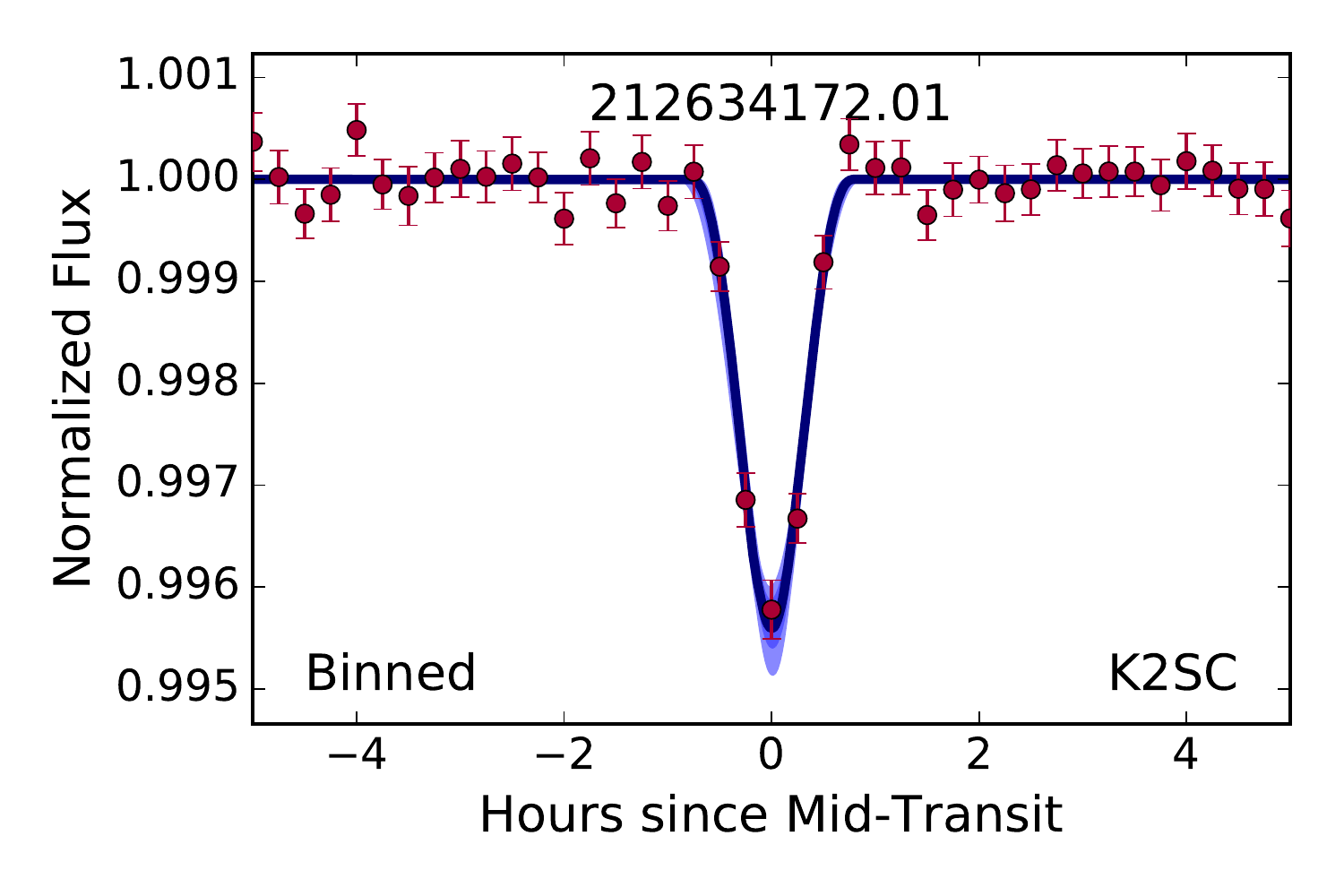} 
\includegraphics[width=0.3\textwidth]{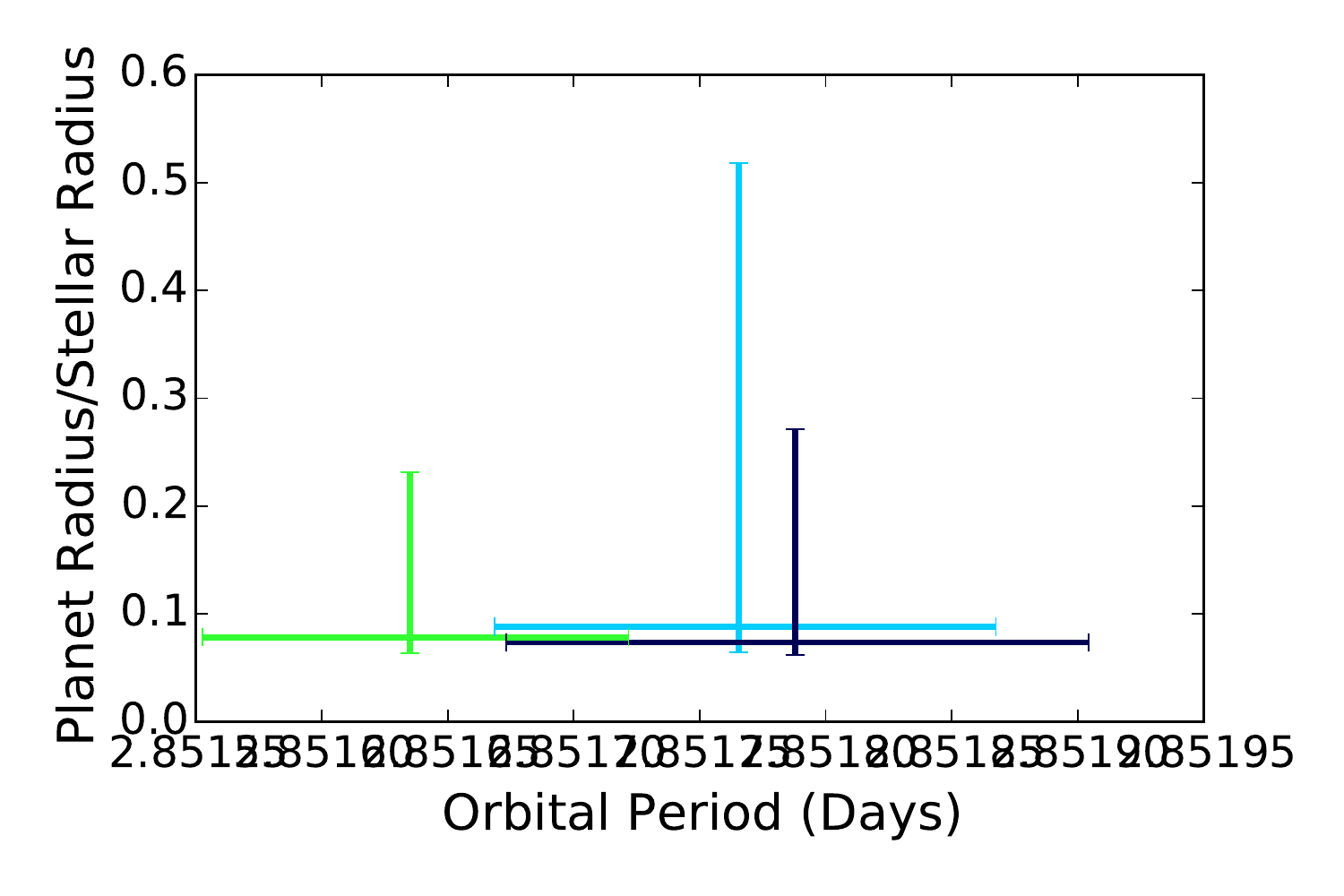} 
\includegraphics[width=0.3\textwidth]{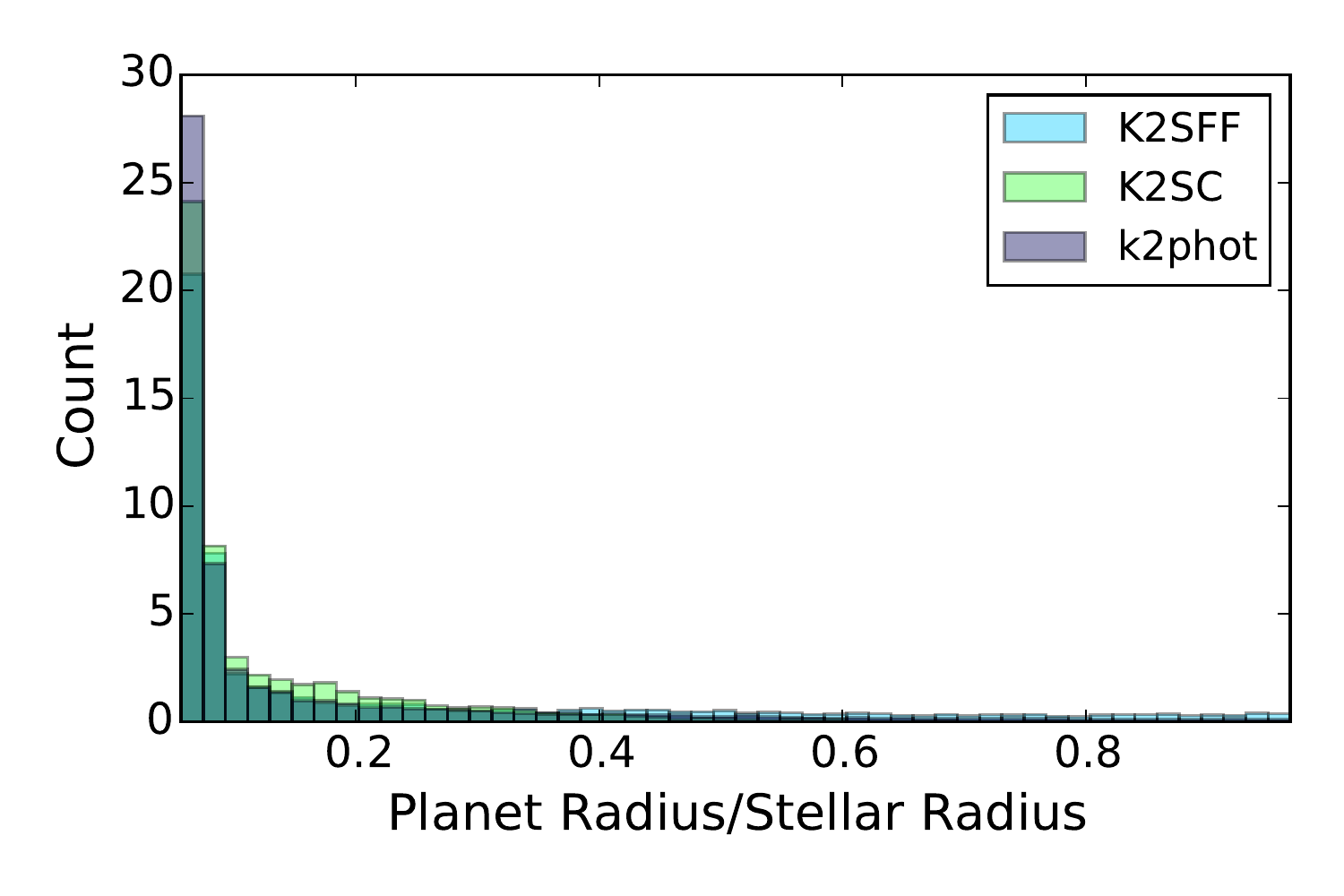} 
\includegraphics[width=0.3\textwidth]{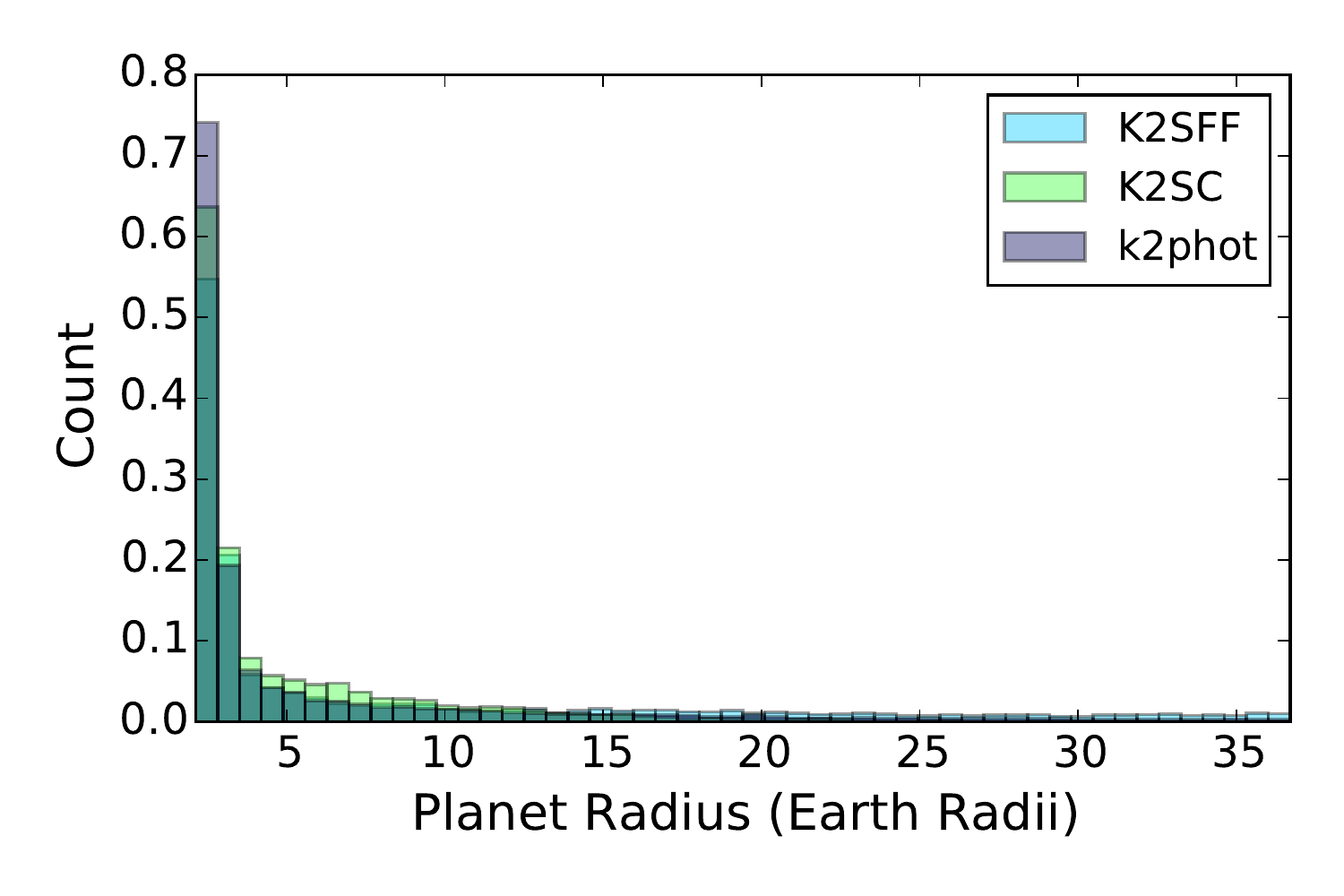} 
\includegraphics[width=0.3\textwidth]{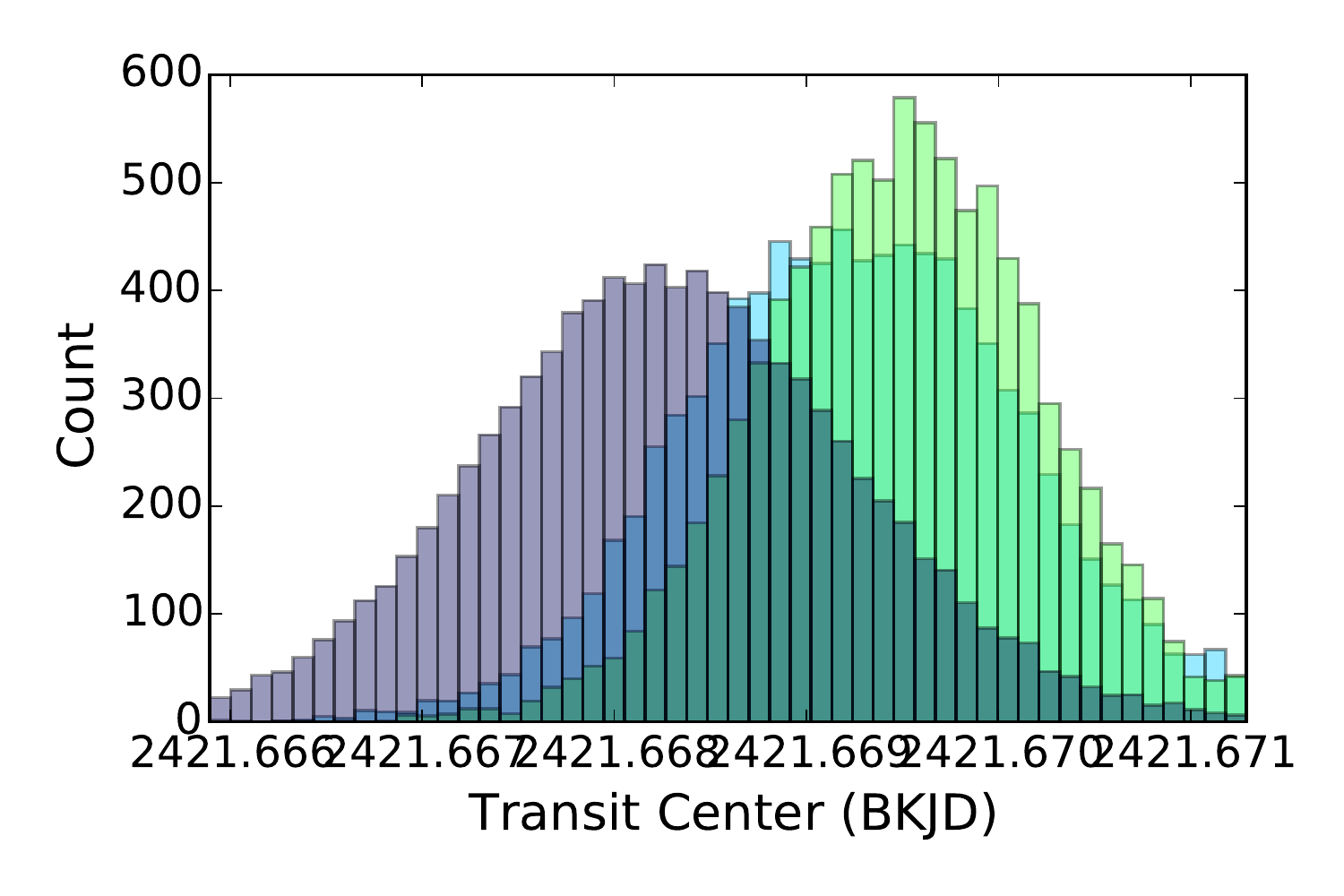} 
\includegraphics[width=0.3\textwidth]{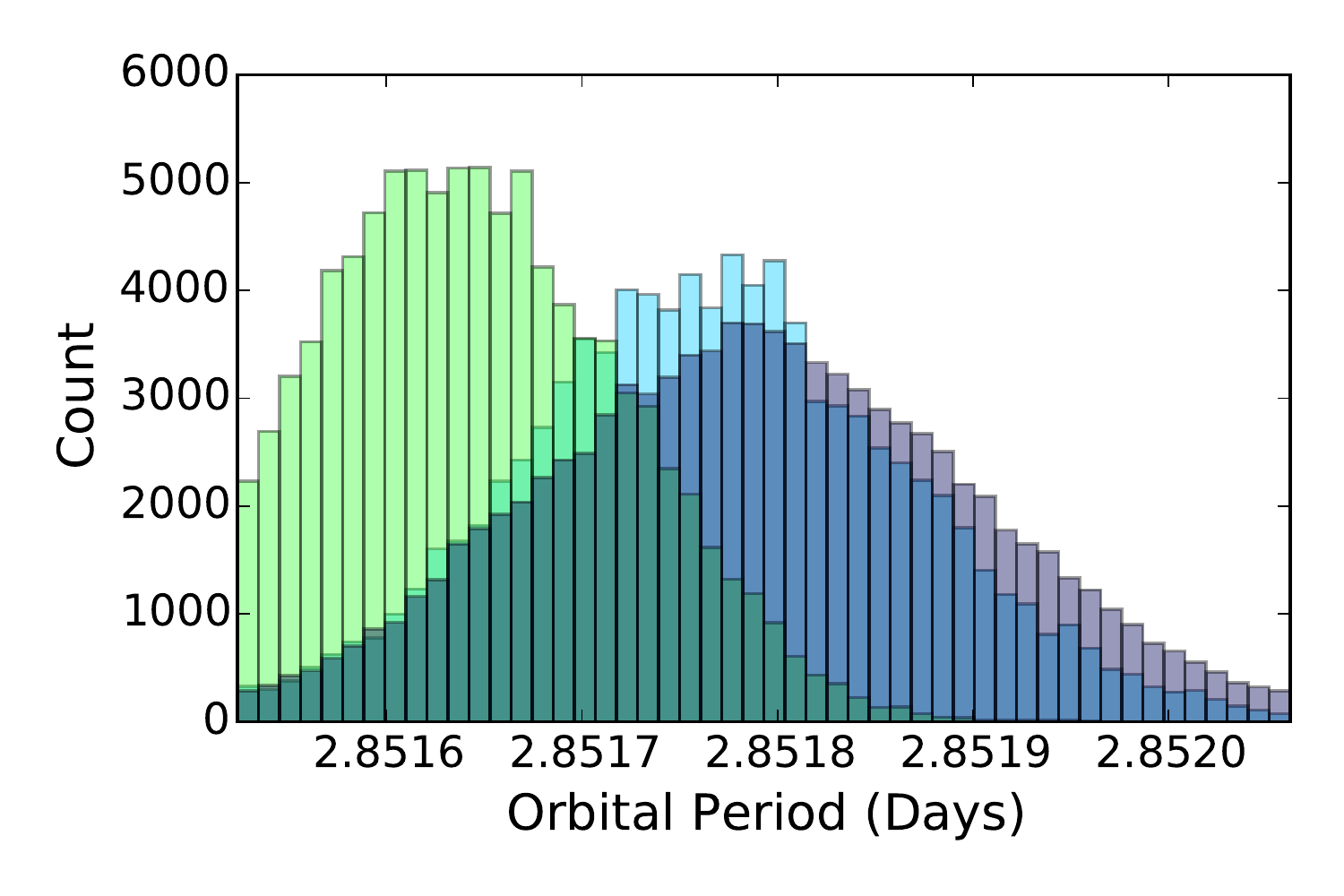} 
\includegraphics[width=0.3\textwidth]{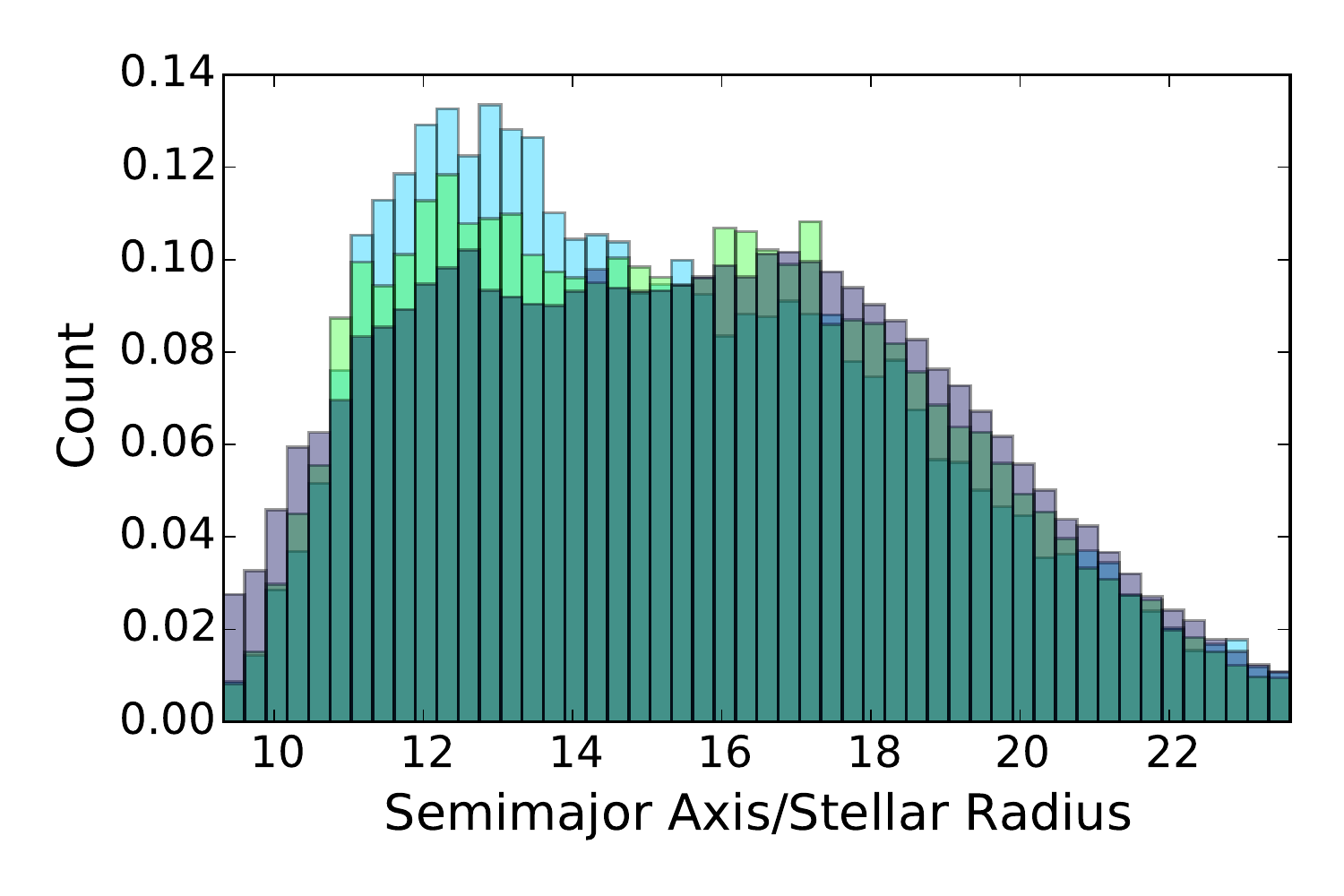} 
\includegraphics[width=0.3\textwidth]{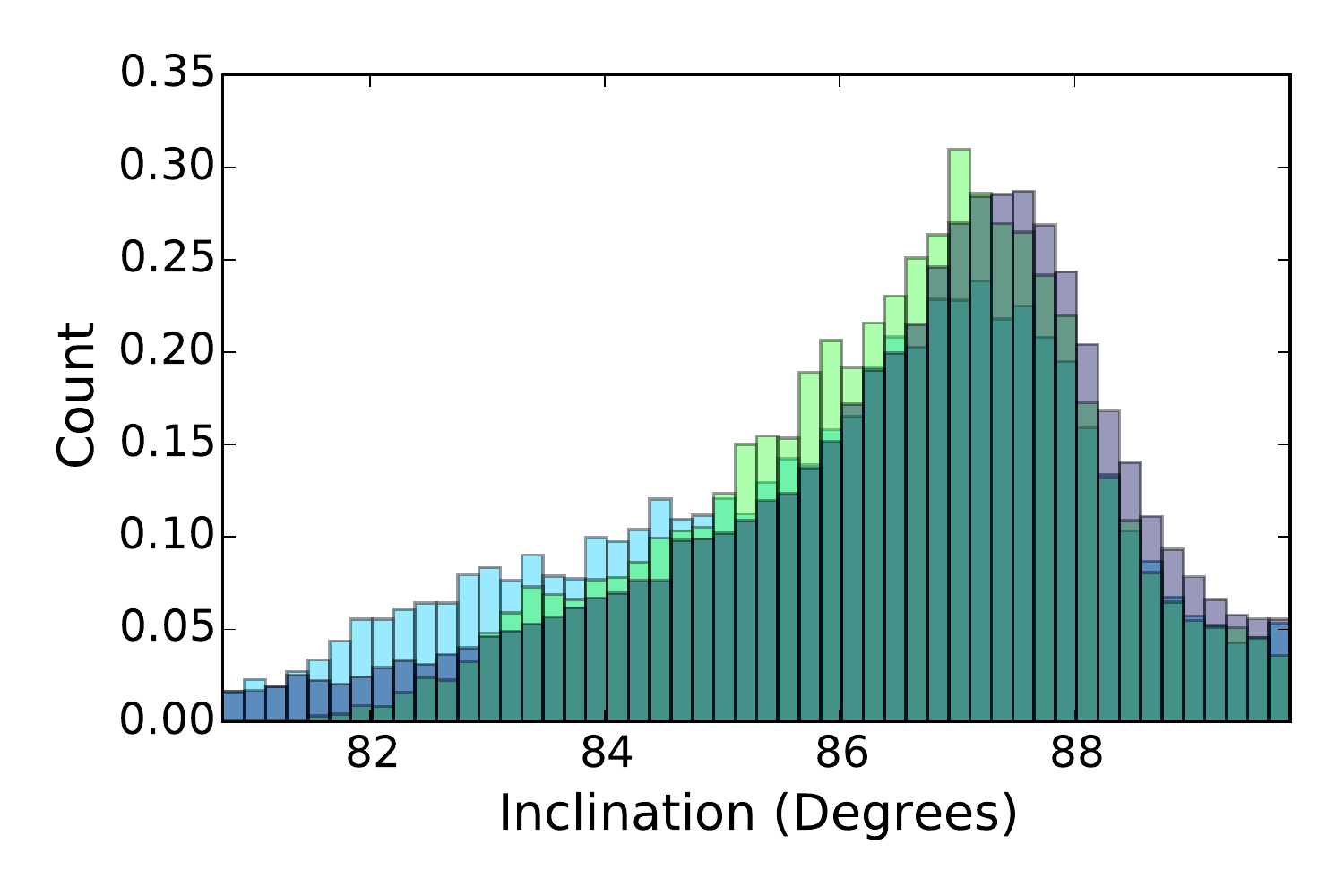} 
\includegraphics[width=0.3\textwidth]{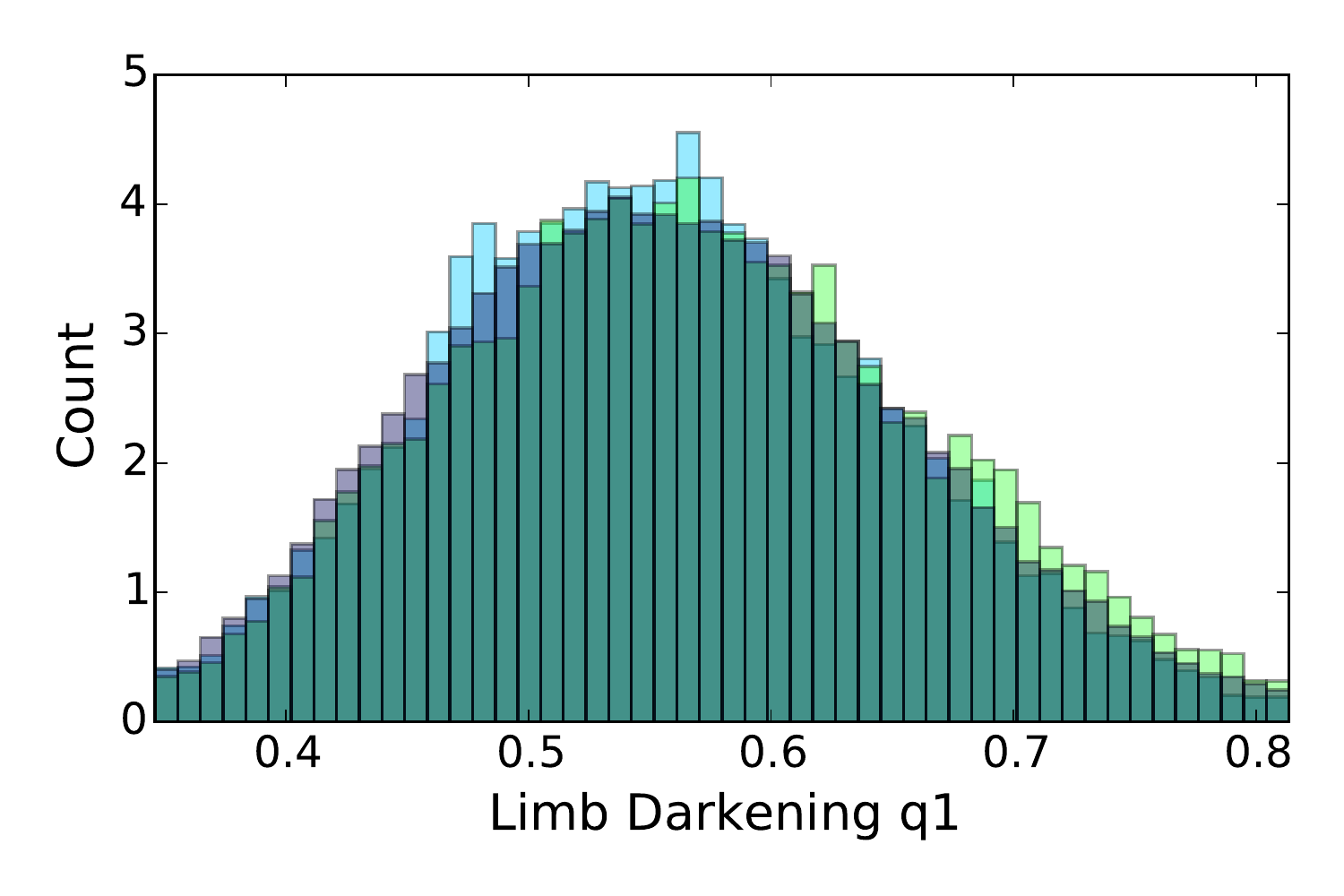} 
\includegraphics[width=0.3\textwidth]{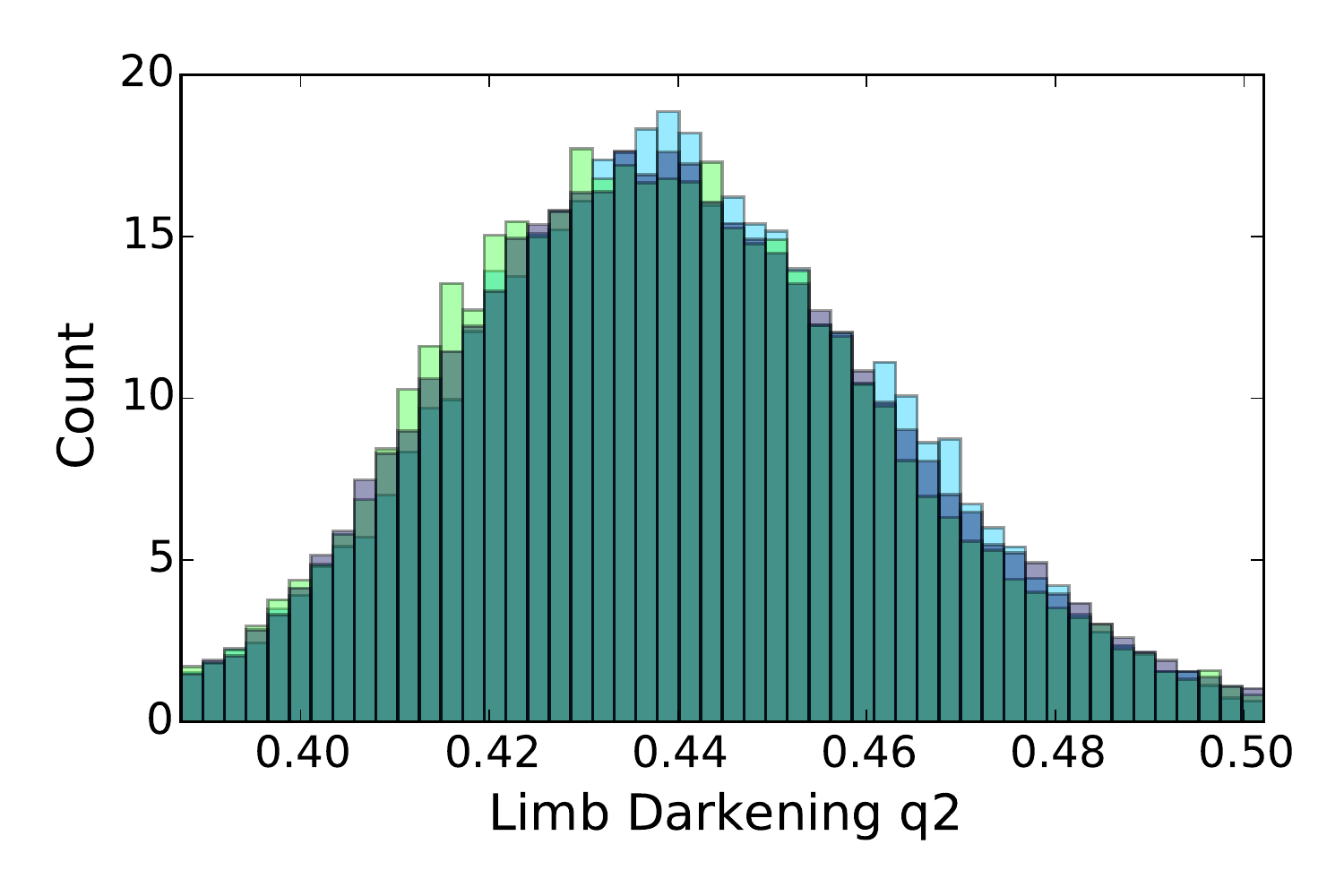} 
\includegraphics[width=0.3\textwidth]{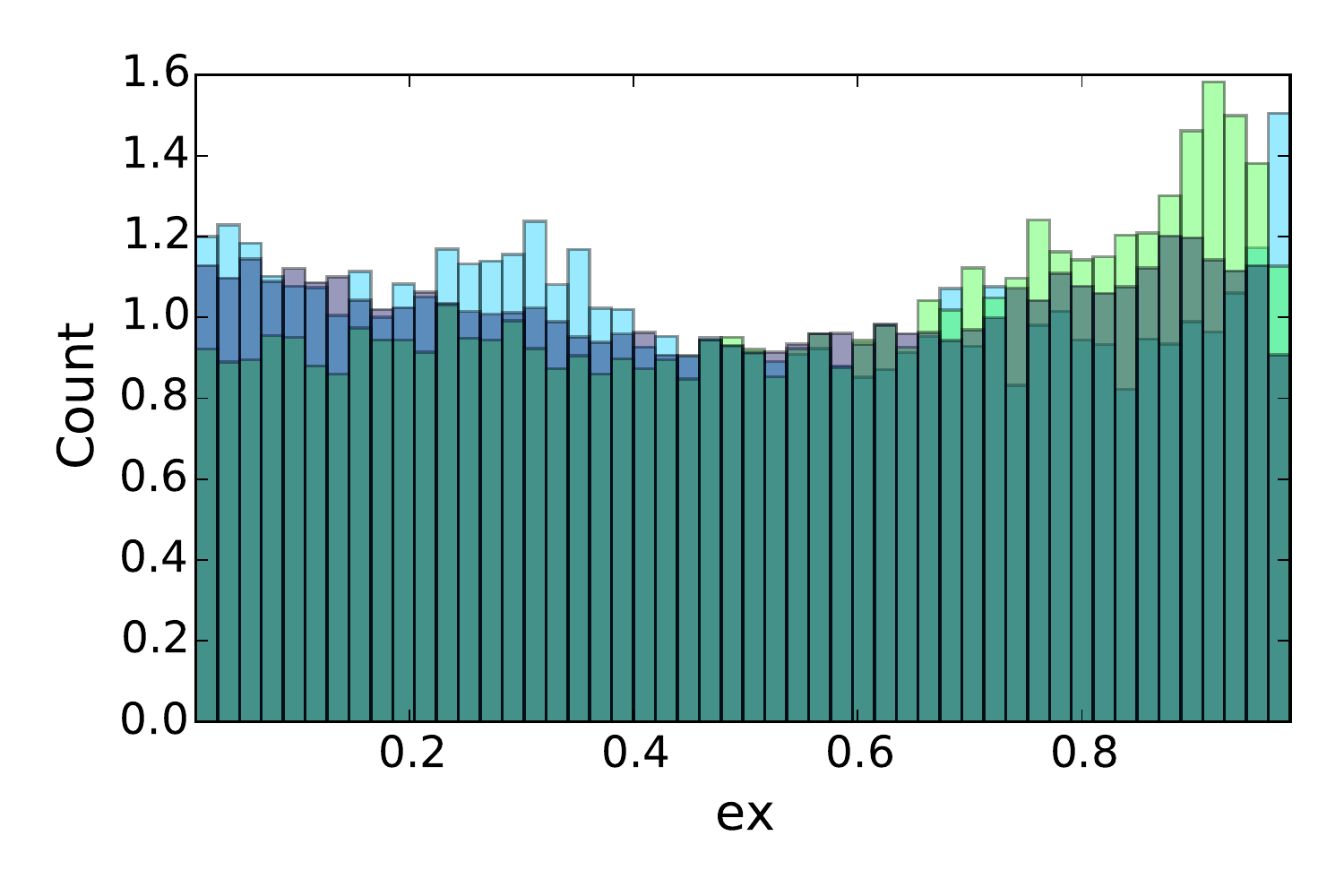} 
\includegraphics[width=0.3\textwidth]{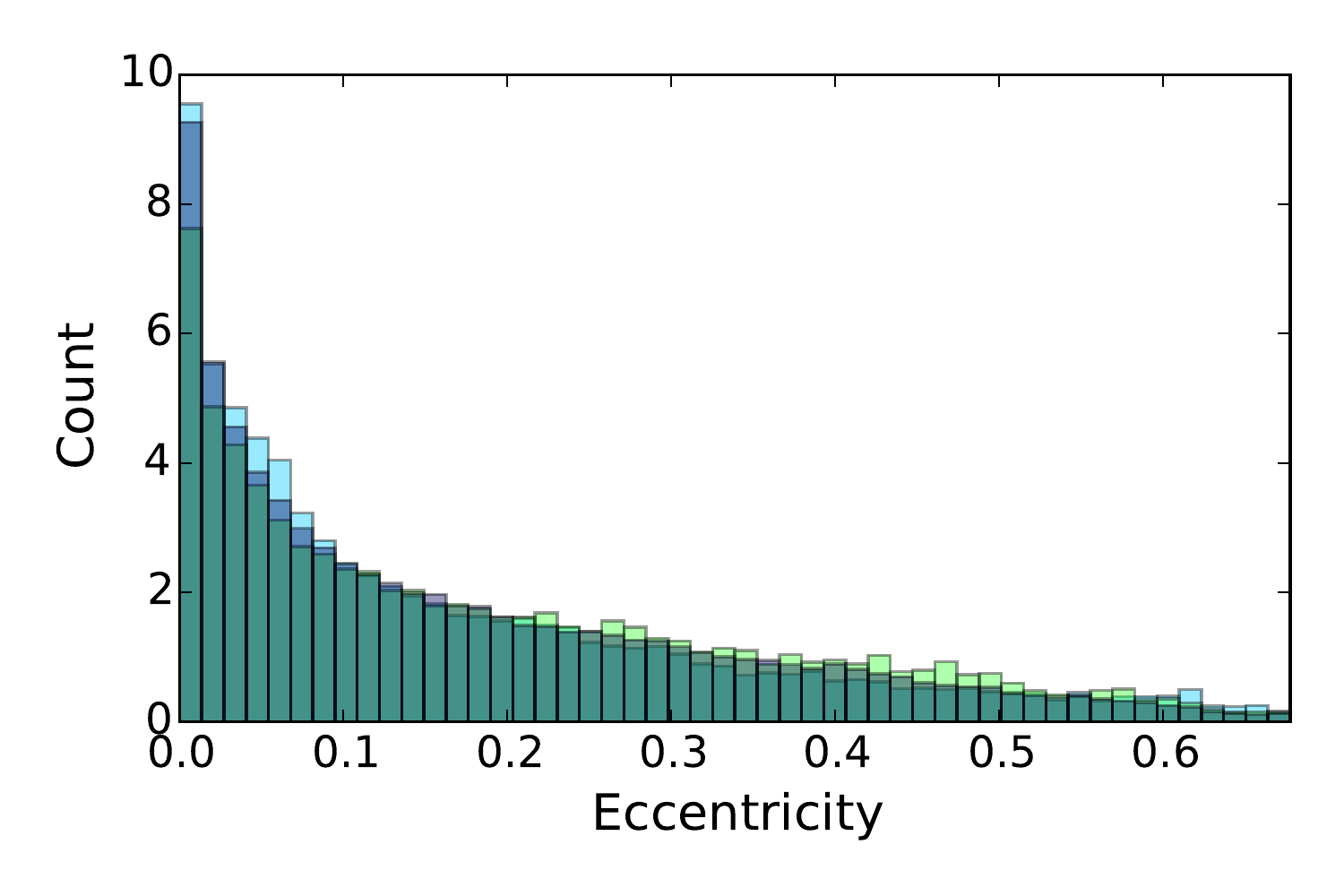} 
\includegraphics[width=0.3\textwidth]{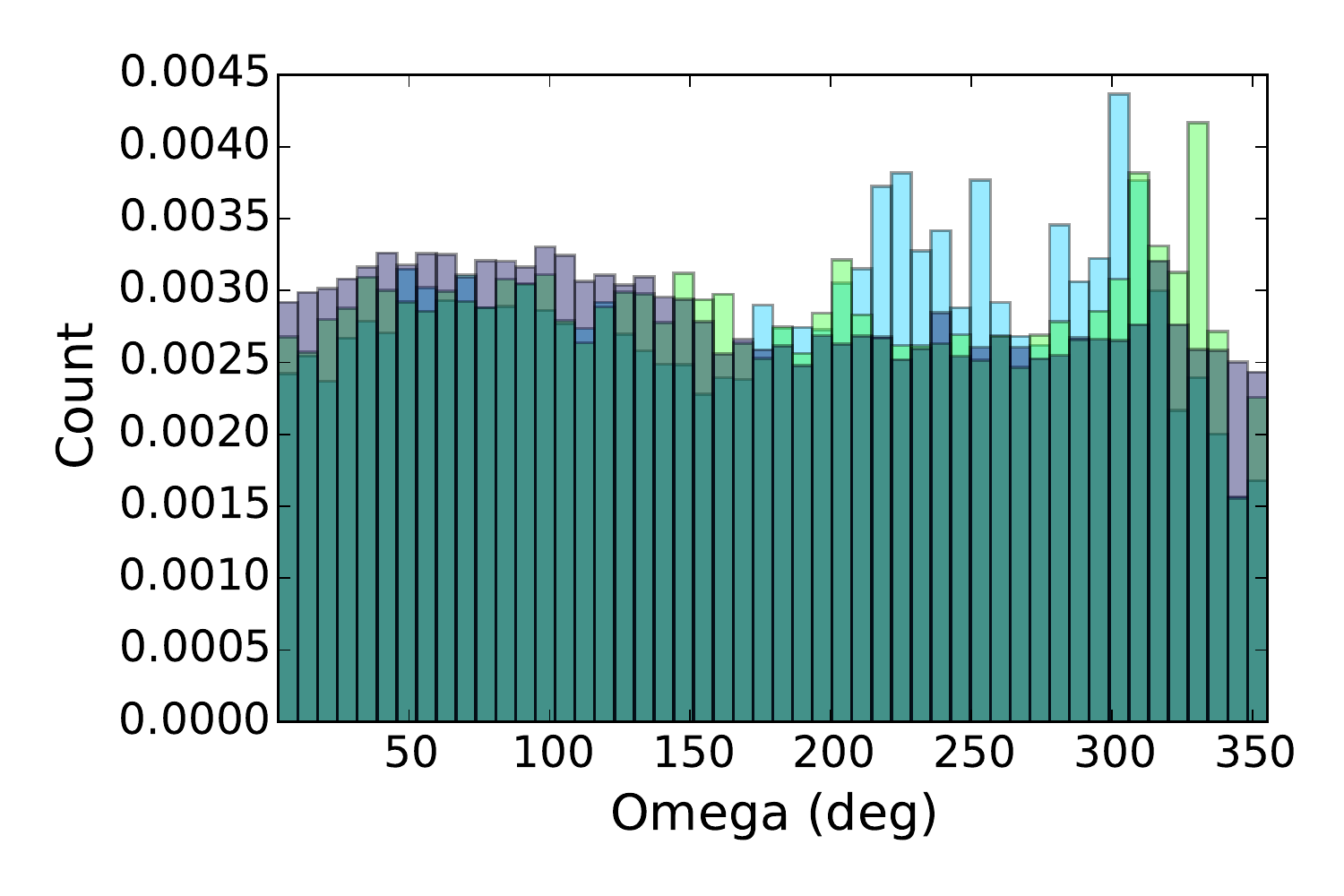} 
\caption{Same as Figure~\ref{fig:vet1} but for the newly detected planet candidate EPIC~212634172.01.}
\label{fig:vet2} 
\end{figure*}

\begin{figure*}[hbp]
\centering
\includegraphics[width=0.3\textwidth]{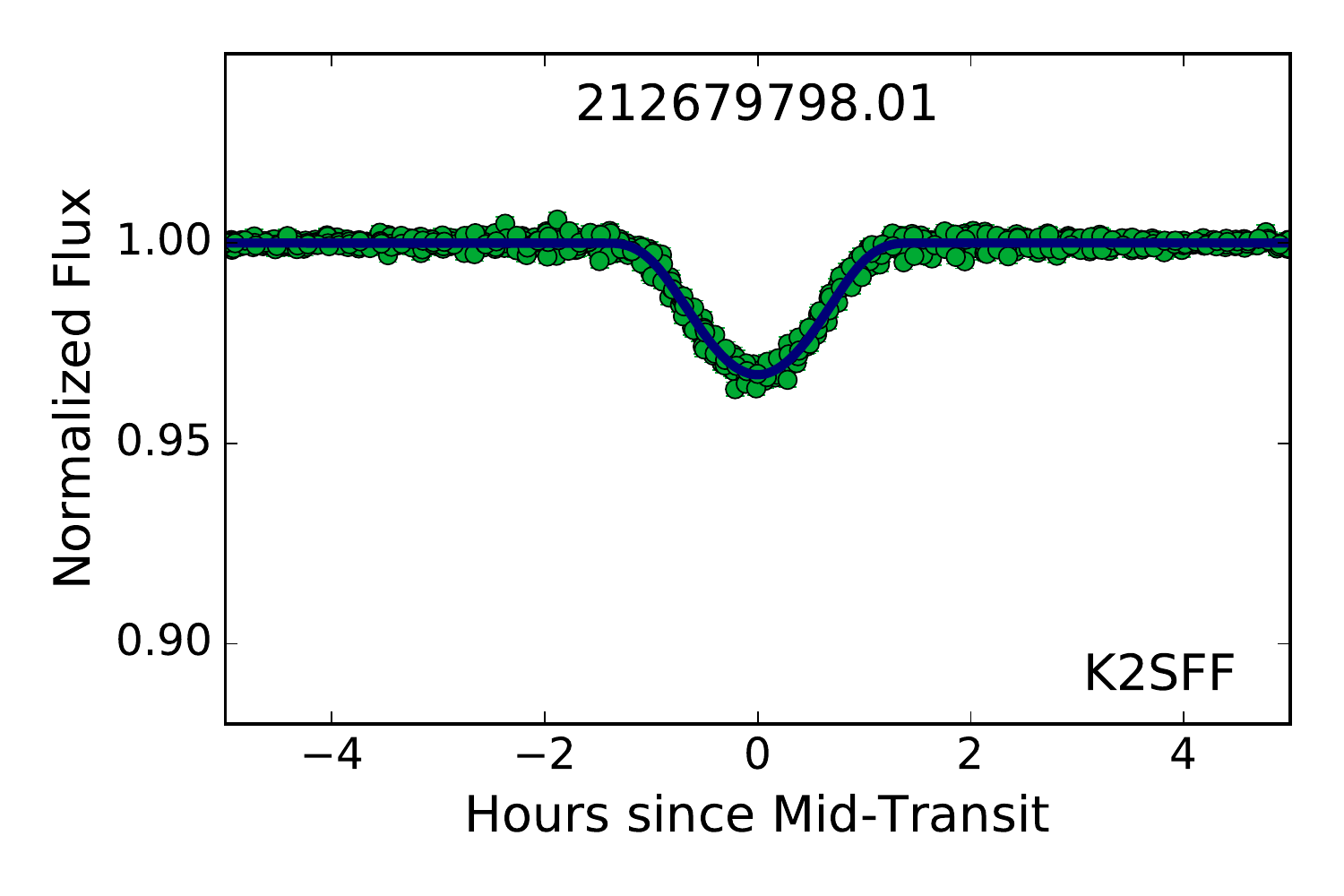} 
\includegraphics[width=0.3\textwidth]{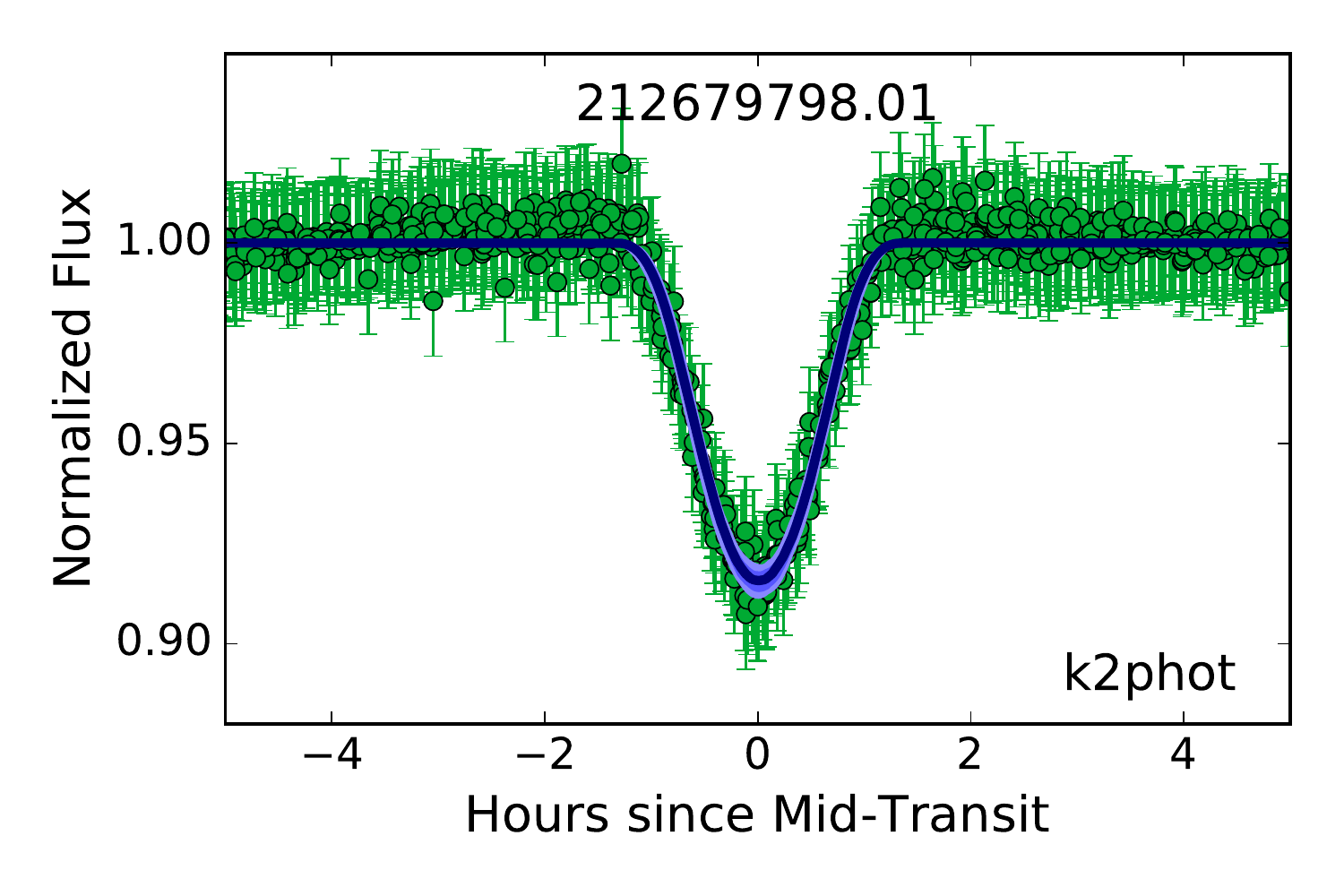} 
\includegraphics[width=0.3\textwidth]{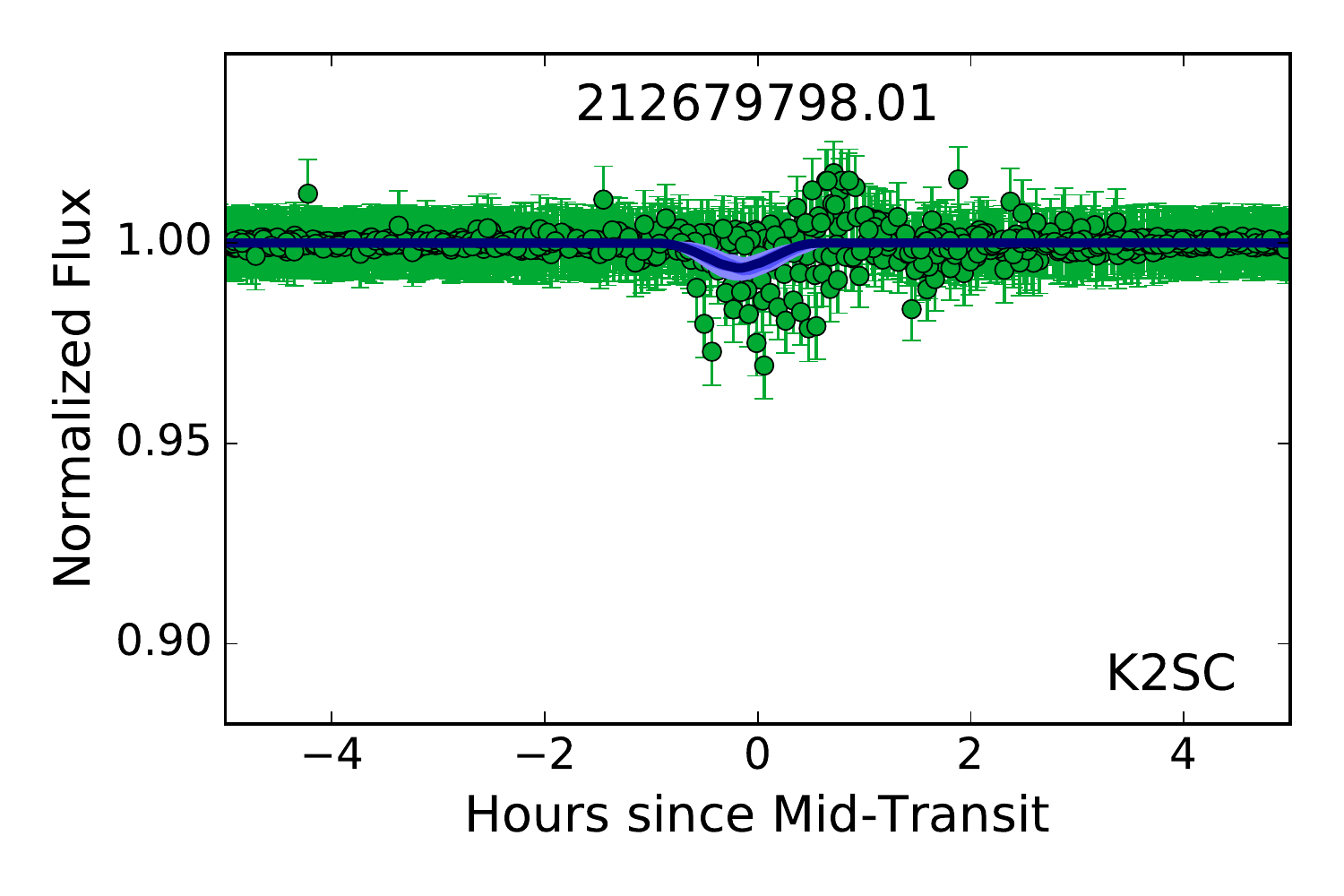} 
\includegraphics[width=0.3\textwidth]{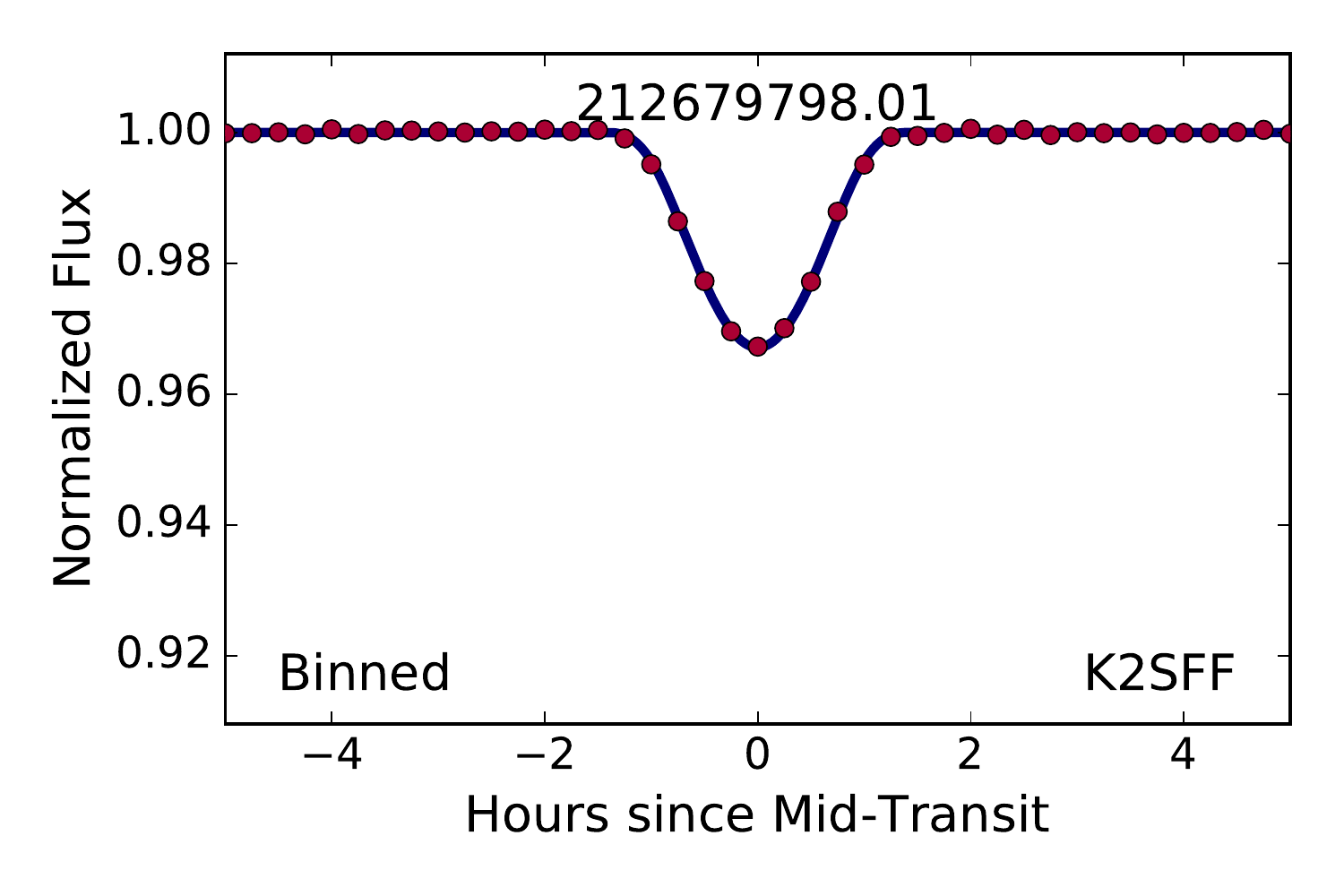} 
\includegraphics[width=0.3\textwidth]{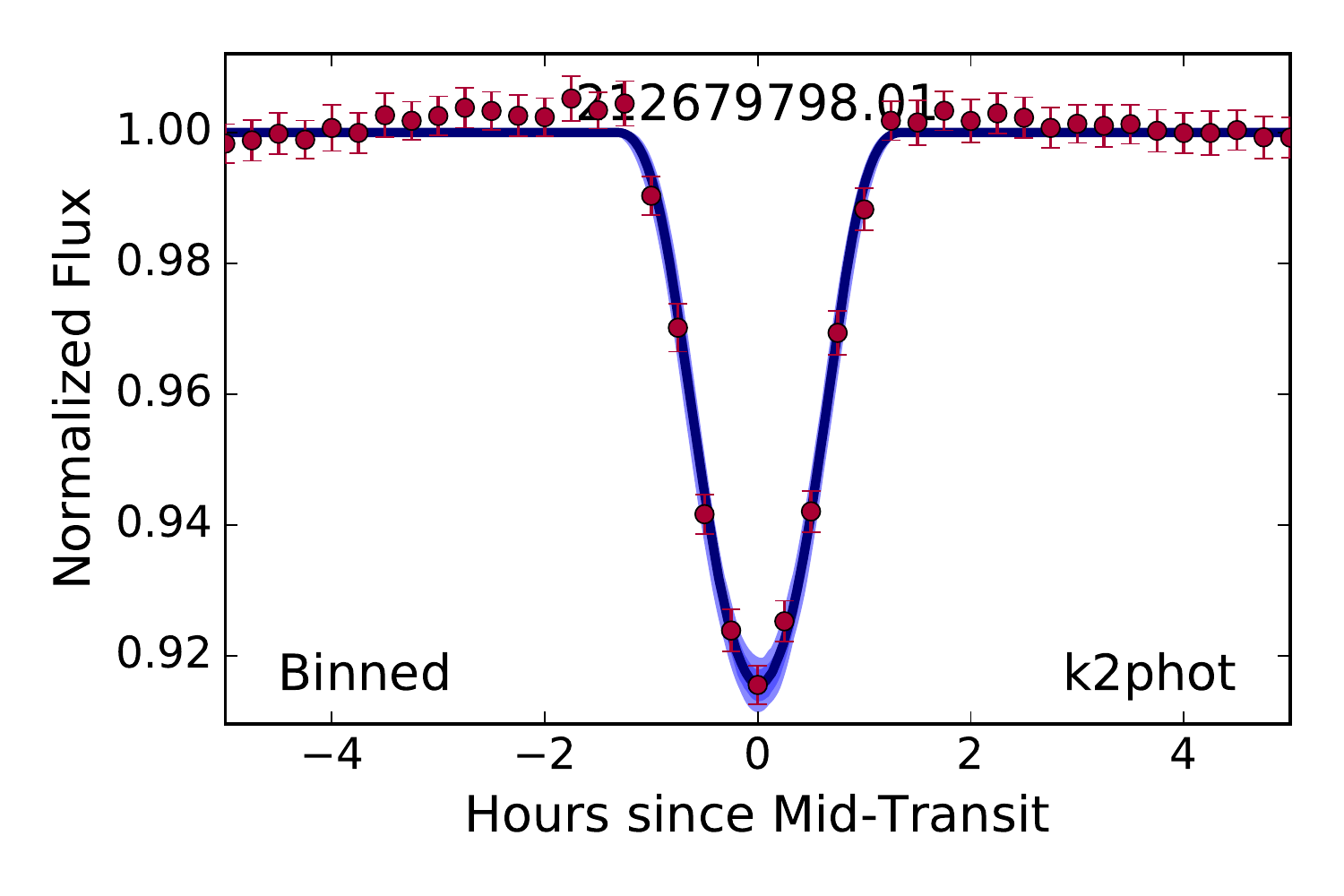} 
\includegraphics[width=0.3\textwidth]{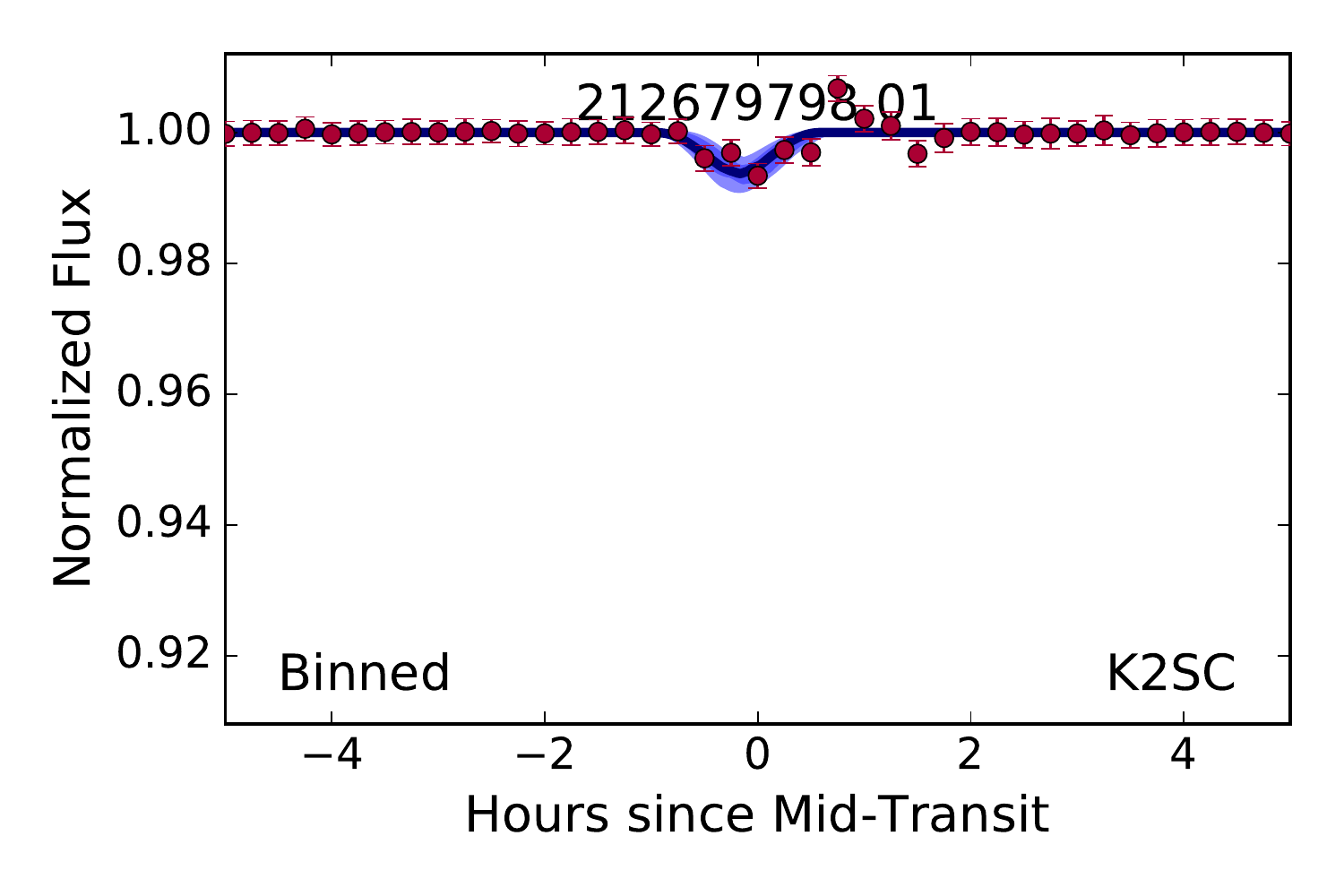} 
\includegraphics[width=1\textwidth]{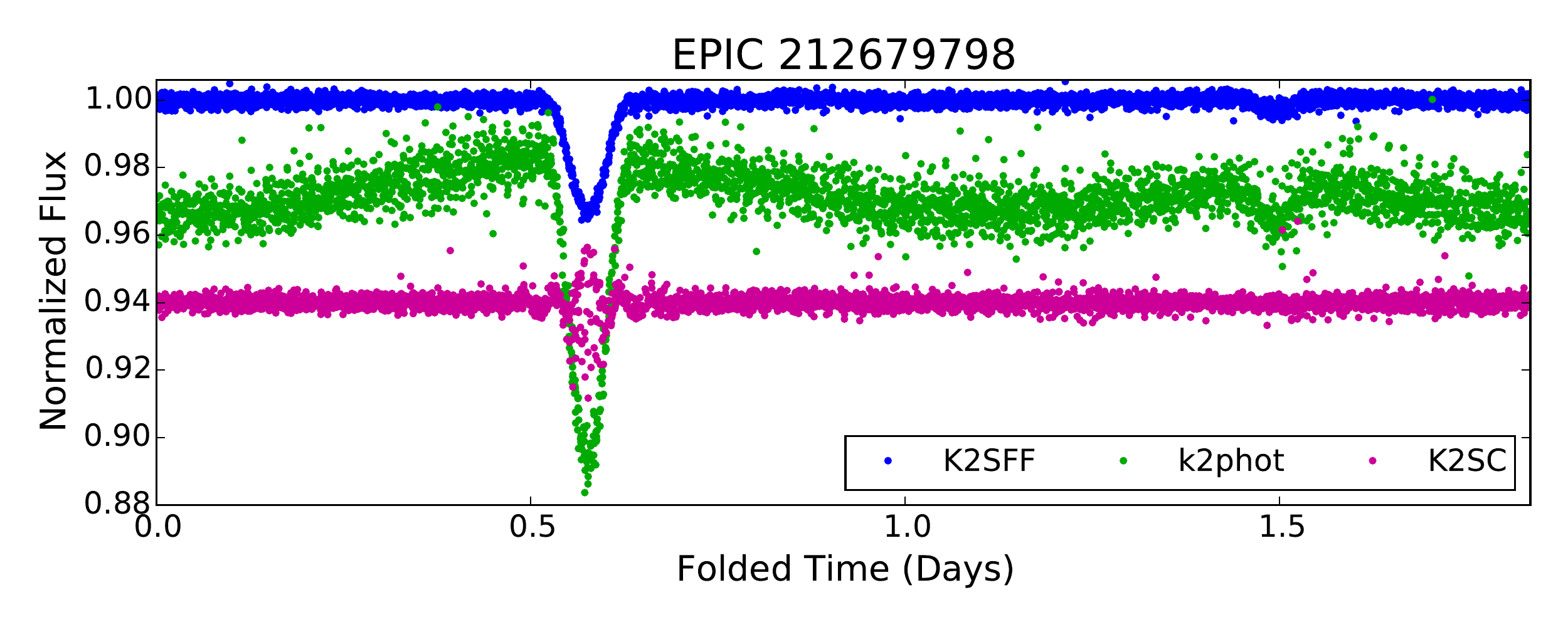} 
\caption{\emph{Top Two Rows: } Same as Figure~\ref{fig:vet1} but for newly rejected false positive EPIC 212679798.01. \emph{Bottom Row: } Phase-folded light curve of EPIC~212679798 shown over the full orbital period of K2OI~212679798.01. The KSFF (blue, top) and k2phot (green, middle) photometry display a clear secondary eclipse near 1.5~days. The K2SC photometry (magenta, bottom) has removed the secondary eclipse and partially flattened the primary eclipse. Note that the depth of the primary and secondary eclipses are also deeper in the k2phot photometry than in the K2SFF photometry. The event depth discrepancy is likely due to the different default photometric apertures used by both pipelines.}
\label{fig:vet3} 
\end{figure*}

As discussed in Section~\ref{sec:vespa}, we will post transit fits and false positive assessments for all K2OIs on the ExoFOP-K2 website. Figures~\ref{fig:vet1}-\ref{fig:vet3} provide an example set of vetting data products. The K2OIs featured are a newly validated planet (EPIC~211680698.01, Figure~\ref{fig:vet1}), a newly detected planet candidate (EPIC~212634172.01, Figure~\ref{fig:vet2}), and a newly rejected false positive (EPIC~212679798.01, Figure~\ref{fig:vet3}).

\end{document}